\documentclass[prd,nofootinbib,11pt]{revtex4}
\usepackage{amssymb}
\usepackage{amsmath}
\usepackage{mathtools}
\usepackage{graphicx}
\usepackage{hyperref}
\usepackage{cleveref}
\usepackage{dsfont}
\usepackage{graphicx}

\usepackage[shortcuts]{extdash}

\usepackage{ulem} 

\newcommand{\new}[1]{#1}
\newcommand{\ii}{i}
\newcommand{\ee}{e}
\newcommand{\dd}{d}

\newcommand{\I}[4]{\ensuremath{\prescript{#1}{#2}{I}^{#3}_{#4}}}
\newcommand{\Ink}[2]{\I{#1}{#2}{n'\kappa'}{n\kappa}}
\newcommand{\iIoo}[2]{\I{1}{00}{#1}{#2}}
\newcommand{\iIio}[2]{\I{1}{10}{#1}{#2}}
\newcommand{\iIoi}[2]{\I{1}{01}{#1}{#2}}
\newcommand{\iIii}[2]{\I{1}{11}{#1}{#2}}
\newcommand{\tIoo}[2]{\I{2}{00}{#1}{#2}}
\newcommand{\tIio}[2]{\I{2}{10}{#1}{#2}}
\newcommand{\tIoi}[2]{\I{2}{01}{#1}{#2}}
\newcommand{\tIii}[2]{\I{2}{11}{#1}{#2}}
\newcommand{\iIoonk}{\I{1}{00}{n'\kappa'}{n\kappa}}
\newcommand{\iIionk}{\I{1}{10}{n'\kappa'}{n\kappa}}
\newcommand{\iIoink}{\I{1}{01}{n'\kappa'}{n\kappa}}
\newcommand{\iIiink}{\I{1}{11}{n'\kappa'}{n\kappa}}
\newcommand{\tIoonk}{\I{2}{00}{n'\kappa'}{n\kappa}}
\newcommand{\tIionk}{\I{2}{10}{n'\kappa'}{n\kappa}}
\newcommand{\tIoink}{\I{2}{01}{n'\kappa'}{n\kappa}}
\newcommand{\tIiink}{\I{2}{11}{n'\kappa'}{n\kappa}}
\newcommand{\pose}{e_\textrm{\tiny pos}}

\let\mathbb\mathds

\usepackage{amsmath}
\usepackage{amssymb}
\usepackage{amsthm}
\usepackage{mathrsfs}
\usepackage{dsfont}
\usepackage{bm}
\usepackage{slashed}

\newcommand{\tr}{\ensuremath{\text{tr}}}

\newcommand{\bigo}{\ensuremath{\mathcal{O}}}

\newcommand{\vect}[1]{\ensuremath{\boldsymbol{#1}}}

\newcommand{\amplitude}{\ensuremath{\mathcal{M}}}

\newcommand{\eqdef}{\ensuremath{:=}}

\usepackage{tkz-graph}
\usetikzlibrary{positioning,calc,arrows}
\usetikzlibrary{positioning,arrows,arrows.meta}
\usetikzlibrary{decorations.pathmorphing}
\usetikzlibrary{decorations.markings}
\usetikzlibrary{calc}
\usetikzlibrary{patterns}
\pgfdeclarelayer{background}
\pgfsetlayers{background,main}
\tikzset{
    expand bubble/.style={
        preaction={draw,line width=10.4pt},
        white,fill,draw,line width=10pt,
    },
}

\newdimen\tmpdimen
\pgfdeclaredecoration{complete sines}{initial}
{
    \state{initial}[
        width=+0pt,
        next state=move,
        persistent precomputation={
            \pgfmathparse{\pgfkeysvalueof{/pgf/decoration/angle step}}%
            \let\anglestep=\pgfmathresult%
            \let\currentangle=\pgfmathresult%
            \pgfmathsetlengthmacro{\pointsperanglestep}%
                {(\pgfdecoratedremainingdistance/\pgfkeysvalueof{/pgf/decoration/number of sines})/360*\anglestep}%
        }] {}
    \state{move}[width=+\pointsperanglestep, next state=draw]{
        \pgfpathmoveto{\pgfpointorigin}
    }
    \state{draw}[width=+\pointsperanglestep, switch if less than=1.25*\pointsperanglestep to final, 
        persistent postcomputation={
        \pgfmathparse{mod(\currentangle+\anglestep, 360)}%
        \let\currentangle=\pgfmathresult%
    }]{%
        \pgfmathsin{+\currentangle}%
        \tmpdimen=\pgfdecorationsegmentamplitude%
        \tmpdimen=\pgfmathresult\tmpdimen%
        \divide\tmpdimen by2\relax%
        \pgfpathlineto{\pgfqpoint{0pt}{\tmpdimen}}%
    }
    \state{final}{
        \ifdim\pgfdecoratedremainingdistance>0pt\relax
            \pgfpathlineto{\pgfpointdecoratedpathlast}
        \fi
   }
}

\tikzset{
fermion/.style={draw=black, postaction={decorate},
    decoration={markings,mark=at position 0.6 with {\arrow[scale=.7,black]{triangle 45}}}},
photon/.style={decorate, draw=black,
    decoration={coil,aspect=0}},
/pgf/decoration/.cd,
    number of sines/.initial=10,
    angle step/.initial=20
}
\usepackage{pgfplots}

\renewcommand{\emph}[1]{{\it #1}}

\makeatletter
\def\@ssect@ltx#1#2#3#4#5#6[#7]#8{%
  \def\H@svsec{\phantomsection}%
  \@tempskipa #5\relax
  \@ifdim{\@tempskipa>\z@}{%
    \begingroup
      \interlinepenalty \@M
      #6{%
       \@ifundefined{@hangfroms@#1}{\@hang@froms}{\csname @hangfroms@#1\endcsname}%
       {\hskip#3\relax\H@svsec}{#8}%
      }%
      \@@par
    \endgroup
    \@ifundefined{#1smark}{\@gobble}{\csname #1smark\endcsname}{#7}%
  }{%
    \def\@svsechd{%
      #6{%
       \@ifundefined{@runin@tos@#1}{\@runin@tos}{\csname @runin@tos@#1\endcsname}%
       {\hskip#3\relax\H@svsec}{#8}%
      }%
      \@ifundefined{#1smark}{\@gobble}{\csname #1smark\endcsname}{#7}%
      \addcontentsline{toc}{#1}{\protect\numberline{}#8}%
    }%
  }%
  \@xsect{#5}%
}%
\makeatother

\begin{document}
\title{\new{Gravitational effects in birefringent quantum electrodynamics}}
\author{Simon Grosse-Holz}
\affiliation{Massachusetts Institute of Technology, Department of Physics, Cambridge, Massachusetts 02139, USA}
\author{Frederic P. Schuller\footnote{Corresponding author: f.p.schuller@utwente.nl}}
\affiliation{Department of Applied Mathematics, University of Twente, 7500 AE Enschede, The Netherlands}
\author{Roberto Tanzi}
\affiliation{Center of Applied Space Technology and Microgravity, University of Bremen,
Am Fallturm,
28359 Bremen,
Germany}

\begin{abstract} 
\new{The most general classical electrodynamics which still respect the linear superposition principle but allow for otherwise arbitrary birefringence, require and imply a refined spacetime geometry described by a fourth-rank tensor field.
Canonical gravitational dynamics for this geometry, if required to co-evolve in causally consistent fashion with the electromagnetic field, were shown to be constructively determined by gravitational closure of the birefringent electromagnetic field equations.
For weak gravitational fields of the resulting birefringent refinement of classical Einstein-Maxwell theory, we show in this article that the corresponding quantum electrodynamics is locally renormalizable at every loop order in gauge-invariant fashion and then employ this result to compute various fundamental processes. Combining quantum field theoretic results in locally essentially flat regions with the global spacetime structure predicted by the refined gravitational dynamics,
we find that the anomalous magnetic moment of the electron, the cross sections of Bhabha scattering, and the hyperfine splitting of the hydrogen all pick up a dependence on position in the gravitational field. Particularly the measurement of the hyperfine line of hydrogen, but quite generally the measurement of any local quantum electrodynamical process, is thus able to inform the search for vacuum birefringence and its effects in a new way, since the gravitational theory allows to predict where the effects will be most pronounced.}
\end{abstract}

\maketitle

\newpage
\tableofcontents

\newpage

\section{Introduction}
\new{This work is part of a larger effort revolving around the gravitational closure mechanism \cite{SSWD}. 
The latter provides, under rather mild physical conditions, a calculational method to obtain a multi-parameter family of gravitational actions for just any tensorial geometry that has been chosen as the geometric background in the formulation of some given matter field theory. The present article now
asks and answers the question of how local quantum effects of the initially stipulated matter theory can be used to further narrow down this family of ensuing gravitational actions.

For definiteness, we study the quantum field theory of general linear electrodynamics \cite{RiveraPhD} interacting with massive Dirac fermions, whose fourth-rank tensorial background geometry we provide with the classical dynamics that are dictated by the said gravitational closure mechanism. 
Since phenomenologically immediately relevant solutions of the ensuing exact gravitational field equations will present refined geometries close to flat metric geometries, we moreover consider only such in this paper. The correspondingly linearized gravitational field equations for the refined background geometry
is unique up to eleven real parameters (rather than only one, namely Newton's constant, in linearized general relativity), which remain to be determined experimentally. 
 
It turns out that a signature implication of  such gravitationally closed general linear electrodynamics is that quantum effects, even if arbitrarily localized, may now depend on the particular geometric background and thus ultimately on the parameters in the gravitational field equations. The simple reason for this is the existence of 
up to five scalar fields that can be constructed in ultralocal fashion from the fourth-rank tensor perturbations of the underlying refined spacetime geometry.
It is precisely these scalar fields that then appear in scattering cross sections and bound state spectra. 

Carefully noting that there is thus a five-fold inifinity of ways in which the refined spacetime geometry can be locally flat, we show the gauge-invariant renormalizability of general linear electrodynamics interacting with Dirac fermions to any loop order. 
Subsequent calculation of cross sections and the hyperfine transition energy of hydrogen in any flat region can thus rely on the same calculation under the technical fiction that the refined spacetime is globally flat, if indeed the quantum process can reasonably be thought to be confined to the respective region. The interesting point, of course, is that any quantum prediction will generically depend on the value of the five scalar fields in the respective region, and thus on the specific gravitational field present there. 

This opens a remarkable way to determine at least some of the eleven gravitational parameters, namely by measurements of quantum electrodynamical effects. We illustrate this by providing four independent ways to determine the same two combinations of new gravitational parameters in not-too-close proximity to a gravitational point mass source. 
In particular, we predict the anomalous magnetic moment of the electron, the hyperfine structure and ionization energy of hydrogen, as well as electron-positron to fermion-antifermion and Bhabha scattering cross sections to be all position dependent and express them in terms of the gravitating point mass, the coordinate distance from it and two combinations of the eleven gravitational parameters.

This article makes significant contact with various previous studies of general linear electrodynamics \cite{Hehl:2002hr,Obukhov:2002xa,Bailey:2004na,Altschul2006,Itin:2009aa} as well as its interacting \cite{deBerredoPeixoto:2006wz,Cambiaso:2014eba} and non-interacting \cite{Rivera2011,Cambiaso:2012vb,Colladay:2014dua} quantum field theory. General linear electrodynamics is only subtly more special than premetric electrodynamics, which is based on a fourth-rank tensor density providing the geometric background structure; for a complete overview see the monograph \cite{hehlbook} and the references therein.  The spacetime kinematics implied by general linear electrodynamics, in particular the definition of observers in whose frame the spacetime field strength and charged currents can be interpreted, follow as a special case from the comprehensive general analysis \cite{RRS} for the physical dispersion relations of any bihyperbolic matter theory. This general analysis is also the kinematical core on which the gravitational closure mechanism \cite{SSWD} is built.

Finally assuming a maximal refinement not only for the geometric background of linear electrodynamics, but also for the quasi-linear non-abelian gauge dynamics and all fermion dynamics in the standard model of particle physics, one arrives at the standard model extension of Colladay and Kostelecky \cite{Colladay:1998fq}; see also the pertinent data tables on Lorentz violation \cite{Kostelecky:2008ts}, which constrain the possible deviations from a Lorentzian spacetime structure tremendously, but in part seem to have been obtained under the assumption that there exists a set of coordinate systems in which these deviations are constant throughout spacetime. 

This assumption of constants deviations cannot be upheld once full gravitational dynamics, for all geometric degrees of freedom, is included into the picture. The thereby predicted deviations from a metric geometry may still be extremely small on average, but at the same time develop appreciable size locally and thus be observable. 
The gravitational dynamics of the fourth-rank tensor geometry predicts precisely where these larger effects will occur. 
The generically resulting non-flatness also has an effect on the geometric degrees of freedom that are detectable by interacting general linear electrodynamics. This is because even small geometric perturbations around a flat metric background generically feature a double trace that becomes measurable when comparing effects in different regions. This is a subtle point, since if all regions were the same, which would be the case if the spacetime was globally flat, the double trace could not be measured since it would be absorbable into redefinitions of fields and charges \cite{Anderson:2004qi}.    
This leaves a crucial signature, for instance on the hyperfine transition energy of hydrogen, in the gravitationally closed scenario we consider in this paper.

The organization of this article is as follows.
We begin by summarizing the relevant features of general linear electrodynamics and of the constructive gravity program in section~\ref{sec:GLED}.
In the same section, we show that the experimental applications considered in this paper, which all pertain to local quantum effects, allow for an effective local, but not global, restriction to a  flat fourth-rank tensor geometry close to a metrically induced one. 
For this case, not only does one know the dynamics of the new geometry (and thus the refined gravitational field generated by a point mass) explicitly, but one can also prove that the general linear electrodynamics can be quantized and renormalized in a gauge-invariant way at every loop order.
The detailed discussion of this quantization and renormalization of general linear electrodynamics interacting with fermions is given in section~\ref{sec:renormalization}.
The following two sections then present applications of the quantized and renormalized general linear electrodynamics and express the result in terms of the refined geometry.
In particular, in section~\ref{sec:scatterings}, we study the scattering of an electron-positron into fermion-antifermion, Bhabha scattering, and we compute the anomalous magnetic moment of the electron.
Notably, the latter is computed to the same loop order as it is known in standard QED on a flat Minkowski background.
The other major application of the quantized and renormalized theory is given in section~\ref{sec:hydrogen}, we study the hyperfine structure of the hydrogen atom and find qualitative and quantitative differences to the results of standard QED. 
Finally, in section~\ref{sec:around-a-point-mass}, we specialize the general results of the previous two sections by inserting the refined gravitational field around a point mass. For the example of this deliberately specific gravitational setting, we make the dependence of the quantum predictions on the pertinent gravitational parameters  explicit and thus link their values to measurable effects.}

Throughout this article, spacetime indices ranging over $0,1,2,3$ are denoted by lower case Latin letters
$a, b, c, \dots$, while purely spatial indices with respect to a particular observer frame, ranging over $1, 2, 3$,  are denoted by lower case Greek letters $\alpha, \beta, \gamma, \dots$. The Levi-Civita tensor densities are normalized to $\epsilon_{0123}=1$ and $\epsilon^{0123}=-1$ and the Minkowski metric $\eta$ in normal coordinates has as non-vanishing components $\eta_{00}=1$ and $\eta_{\alpha\alpha}=-1$. 

\section{Classical general linear electrodynamics and its gravity} \label{sec:GLED}
In this initial section, we discuss classical general linear electrodynamics and its gravitational closure. Qualitatively, the latter will be relevant throughout the paper, although quantitatively, it will not enter the discussion again before section \ref{sec:around-a-point-mass}. The gravitational closure mechanism asserts that three simple causality and energy conditions on any given linear {\it matter dynamics} enable one to derive, rather than stipulate, the  diffeomorphism-invariant classical {\it gravitational dynamics} that the underlying geometry must satisfy in order to canonically co-evolve with the given matter fields. Indeed, that required gravitational Lagrangian density arises from the solution of a countable set of partial differential equations whose coefficients are wholly determined from the coefficients of the given matter action. 

For the general linear electrodynamics, whose interacting quantum field theory we undertake to study in this article, we identify a fourth-rank tensor field, rather than the usual Lorentzian metric, as the relevant underlying geometry and provide some important implications in the first subsection below. We then briefly discuss the gravitational closure mechanism in general and provide, as a concrete example, the fourth-rank tensor field that arises as the solution of the weak field gravitational field equations around a point mass. We finally discuss how any solution of the linearized field equations for the fourth-rank tensor, as they have been obtained by gravitational closure, can be employed to predict concrete local quantum electrodynamics effects, as is the purpose of this paper. 

\subsection{General linear electrodynamics and its refined background geometry} \label{sec:EDyn}
The most general theory of linear electrodynamics sourced by a conserved vector current $j$ on a four-dimensional smooth manifold is given by the action 
\begin{equation}\label{Sstrong}
S[A] = \int\dd^4x\,\,\omega(x)\!\left[-\frac{1}{8}  G^{abcd}(x)F_{ab}(x)F_{cd}(x) + j^a(x) A_a(x)\right]\,,
\end{equation}
where $A$ is the one-form abelian gauge potential and $F = \dd A$ the associated field strength. 
The fourth rank contravariant tensor field $G$ featuring in this action is restricted, without loss of generality, to possess the same algebraic symmetries as a curvature tensor,
\begin{equation} \label{area-metric-symmetries}
G^{abcd} = G^{[ab][cd]} = G^{cdab} \qquad\text{and}\qquad G^{abcd} + G^{acdb} + G^{adbc}=0\,,
\end{equation}
and can physically be thought of as the constitutive tensor that relates the electromagnetic inductions to the field strengths \cite{postbook,hehlbook}, while the weight-one scalar density $\omega$  is a further background field degree of freedom, independent of the constitutive tensor and required in order to make the integral measure well-defined under general smooth changes of coordinate chart. \new{The ubiquitous appearance of the scalar density $\omega$ in this section, and in our later perturbative treatment its perturbation $e$ from unity, ensures that all our component equations involving partial derivatives are independent under general coordinate transformations. An elegant covariant derivative, as it is induced and useful in metric geometry, is not available but also not needed.}

We will refer, for short, to a manifold $(M,G,\omega)$ equipped with such $G$ and $\omega$ as an area metric manifold \cite{SWW}. \new{In other contexts, it is more convenient to encode the pair $(G,\omega)$ more compactly in a single fourth rank tensor $\hat G^{abcd} := G^{abcd} + \omega^{-1} \epsilon^{abcd}$, from which the density $\omega$ can be readily extracted by complete contraction of $\hat G^{abcd}$ with the tensor density $\epsilon_{abcd}$ of weight minus one. In the context of dimensional regularization, however, this spurious appearance of the tensor density $\epsilon_{abcd}$ is technically inconvenient, whence we chose to consider the entirely equivalent view that treats the algebraically independent parts $G$ and $\omega$ separately. Note that use of $\hat G$ instead of $G$ and $\omega_{\hat G}:= (-\tfrac{1}{24}\epsilon_{abcd}\hat G^{abcd})^{-1}$ instead of $\omega$ in the action (\ref{Sstrong}) only produces an additional surface term.}

 
Not just any such area metric geometry, however, gives rise to a physically viable theory of electrodynamics. Indeed, elementary physicality requirements  \cite{RRS} on the implied dispersion relation for electromagnetic waves impose key conditions on the area metric $G$, which also play also a central role in the gravitational closure procedure discussed in the next subsection. 
The two key mathematical objects for this discussion are identified as follows. Through variation of the general linear electrodynamics action with respect to the gauge potential $A$, one finds the equations of motion
\begin{equation}\label{eq:motion}
-\frac{1}{\omega}\partial_b(\omega G^{abcd}\partial_c A_d) = j^a\,,
\end{equation}
whose geometric-optical limit restricts the wave covector of light rays to satisfy a quartic dispersion relation
\begin{equation}\label{masslessdisp}
P^{abcd}(x) k_a k_b k_c k_d = 0\qquad\textrm{ for } k\in T_x^*M\,,
\end{equation}
rather than the familiar quadratic one that is enforced in a Lorentzian metric background geometry. 
The polynomial defined by the left hand side arises as the principal polynomial of the involutive form \cite{MScWierzba} of the partial differential matter field equation (\ref{eq:motion}). Indeed, after also removing the abelian gauge ambiguity (in one way or another, see \cite{rubilar2002generally,itin2009light}), one obtains the explicit expression for the totally symmetric tensor field 
\begin{equation}\label{eq:Ppolarization}
{P}^{abcd} = -\frac{1}{24}\omega^2\epsilon_{mnpq}\epsilon_{rstu}G^{mnr(a}G^{b|ps|c}G^{d)qtu}\,
\end{equation}  
that defines the quartic dispersion relation above, up to an undetermined factor of an everywhere non-zero spacetime function. 
If the area metric structures $G$ and $\omega$ are chosen such that the electromagnetic field equations have a well-posed initial value problem, it follows that the principal polynomial is hyperbolic \cite{gaarding1951linear}, which indeed is also the first of the aforementioned conditions on the geometry to give rise to a physical dispersion relation. We make use of the hyperbolicity requirement in conjunction with the freedom to choose the undetermined overall function, by requiring the sign convention
\begin{equation}\label{signconv}
P^{abcd}(x) h_a h_b h_c h_d > 0
\end{equation} 
for any covector $h$ in the non-empty hyperbolicity cone \cite{gaarding1951linear} of any such polynomial at any spacetime point $x$. In any case, the quartic structure of the massless dispersion relation (\ref{masslessdisp}) generically admits two different future-directed light rays in the same spatial direction, which effect is known as birefringence. Either of them corresponds to a particular wave covector $k$, which in turn 
is related by \cite{punzi2009propagation}
\begin{equation}
G^{abcd}(x)k_a k_c a_d = 0\,
\end{equation}
to the polarization $a$ of the same light ray at that spacetime point. The quartic dispersion relation finally dictates the action for the worldline $x(\lambda)$ of a light ray, namley
\begin{equation}\label{lightpathaction}
S_\textrm{\tiny light ray}[x,\mu] := \int d\lambda\, \mu \,P^\#_{abcd}(x)\, \dot x^a\, \dot x^b\, \dot x^c\, \dot x^d\,,
\end{equation}
where the so-called dual polynomial defined by the above integrand is given by the totally symmetric tensor field
\begin{equation}
P^\#_{abcd} =  -\frac{1}{24}\omega^{-2}\epsilon^{mnpq}\epsilon^{rstu}G_{mnr(a}G_{b|ps|c}G_{d)qtu}\,,
\end{equation}
as was shown in \cite{RRS}, by employing an interplay of results from the convex analysis of hyperbolic polynomials and real algebraic geometry. 

\new{Imposing the hyperbolicity of this dual polynomial at each point in spacetime, in addition to the already required hyperbolicity of the principal polynomial, is the second of the aforementioned conditions for a physical dispersion relation and allows to determines, in particular, the action for massive point particles in terms of $P$. For a thorough discussion and proofs, see \cite{RRS}. 
The combined requirements of hyperbolicity of both $P$ and its dual $P^\#$ in conjunction with the signature convention (\ref{signconv}) result in a significant restriction of the possible constitutive tensors $G^{abcd}$, namely from 23 possible algebraic classes down to only seven \cite{SWW}. This is precisely the same mechanism as the one that restricts any metric that can underlie standard Maxwell theory down to only one class (namely metrics of mostly-minus Lorentzian signature) from the five algebraic classes corresponding to the signatures that a non-degenerate symmetric bilinear map can have in four-dimensions. Note that in the metric case, hyperbolicity of the principal polynomial $P^{ab}=g^{ab}$ already implies hyperbolicity of the dual dual polynomial given by  $P^\#_{ab} = g_{ab}$.   }

\subsection{\new{Weak gravitational closure of general linear electrodynamics}} \label{subsec:grav_closure}
In this paper, we consider quantum effects of electrodynamics coupled to Dirac fermions on a refined background provided by an area metric manifold $(M,G,\omega)$. In order to make concrete predictions, we thus need to know exactly how this refined background actually looks like in various  situations of physical interest, e.g., around a heavenly body. In order to achieve that, we need to extend the given matter dynamics by additional dynamics for the employed background geometry. For Maxwell theory on a Lorentzian manifold, this closure of the dynamical equations would be afforded by the Einstein-Hilbert action, while now, with the fourth-rank tensor $G$ and the scalar density $\omega$ taking the role previously played by the metric and the associated scalar density $\sqrt{-g}$, one clearly needs a new action for this refined area metric geometry instead.

Remarkably, the appropriate dynamics for any tensorial background, as it is employed in given matter field equations, must not be stipulated but can be obtained constructively. Indeed, the hardly negotiable requirement that both the matter and the geometry of a spacetime canonically evolve together leaves one with little choice for diffeomorphism-invariant gravitational dynamics that can provide evolution equations to the coefficients of the given system of matter field equations. 

More precisely, as was shown in \cite{SSWD}, for any diffeomorphism invariant matter action $S_\text{\tiny matter}$ whose integrand depends locally on some tensorial matter field $A$ and ultralocally on a geometric background described by some tensor field $G$ of arbitrary valence, i.e.,
\begin{equation}\label{intromatter}
S_\text{\tiny matter}[A;G) 
= \int \mathrm d^4 x \,\mathscr{L}_\textrm{\tiny matter}(A(x),\partial A(x), \dots, \partial^\textrm{\tiny finite}\!A(x); G(x))\,,
\end{equation}
and whose equations of motion feature a principal polynomial that is hyperbolic, has a hyperbolic dual and satisfies an energy condition --- see \cite{RRS} for these three conditions and their physical relevance --- one may calculate four geometry-dependent coefficients $E^A{}_\mu$, $F^A{}_\mu{}^{\nu}$, $M^{B\mu}{}$ and $p^{\alpha\beta}$ that enter the so-called gravitational closure equations, which is the countable set of linear homogeneous partial differential equations derived in \cite{SSWD} whose solution $\mathscr{L}_\textrm{\tiny geometry}$ provides the extension
\begin{equation}\label{Sclosed}
S_\text{\tiny closed}[A;G] 
= S_\text{\tiny matter}[A;G) + \int \mathrm d^4 x\, \mathscr{L}_\textrm{\tiny geometry}(G(x),\partial G(x), \dots,\partial^\textrm{\tiny finite}G(x))\,
\end{equation}
of the originally given matter field dynamics by inclusion of gravitational dynamics for the background geometry. This way, the corresponding system of matter field equations is closed gravitationally. For the physically important issue of how the energy-momentum-stress tensor arises  from a variation with respect to an area metric and the associated Bianchi identity it shares with the associated area metric spacetime geometry, see  \cite{Schuller:2017dfj}.   

In the context of the present paper, this allows us to solve the gravitational field equations that follow, by this mechanism, from general linear electrodynamics for any situation of physical interest. In particular, we are able to obtain {\it concrete} predictions not only about  possible quantum electrodynamics effects, but indeed about where and when they occur with what strength. 


Since we will conduct our analysis of quantum electrodynamics coupled to fermions in the presence of a weak gravitational field in this article, we consider linear perturbations of the form
\begin{equation}\label{omegaGdecomp}
 \omega = 1 + h \qquad\textrm{and}\qquad G^{abcd} = \eta^{ac} \eta^{bd} - \eta^{ad} \eta^{bc} + H^{abcd} 
\end{equation}
around a flat metric background $\eta$ described in coordinates where the components of the Lorentzian metric take their Minkowskian normal form. It is useful to perform a $(3+1)$-decomposition of the perturbations $h$ and $H$ with respect to a foliation whose leaves are flat spacelike hypersurfaces of the flat metric manifold around which we perturb. Identifying these leaves with the various hypersurfaces labelled by a constant zeroth coordinate with respect to a global normal coordinate system for $\eta$, the decomposition takes the form \cite{SSSW} \begin{eqnarray}
h &=&  \frac{3}{2} (\widetilde F - A) \,,\\
H^{0\beta0\delta} &=& (2A-\widetilde F) \gamma^{\beta\delta} - \Delta^{\beta\delta} F - 2 \partial^{(\beta} F^{\delta)} - F^{\beta\delta} \,,\\
H^{0\beta\gamma\delta} &=& \epsilon^{\gamma\delta}{}_\nu \Big[\epsilon^{\beta\nu\rho}(\partial_\rho B + B_\rho)  + \Delta^{\beta\nu} C + 2 \partial^{(\beta} C^{\nu)} + C^{\beta\nu} \Big]\,,\label{HelmH}\\
H^{\alpha\beta\gamma\delta} &=& \epsilon^{\lambda\alpha\beta} \epsilon^{\kappa\gamma\delta} \Big[(3 \widetilde F + \widetilde E) \gamma_{\lambda\kappa} + \Delta_{\lambda\kappa} E + 2 \partial_{(\lambda} E_{\kappa)} + E_{\lambda\kappa}\Big]\,,
\end{eqnarray}
where $\widetilde F, \widetilde E, F, E, C$ are one-parameter families of scalar fields, $F^\alpha, E^\alpha, C^\alpha$ such families of solenoidal vector fields and $F^{\alpha\beta}, E^{\alpha\beta}, C^{\alpha\beta}$ such families of transverse and $\gamma$-traceless symmetric tensor fields on three-dimensional flat Euclidean space, where the parameter identifying a particular member of the family coincides with the label of the unique leaf containing the point at which the left hand sides are evaluated. Indices are raised and lowered with the Euclidean metric $\gamma$ and the trace-removed Hesse operator $\Delta^{\alpha\beta} := \partial^\alpha \partial^\beta - \tfrac{1}{3}\gamma^{\alpha\beta} \Delta$. The diffeomorphism invariance of the full theory translates to the physically irrelevant gauge transformations  
\begin{eqnarray}
\Delta_\xi \widetilde F = - \tfrac{2}{3} \Delta L,\quad \Delta_\xi A = \dot T,\quad\Delta_\xi \widetilde E &=& \tfrac{2}{3} \Delta L,\quad\Delta_\xi B = \dot L - T,\quad \Delta_\xi F = - 2L,\\
\Delta_\xi E= 2L,\quad \Delta_\xi B^\alpha = \dot L^\alpha,&\!\!\!& \Delta_\xi F^\alpha = - L^\alpha, \quad \Delta_\xi E^\alpha = L^\alpha
\end{eqnarray} 
of the three-dimensional scalar, solenoidal vector and transverse tracefree tensor fields, 
where $T$ and $L^\alpha$ are arbitrary smooth field components, which can be chosen in terms of the perturbation fields themselves in order to remove four degrees of freedom.

Perturbative application of the gravitational closure mechanism to the matter action (\ref{Sstrong}) yields the linearized gravitational field equations for the background geometry (\ref{omegaGdecomp}) directly in terms of the scalar, solenoidal vector and tensor modes above, see \cite{SSSW} for their explicit form and dependence on field-generating matter. As is shown there, these linearized field equations contain eleven undetermined parameters $\kappa_1,\dots,\kappa_{11}$, which arise as constants of integration when solving the pertinent gravitational closure equations. As in the case of general relativity, where Newton's constant and the cosmological constant arise this way, one must perform sufficiently many experiments to fix these parameters before the theory is quantitatively fully predictive. This is because particular solutions of the field equations of course depend on the parameters.  

The results on quantum effects, which we derive in this article, can be evaluated on any solution of the linearized gravitational field equations. In section \ref{sec:around-a-point-mass}, we will demonstrate this explicitly for the particular solution that describes a weak static gravitational field around a point mass $M$. In the gauge where $F=B=0$, the 
spacetime perturbations are given by
\begin{eqnarray} \label{am-around-point-mass}
h &=&  A- \frac{3}{4} \widetilde U -\frac{3}{4} \widetilde V \,,\\
H^{0\beta0\delta} &=& \left(2 A -\frac{1}{2} \widetilde U -\frac{1}{2}\widetilde V \right) \gamma^{\beta \delta} \,,\\
H^{0\beta\gamma\delta} &=& 0\,,\label{HelmH2}\\
H^{\alpha\beta\gamma\delta} &=& \Big(\widetilde U + 2 \widetilde V \Big) \big(\gamma^{\alpha \gamma} \gamma^{\beta \delta}-\gamma^{\alpha \delta} \gamma^{\beta \gamma} \big)\,,
\end{eqnarray}
for scalar modes
\begin{align}  
A(r) &\eqdef \frac{M}{4 \pi r} \left\lbrack -\frac{\kappa}{2}  -\frac{\beta+3\gamma}{4} e^{-\sqrt{\mu}r} \right\rbrack\,,\\
\widetilde U(r) &\eqdef \frac{M}{4 \pi r} \left\lbrack -2 \kappa + \beta e^{-\sqrt{\mu}r} \right\rbrack\,,\\
\widetilde V(r) &\eqdef \frac{M}{4 \pi r} \left\lbrack \gamma e^{-\sqrt{\mu}r} \right\rbrack
\label{am-around-point-mass-end}
\end{align}
with $\kappa$, $\beta$, $\gamma$, $\mu$ being four independent particular combinations of eleven constants of integration, which emerge during the closure process and must be determined experimentally. 

\subsection{\new{Ultralocal structure of area metric manifolds}}
One instructive way to view the refined geometry $(G,\omega)$ presented by an area metric manifold as compared to a metric manifold is to consider the geometry at strictly one point.
While Sylvester's theorem guarantees that the $16$-parameter group $GL(4,\mathbb{R})$ of frame and coframe transformations at the chosen point suffices to take the components $g^{ab}$ of a Lorentzian inverse metric tensor of mainly minus signature to the standard form $\textrm{diag}(1,-1,-1,-1)^{ab}$, it is clear that $GL(4,\mathbb{R})$ cannot suffice to take the $20$ independent components $G^{abcd}$ of the area metric tensor plus the scalar density $\omega$ to one such normal form.
Indeed, a generic area metric contains ultralocal information (constructed from the area metric at a point, but not its derivatives) amounting to up to five scalars \cite{SWW}. 

For the purposes of the present article, it is interesting to consider the implications of this fact for the form taken by the principal tensor (\ref{eq:Ppolarization}) in an observer co-frame $\epsilon^0, \epsilon^1, \epsilon^2, \epsilon^3$.
The latter is characterized by the normalization and orthogonality conditions 
\begin{equation} \label{observer-frame-cond}
 P(\epsilon^0,\epsilon^0,\epsilon^0,\epsilon^0)=1
 \qquad \text{and} \qquad
 P(\epsilon^0,\epsilon^0,\epsilon^0,\epsilon^\alpha)=0 \quad\textrm{ for } \alpha=1,2,3 \,.
\end{equation}
Physically, these conditions are best interpreted in the unique dual frame $e_0, e_1, e_2, e_3$:  
The first one is equivalent to the requirement that a clock carried by an observer whose worldline tangent at the given point agrees with $e_0$ measures proper time, while the 
last three conditions ensure that the three vectors $e_1, e_2, e_3$ span the purely spatial directions seen by that same observer.
One can show that it is always possible to find such an observer frame at a given point for a generic area metric. However, according to the above discussion, obviously neither the area metric nor the principal tensor $P$ can be entirely trivialized  by such an ultralocal choice of frame. 

In order to identify the remaining information of interest for our purposes, we consider the here relevant case of an area metric and scalar density taking the form (\ref{omegaGdecomp}), both differing from those induced by a flat Lorentzian metric background only by a linear area metric perturbation.
For the purpose of showing precisely the order in the perturbative expansion, we make use of a small control parameter $\lambda$ in this subsection, so that the area metric~(\ref{omegaGdecomp}) becomes
\begin{equation} \label{omegaGdecomp-lambda}
 \omega = 1 + \lambda \, h
 \qquad\textrm{and}\qquad
 G^{abcd} = \eta^{ac} \eta^{bd} - \eta^{ad} \eta^{bc} + \lambda \, H^{abcd} \,.
\end{equation}
For such an area metric, the principal polynomial~(\ref{eq:Ppolarization}) becomes
\begin{equation}
  P^{abcd} := P_0^{abcd} + \lambda \, P_1^{abcd} + \lambda^2 \,  P_2^{abcd} + \mathcal{O}(\lambda^3) 
\end{equation}
in terms of the zeroth, first and second order contributions 
\begin{eqnarray}
  P_0^{abcd} &=& \eta^{(ab}\eta^{cd)}\,,\\
  P_1^{abcd} &=& 2 \eta^{(ab}\eta^{cd)} h + \eta^{(ab}H^{cd)}\,,\\
  P_2^{abcd} &=& \eta^{(ab}\eta^{cd)} h^2 + 2 \eta^{(ab}H^{cd)} h + \tfrac{1}{2} H^{(ab} H^{cd)} - \tfrac{1}{2} H^{2\,(abcd)}\,,
\end{eqnarray}
where we have employed the shorthand notations
\[
 H^{ab}:=H^{ambn} \eta_{mn}
 \qquad \text{and} \qquad
 H^{2\, abcd}:= H^{ambr} H^{cnds} \eta_{mn} \eta_{rs} \,. 
\]

It is now possible to identify a frame in which not only the two conditions~(\ref{observer-frame-cond}) are satisfied to first order in the perturbation, but where also the principal polynomial~(\ref{eq:Ppolarization}) itself coincides with the metric-induced one to first order.
To do so, consider the $GL(4,\mathbb{R})$ transformation 
\begin{equation} \label{frame-transf}
  e^{a'}{}_a = (1-\tfrac{1}{2} \lambda \, h) \delta^{a'}{}_a - \tfrac{1}{4} \lambda H^{a'mk n} g_{mn} g_{ak}\,,
\end{equation}
from a generic frame at the point under consideration to the a new frame, in which one has
\begin{equation}\label{perta}
\omega = 1 + \lambda \, e
\qquad \textrm{ and } \qquad
G^{abcd} = \eta^{ac} \eta^{bd} - \eta^{ad} \eta^{bc} + \lambda \, E^{abcd}
\end{equation}
for now restricted perturbations $E^{abcd}$ and $e$ linked by the 10 conditions
\begin{equation}\label{pertconditionsa}
    E^{ab} = -2 \eta^{ab} e +\mathcal{O}(\lambda) \,.
\end{equation}
In this frame and in terms of this decomposition, the components of the principal polynomial now read
\begin{eqnarray}
  P_0^{abcd} &=& \eta^{(ab}\eta^{cd)}\,,\label{normform1}\\
  P_1^{abcd} &=& 0\,,\\
  \label{principal-polynomial-2nd}
  P_2^{abcd} &=& - \eta^{(ab}\eta^{cd)} e^2 - \tfrac{1}{2} E^{2\,(abcd)}+\eta^{(ab}\delta E^{cd)}\,,\label{normform3}
\end{eqnarray}
where $\delta E^{ab}\eqdef \big(E^{ab}+2\eta^{ab} e \big)/\lambda$, which is $\mathcal{O}(1)$ in $\lambda$ because of~(\ref{pertconditionsa}).
We remark in passing that it is also possible to explicitly find a frame in which the two conditions~(\ref{observer-frame-cond}) are satisfied at second order, namely by means of a second transition map
\begin{equation} \label{frame-transf-2nd}
 \tilde{e}^{a'}{}_{a}=\delta^{a'}_{a}+\lambda^2 \, \delta \tilde{e}^{a'}{}_{a}
\end{equation}
that has to be performed after~(\ref{frame-transf}) and satisfies the conditions
\begin{equation}
 \delta \tilde{e}^{0}{}_{0}=-\frac{P_2^{0000}}{4}
 \qquad \text{and} \qquad
 \delta \tilde{e}^{\alpha}{}_{0}+ \eta^{\alpha \beta} \delta \tilde{e}^{0}{}_{\beta}=-P_2^{000\alpha} \,,
\end{equation}
but in order to derive the results of the present paper, it will fully suffice to consider~(\ref{frame-transf}) without the subsequent transformation~(\ref{frame-transf-2nd}).


The ultralocally always achievable normal form (\ref{normform1})--(\ref{normform3}) for the principal tensor components explicitly reveals two important facts.
One is that, at first order, any effective perturbation of the scalar density is entirely determined by the effective perturbation of the constitutive tensor by means of~(\ref{pertconditionsa}), where the double trace $E^{ab}{}_{ab}$ plays a non-trivial r\^ole as soon as  as the area metric spacetime fails to be globally flat; this important point will be discussed in the next subsection.   
The other important fact follows from the observation that the solutions of the massive dispersion relation $P^{abcd} p_a p_b p_c p_d=m^4$ do not contain corrections linear in the perturbations $E^{abcd}$ and $e$, which is in stark contrast to the massless dispersion relation $P^{abcd} k_a k_b k_c k_d=0$, which picks up linear contributions from~(\ref{principal-polynomial-2nd}).
While expressed here in the general language of the principal polynomial, the above results are not new; see \cite{Kostelecky:2001mb,Kostelecky:2002hh} for a thorough discussion of such signatures of birefringence on an area metric background. 
Having made these observations, we now remove, in the remainder of the paper, the explicit appearance of the control parameter $\lambda$ from equations.
All this boils down, for the purposes of the present paper, to the \textit{massive} dispersion relation being given by the metric one to relevant order, while the massless dispersion relation is indeed not needed for any of the results we derive.

\subsection{\new{Effectively flat regions}} \label{sec:flat-regions}
Flatness of a region is synonymous, for any type of tensorial geometry $G$, to the existence of coordinate charts covering that region such that the coefficients of the tensor field $G$ are constant with respect to each individual chart. In other words, there is a collection of charts where all partial derivatives of the geometric structure components vanish, so that any tensorial quantities built from these derivatives must vanish with respect to these coordinates and then of course will vanish with respect to any other. In this fashion, Darboux's theorem shows that any symplectic manifold is flat, but for metric geometry the vanishing of the Riemann tensor is a necessary and sufficient criterion for flatness. 

Within a flat region, the ultralocal discussion of the previous section extends from any one point to the entire region for any tensorial geometry. This is because one is always able to find coordinates that take the components of the principal tensor to the normal form (\ref{normform1})--(\ref{normform3}) with $E^{abcd}$ being constant throughout each coordinate chart domain of the flat region, but not beyond.

Only in metric and symplectic geometry is one flat region of some globally non-flat manifold like any other of its flat regions. For while these regions might be far apart and the coefficients of the geometric structure in question constant with respect to different sets of coordinate charts, there are no ultralocal scalars associated with a metric and due to the local flatness also no merely local ones. This is, as we noted at the beginning of the previous section, distinctly different for an area metric manifold. Correspondingly, two different flat regions of an area metric manifold will generically differ by the values taken by the respective ultralocal scalars: while constant throughout each flat region with respect to a suitable coordinate system, these scalars may well differ between the two regions. This directly translates to the values of the restricted perturbations (\ref{pertconditionsa}), see Fig. \ref{fig:quasi-flat-regions}. 

\begin{figure}[hh]
\begin{tikzpicture}
    \coordinate (p1) at (0, 0);
    \coordinate (p2) at (1, 1);
    \coordinate (p3) at (2, 0.5);
    
    \node at (-0.1,0.9) {$R_1$};
    \node at (1,0.5) {$E^{abcd}_1$};
    \begin{pgfonlayer}{background}
      \path[expand bubble]plot [smooth cycle,tension=1] coordinates {(p1) (p2) (p3)};
    \end{pgfonlayer}
  
    \coordinate (r1) at (5, 0.5);
    \coordinate (r2) at ($(r1)+(1.1, 0.7)$);
    \coordinate (r3) at ($(r1)+(1.6, -0.4)$);
    
    \node at ($(r1)+(1.8,0.7)$) {$R_2$};
    \node at ($(r1)+(0.9,0.1)$) {$E^{abcd}_2$};
    \begin{pgfonlayer}{background}
      \path[expand bubble]plot [smooth cycle,tension=1] coordinates {(r1) (r2) (r3)};
    \end{pgfonlayer}  
    
  \end{tikzpicture}
 \caption{Two 
 flat regions $R_1$ and $R_2$ of a globally non-flat area metric manifold whose geometry only weakly differs from that induced by a flat Lorentzian manifold: with respect to suitable coordinate systems, the components $E^{abcd}(x)$ of restricted area metric perturbations  have the 
 constant values $E_1^{abcd}$ for all $x$ in $R_1$ and the 
 constant values $E_2^{abcd}$ for all $x$ in $R_2$. Notably, however, $E_1^{abcd} = E_2^{abcd}$ cannot be achieved in general.}
 \label{fig:quasi-flat-regions}
 \end{figure}
 
One particular consequence of this failure of two separate flat regions to have exactly the same geometry (unless indeed the entire manifold is flat, in which case they trivially do share the same geometry) is that the double trace $e$ (and the thus by (\ref{pertconditionsa}) uniquely determined single trace part $E^{ab}$) of restricted area metric perturbations will generically vary between separate flat regions. Due to this, they are {\it not} absorbable into a redefinition of fields and constant fermion masses and charges in the soon-to-be-considered action (\ref{action-gled-fermion}), as they would be on a globally flat area metric manifold. Indeed, they can be determined experimentally when comparing effects in separate flat regions. This is all the more meaningful since we can independently predict the respective values of these traces in different spacetime regions due to our knowledge of the weak gravitational field equations for the area metric underlying general linear electrodynamics. 

Physically, a small region where the perturbation of the geometry varies sufficiently little will be experimentally indistinguishable from being flat. How large such a region is and whether the considered experiment can be thought to be confined to that region depends on both the spacetime geometry and the specific physical effects one considers. 
For the weakly area metric geometry around a point mass $M$, as it follows from the gravitational closure of general linear electrodynamics and is given at the end of section \ref{subsec:grav_closure}, for instance, each of the components of the area metric is of the form $f (r) \eqdef M/r \, \big(C_1 + C_2\, e^{-r/\ell} \big)$, for a length scale $\ell \eqdef 1/\sqrt{\mu}$ and some constants $C_1$ and $C_2$.
Now very roughly assuming a typical size $\Delta r$ for the processes we investigate, we encounter a relative variation of the function $f(r)$ between $r$ and $r+\Delta r$ of about
\begin{equation} \label{functions-variation1}
 \left|\frac{\Delta f}{f} \right|_r
 =\left| \frac{\Delta r}{r} \right| \cdot \left| \, 1 + \frac{r}{\ell} \, e^{- r/\ell} \, \frac{C_2}{C_1+ C_2 \, e^{-r/\ell}} \, \right| \,.
\end{equation}
Now, we note that the area-metric contribution $f(r)$ differs in general from the form of a standard metric contribution $\bar f(x) = MC/r$, where $C$ is a constant.
However, for the situations that we consider in this paper, we expect the area metric not to produce big deviations from the standard metric solution. 
One can identify three situations in which the area metric contribution is close enough to the standard metric one, so that $(f- \bar f )/ \bar f \ll 1$.
Namely, (i) when $|C_1/C_2|\ll 1$, in which case $C \approx C_1$, (ii) when $r \gg \ell$, in which case again $C \approx C_1$, and (iii) when $r\ll \ell$, in which case $C \approx C_1+C_2$.
In all these three situations (i)--(iii), one can show that the second summand in the right-hand side of~(\ref{functions-variation1}) is negligible, so that one arrives at the approximate equality
\begin{equation} \label{functions-variation2}
 \left|\frac{\Delta f}{f} \right|_r
 \approx \left| \frac{\Delta r}{r} \right| \,.
\end{equation}
Now consider the processes studied in this paper, such as the scatterings of section~\ref{sec:scatterings} and the hydrogen spectroscopy of section~\ref{sec:hydrogen}. The typical size $\Delta r$ for these, including the detector, is not more than some tens of meters.
At the same time, the minimal distance from the point mass $r$ is not, even for compact objects, less than some kilometers and more that $6000$ kilometers for Earth.
For these reasons, we can safely consider $\Delta r/r \ll 1$.
This fact, together with~(\ref{functions-variation2}), implies that indeed $|\Delta f /f |_r \ll 1$. 
In other words, we are justified in assuming that the variations~(\ref{functions-variation1}) are small enough for the area metric to be considered constant.

The remaining and possibly quite appreciable large-scale variation of the area metric geometry then only becomes relevant when comparing the results of two experiments performed in distant regions.
In this case, we can assume that each one of the two experiments take place in a flat region where the area metric is constant, but the specific values of the area metric in the respective regions are different, but indeed predicted by the gravitational field equations. We will see this effect at work in section \ref{sec:around-a-point-mass}.


\section{Quantization and renormalization of general linear electrodynamics} \label{sec:renormalization}

In this section, we begin the study of quantum electrodynamics on a weak area metric background. We consider quantum processes that are sufficiently localizable for us to be justified in considering the spacetime as a patchwork of effectively flat regions which are small enough with respect to the variation of the area metric and big enough with respect to the typical size of the quantum process. 

This affords us the possibility to perform a quantum field theoretic calculation of such processes under the assumption of an underlying flat area metric geometry, which will indeed yield locally reliable results. But since different effectively flat regions may well differ geometrically, as discussed in the previous section, we will still be able to see the imprint of a globally non-flat area metric geometry in local quantum processes. 
Technically, the weakness of the refined gravitational field and the localizability of the quantum electrodynamic effects on such a background boils down to two simple facts. First, the area metric can be assumed to be a small constant perturbation of a flat background metric in each one of these regions.
Second, as long as only sufficiently localized quantum effects are considered, these regions can be formally considered to have an infinite size.

Under these assumptions, we show that quantum electrodynamics with area-metric deviations from a Lorentzian background is renormalizable and gauge invariant to arbitrary loop-order.
To this end, we first introduce Dirac fields on the modified background, then prepare dimensional regularization by extending the theory to arbitrary dimensions and finally quantize \`a la Batalin-Vilkovisky.
The Feynman rules, which we identify in the course of these more formal developments, will be practically relevant for the calculations of scattering amplitudes and vertex corrections in section \ref{sec:scatterings}. We will show how to compare the results obtained in two different effectively flat regions in the case of an area-metric perturbation around a gravitating point mass in section~\ref{sec:around-a-point-mass}.

\subsection{Coupling to massive Dirac fields on flat area metric manifolds}
In order to study an interacting theory of general linear electrodynamics, we introduce massive Dirac fields on a flat area-metric background.
At first, we consider an area metric that may deviate arbitrarily from a metric manifold.
To this end, we consider a field $\psi$ that takes values in some finite-dimensional hermitian inner product space $V$ and satisfies a first derivative order field equation of the form
\begin{equation} \label{generalized-Dirac-equation}
 \Big[ i \gamma^{a} \partial_{a} -m \Big] \psi (x)=0 \,,
\end{equation}
where $m>0$ is the mass of the field and the $\gamma^a$ are four endomorphisms on $V$ chosen such that the above equation possesses the same initial value surfaces as those for general linear electrodynamics. 
This is ensured~\cite{RiveraPhD} if the endomorphisms $\gamma^a$ satisfy the {\it quaternary algebra}
\begin{equation} \label{generalized-Dirac-algebra}
 \gamma^a \gamma^b \gamma^c \gamma^d = P^{abcd}\; \textrm{id}_{V} 
\end{equation}
together with the three trace conditions
\begin{equation} \label{generalized-Dirac-algebra-trace}
 \tr \big[ \gamma^a \big]=0 \,, \qquad \tr \big[ \gamma^a  \gamma^b \big]=0 \,, \qquad  \tr \big[ \gamma^a \gamma^b \gamma^c \big]=0\,.
\end{equation}
Any quadruplet of such matrices $\gamma^a$ thus presents a refinement of the Dirac algebra to an area metric background, and we will thus refer to them as {\it refined Dirac matrices}.
If one can find, additionally, a hermitian endomorphism $\Gamma=\Gamma^{\dagger}$ satisfying
\[
 (\Gamma^{\dagger})^{-1} (\gamma^a)^{\dagger}\Gamma^{\dagger}=\gamma^a\,,
\]
one may also provide an action from which one obtains the refined Dirac equation (\ref{generalized-Dirac-equation}) by variation with respect to $\psi$, or equivalently,  $\bar{\psi} \eqdef \psi^{\dagger} \Gamma$, namely
\begin{equation} \label{fermions-action}
  S[\psi, \bar{\psi}]=\int d^{\,4}x \;  \bar{\psi} (i \gamma^a \partial_a -m) \psi \,.
\end{equation}

Explicit representations of this refined Dirac algebra are hard to obtain in general, but have been provided for special cases of area metrics in \cite{RiveraPhD}. Fortunately, for area metrics that arise as small linear perturbations of a flat metric background, as we wish to consider them here, the Dirac algebra implied by the principal polynomial (\ref{eq:Ppolarization}) effectively reduces to the standard Dirac algebra. This is because, at least for the massive Dirac fermions of interest here, it suffices to consider the dispersion relation (and thus the principal polynomial) to zeroth and first order in the area metric perturbation. To this order, however, the principal polynomial is seen from (\ref{normform1})--(\ref{normform3}) to be simply induced by the flat Minkowski metric. Thus only the electromagnetic gauge potential feels the deviation from a metric geometry by virtue of the dynamics (\ref{Sstrong}), while the massive Dirac field does not.
In the case of a globally non-flat area metric, the above representation is locally still applicable, as long as we may assume to be inside an effectively flat region.

The coupling of a massive Dirac field with the gauge potential of general linear electrodynamics is straightforwardly achieved by use of a charged conserved current $j^a :=  Q \pose \bar\psi \gamma^a \psi$, where $\pose >0$.
We thus obtain, for a small area-metric perturbation inside an effectively flat region in an observer frame, the total action
\begin{equation} \label{action-gled-fermion}
 S[\psi,A]=\int d^{\,4} x\, (1+e)\! \left[ -\frac{1}{8} (\eta^{ac} \eta^{bd} - \eta^{ad} \eta^{bc} + E^{abcd}) F_{ab} F_{cd} +
 \bar{\psi} (i \gamma^a \partial_a - Q \pose \gamma^a A_a -m) \psi \right] 
\end{equation}
for general linear electrodynamics minimally coupled to a Dirac field of charge $Q \pose$, where the perturbations $e$ and $E^{abcd}$ satisfy the conditions~(\ref{pertconditionsa}) and $\gamma^{(a} \gamma^{b)} = \eta^{ab}$.
This action provides the starting point of our analysis of the quantum effects produced by a small linear area-metric perturbation inside an effectively flat region.
In the remainder of this section, we show how to dimensionally regularize the action, quantize it \`a la Batalin-Vilkovisky and show that it is renormalizable at every loop order in a gauge-invariant way.

\subsection{Dimensional regularization} \label{sec:weakly-birefringent-spacetime-complex-dim}
The action (\ref{action-gled-fermion}) readily generalizes to integer dimensions $D \geq 3$, 
with the only difference that $A$, $F$, $E$ and $\eta$ now come as their $D$-dimensional versions, and $\gamma^a$ and $\psi$ are the $2^{\lfloor D/2 \rfloor}$\=/dimensional representations of the Dirac gamma matrices and the Dirac spinor, respectively. Note that there is no explicit appearance of the dimension in the  Lagrangian. 

In contrast, the four-dimensional perturbation conditions (\ref{pertconditionsa}) do pick up an explicit dependence on the dimension $D$. Indeed, while the $D$-dimensional perturbation still takes the form
\begin{equation}\label{pert}
G^{abcd} = \eta^{ac} \eta^{bd} - \eta^{ad} \eta^{bc} + E^{abcd} \qquad \textrm{ and } \qquad \omega = \sqrt{-\det \eta_{ab}} + e\,,
\end{equation}
the restrictions on the corresponding $D$-dimensional perturbations $E^{abcd}$ and $e$ now read
\begin{equation} \label{pertconditionsb}
   E^{amb}{}_{m}=\frac{1}{D} E^{mn}{}_{mn}\,  \eta^{ab} \qquad\textrm{ and } \qquad e = -\frac{1}{2D} E^{mn}{}_{mn} \,,
 \end{equation}
as one shows exactly along the same lines as before for the physical case of four dimensions. The principal polynomial, up to second order corrections, is then again simply the Minkowski metric, now in $D$ dimensions,
$$P^{ab} = \eta^{ab}\,,$$
without explicit dependence on $D$.

The above distinction, between implicit and explicit dependence on the spacetime dimension, plays a role for the extrapolation of the above expressions to $D$ complex dimensions. For in the practical implementation of dimensional regularization, one may treat the implicit dimensional dependence of the various tensorial objects (in our case the electromagnetic field $A_a$, Minkowski metric $\eta^{ab}$, perturbations $E^{abcd}$ and $e$, density $\omega$, Dirac matrices $\gamma^a$ and Dirac spinors $\psi$) entirely formally. The only exception would be presented by tensors and tensor densities (such as $\epsilon^{a_1 a_2 \dots a_D}$) whose rank varies with $D$, but this case does not occur in this article. In contrast, any explicit dependence (such as the value of the trace $\eta^{ab} \eta_{ab} = D$ or the above relations (\ref{pertconditionsb}) between the various perturbation components) is simply extended from the calculated expressions for integer dimensions to complex $D$. We will use the above quantities with this understanding whenever we dimensionally regularize in this article.

\subsection{Batalin-Vilkovisky quantization}\label{sec_BVquant}
We now quantize the theory in the Batalin-Vilkovisky formalism \cite{Batalin:1981jr,Batalin:1984jr}. The proof of renormalizability and gauge invariance to every loop order is comparatively straightforward there.
The procedure is to first parametrize the infinitesimal version of the $U(1)$ gauge symmetry 
\begin{equation*}
 \psi'(x)= e^{-iQ\pose\Lambda(x)} \psi(x) \,, \quad
 \bar{\psi}'(x)= e^{iQ\pose\Lambda(x)} \bar{\psi}(x) \,, \quad
 A'_n(x)=A_n(x)+\partial_n \Lambda (x) \,
\end{equation*}
of the classical action (\ref{action-gled-fermion}) through an anti-commuting field $C(x)$, such that $\Lambda(x)=\theta C(x)$ for a Grassmann number $\theta$.
The infinitesimal gauge transformations thus become
\begin{equation} \label{gauge-transformation-BRST}
 s \psi=-iQ\pose C(x) \psi(x) \,, \qquad  s \bar{\psi}(x)=-iQ\pose \bar{\psi}(x) C(x) \,, \qquad s A_n(x)= \partial_n C (x) \,,
\end{equation}
where the generator $s$ is defined as 
\[
 \theta s \phi (x) \eqdef \phi'(x)-\phi(x)  \qquad \text{for} \quad \phi=\psi,\, \bar{\psi},\, A_n\,.
\]
Another anti-commuting field $\bar{C}(x)$ and an auxiliary field $B(x)$ are introduced to later fix the gauge. It is convenient to collect all these fields in the sextuple $\Phi=(\psi,\bar{\psi},A,C,\bar{C},B)$.
For every field $\Phi^{\alpha}$, one then introduces a BRST source $K_\alpha$, with the opposite statistics (commuting if the field is anti-commuting, and vice versa).
The fields and the BRST sources are conjugate variables with respect to the anti-commuting bracket
\[ 
 (F,G):=\int d^{D} \!x \; \omega \! \left( \frac{\delta_r F}{\delta \Phi^\alpha (x)} \frac{\delta_\ell G}{\delta K_\alpha (x)}
 -\frac{\delta_r F}{\delta K_\alpha (x)} \frac{\delta_\ell G}{\delta \Phi^\alpha (x)} \right)
\]
acting on any two functionals $F(\Phi,K)$ and $G(\Phi,K)$ of the fields and sources. The subscripts $r$ and $\ell$ on the derivatives indicate whether the differentiated variable is sorted to the right or to the left of its prefactor, which is relevant for Grassmann variables.
Next one extends the classical action~(\ref{action-gled-fermion}) in $D$ complex dimensions to an action $S_\textrm{\tiny BV}[\Phi,K]$ that satisfies the master equation $(S_\textrm{\tiny BV},S_\textrm{\tiny BV})=0$ subject to the four boundary conditions
\begin{equation*} 
 S_\textrm{\tiny BV}[\psi,\bar\psi,A,0,\dots,0] = S[\psi,\bar\psi,A] \qquad \textrm{ and }\qquad \frac{\delta_r S_\textrm{\tiny BV}}{\delta K_{\tilde\alpha}}(\Phi,0) = - s \Phi^{\tilde\alpha} 
 \end{equation*}
for $\tilde\alpha$ restricted to $1,2,3$.
Thus one finds the extended action of QED on an effectively flat region of a general linear background to be 
\begin{equation} \label{extended-action}
 \begin{aligned}
  S_\textrm{\tiny BV}[\Phi^\alpha,K_\alpha]=& \int d^{D}\! x\, \omega \left[ -\frac{1}{4} F_{ab} F^{ab} -\frac{1}{8} F_{ab} F_{cd} E^{abcd} 
  +\bar{\psi} (i\slashed{\partial}-m-Q\pose \slashed{A}) \psi \right]+ \\
  +&\int d^{D}\! x\, \omega \left[ -\frac{\lambda}{2} B^2+ B \partial^n A_n-\bar{C} \square C \right]+ \\
  +&\int d^{D}\! x\, \omega \,\Big[ K_A^n \partial_n C +iQ\pose\bar{\psi} C K_{\bar{\psi}} +iQ\pose K_\psi C \psi -BK_{\bar{C}} \Big] \,.
 \end{aligned}
\end{equation}
The first line is the classical action for birefringent electrodynamics with a minimally coupled fermion, the second line is the gauge fixing sector and the last line ensures the boundary conditions above.
Since this whole construction only builds on the gauge symmetry of the theory, it is not surprising that the only non-metric term in the extended action above is the linear one inherited from the classical action.

The extended action~(\ref{extended-action}) becomes the bare extended action if all the fields, coupling constants and external sources are replaced by their bare counterparts, which we denote either with a subscript or a superscript $B$.
After integrating the auxiliary field $B(x)$ out of the path integral, the ghosts $C(x)$ and $\bar{C}(x)$ decouple and after setting to zero also the BRST sources, the remaining action reads 
\begin{equation}\label{qed-action}
 S_B[\psi_B, A_B]\!=\!\int\!d^{D}\!x (1+e_B)\!\left[
 -\frac{1}{4} F_{ab}^B F^{ab}_B -\frac{1}{8} F_{ab}^B F_{cd}^B E^{abcd}_B
+\bar \psi_B (i\slashed{\partial}\!-\!m_B\!-\!Q\pose^B\slashed{A}_B) \psi_B 
 \!+\!\frac{1}{2 \lambda_B} (\partial_a A^a_B)^2
 \right] \,,
\end{equation}
with the usual Lorenz gauge-fixing term. Thus one directly reads off the Feynman rules. First we find the {\it photon propagator}  
\begin{equation} \label{photon-propagator}
\begin{tikzpicture}[baseline={([yshift=-2mm]current bounding box.center)}]
\coordinate[label=left:$a$] (e1) at (-.75,0);
\coordinate[label=right:$b$] (e2) at (.75,0);
\coordinate[label=above:$q$] (lab) at (0,0);
\path [sines/.style={
        line join=round, 
        draw=black, 
        decorate, 
        decoration={complete sines, number of sines=5, amplitude=2mm}},postaction={sines}] (e1) -- (e2);
\end{tikzpicture}
=-\frac{i}{q^2+i\epsilon}\left[ \eta_{ab}+ (\lambda-1) \frac{q_a q_b}{q^2+i\epsilon} -E_{arbs} \frac{q^r q^s}{q^2+i\epsilon} \right]\,.
\end{equation}
The photon polarizations 
can be found explicitly only when the specific form of the perturbation $E^{abcd}$ is known, \new{ see~\cite{SME-scattering,Kostelecky:2001mb,Kostelecky:2002hh},} where the perturbation $E$ is denoted by $2 k_F$ and does not satisfy the restrictions~(\ref{pertconditionsb}).
However, the diagrams we consider in section~\ref{sec:scatterings} only contain fermions as ingoing and outgoing particles, thus we do not have to worry about photon polarizations. Secondly, we find the {\it fermion propagator} 
\begin{equation} \label{fermion-propagator}
\begin{tikzpicture}[baseline={([yshift=-2mm]current bounding box.center)}]
\coordinate[label=left:$\alpha$] (e1) at (-.75,0);
\coordinate[label=right:$\beta$] (e2) at (.75,0);
\coordinate[label=above:$p$] (lab) at (0,0);
\draw[fermion] (e2) -- (e1);
\end{tikzpicture}
=\frac{i(\slashed{p}+m)^{\alpha}_{\;\; \beta}}{p^2-m^2+i\epsilon}\,,
\end{equation}
and finally the {\it vertex}
\begin{equation} \label{vertex}
\begin{tikzpicture}[baseline={(current bounding box.center)}]
\coordinate (aux) at (0,0);
\coordinate[label=left:$\alpha$] (e1) at (-.75,.5);
\coordinate[label=right:$\beta$] (e2) at (.75,.5);
\coordinate[label=below:$n$] (e3) at (0,-0.6);

\draw[fermion] (aux) -- (e1);
\draw[fermion] (e2) -- (aux);
\path [sines/.style={
        line join=round, 
        draw=black, 
        decorate, 
        decoration={complete sines, number of sines=2, amplitude=2mm}},postaction={sines}] (aux) -- (e3);
\end{tikzpicture}
=-i Q \pose \,\mu^{\varepsilon/2} (\gamma^n)^{\alpha}_{\;\; \beta}\,.
\end{equation}
The factor $\mu^{\varepsilon/2}$ in the vertex, employing an arbitrary mass scale $\mu$, has been introduced in order for the charge to have correct physical dimension in complex $D = 4-\varepsilon$ spacetime dimensions.
We immediately notice that both, the fermion propagator and the vertex, are unchanged compared to standard QED on flat Minkowski spacetime.
This is due to the fact that the Dirac algebra remains unchanged, which also implies that the free Dirac spinors are the usual $u(\vect{p},s)$ and $v(\vect{p},s)$.

\subsection{Gauge invariance and renormalizability at every loop} \label{sec:gauge-invariance-renormalization}
It is now straightforward to establish that the maximal linear extension of quantum electrodynamics, from the usual Lorentzian metric background to an effectively flat region of a general linear background, is renormalizable and thus physically meaningful.

For redefining the fields, the couplings and the external sources in the bare action as
\begin{equation} \label{field-coupling-redefinitions}
 \begin{aligned}
  (\Phi^\alpha)_B &= (Z_{\Phi^\alpha})^{1/2} \Phi^\alpha \,,	\quad & (K_\alpha)_B &= Z_{K_\alpha} K_\alpha  \,,	\quad
  & E^{abcd}_B &= Z_E E^{abcd} \,,\\
  m_B &= Z_m m \,,		\quad & Q \pose^B &= Z_{\pose} \mu^{\varepsilon/2} Q \pose \,,	\quad & \lambda_B &= Z_\lambda \lambda \,,
 \end{aligned}
\end{equation}
where no sum over $\alpha$ is understood, one obtains the renormalized action, whose classical sector is
\begin{equation} \label{renormalized-classical-action}
  S= \int d^{D}\! x\, \omega \left[ -\frac{1}{4} Z_A F_{ab} F^{ab} -\frac{1}{8} Z_A Z_E F_{ab} F_{cd} E^{abcd}  
  +Z_{\psi} \bar{\psi} (i\slashed{\partial}-Z_m m-Z_{\pose} Z_A^{1/2} \mu^{\varepsilon/2} Q \pose \slashed{A}) \psi \right] \,.
\end{equation}
At the end of this section, we show that every divergence can be eliminated by a suitable choice of the renormalization constants $Z$, of which we now determine the independent ones.
Since the ghosts decouple and one cannot build any one-particle-irreducible diagrams with exterior legs provided by the auxiliary field $B$ or the source $K_\alpha$, the gauge fixing sector and the sources sector do not renormalize.
This fact and the Ward identity~(\ref{Ward-identity-vertex-fermion-self-energy}), detailed later on in this subsection, imply the relations
\[
 Z_A=Z_{\pose}^{-2} \,, \qquad 	Z_\lambda=Z_{\pose}^{-2} \,,	\qquad 	Z_{K_\alpha}=Z_{\pose}^{-1}(Z_{\Phi^\alpha})^{-1/2} 
\]
between the renormalization constants.
Apart from the field and coupling redefinitions~(\ref{field-coupling-redefinitions}), we also allow the scalar density perturbation $e$ to be renormalized by 
$$e_B= Z_e\, e\,.$$ 
Requiring, however, that~(\ref{pertconditionsb}), the second restriction on the perturbation, holds after renormalization, immediately yields $Z_e=Z_E$. Thus the scalar density perturbation renormalizes exactly as the tensor perturbation. 
In summary, the only renormalization constants left to be determined are $Z_\psi$, $Z_m$, $Z_{\pose}$ and $Z_E$.

Before we can prove the renormalizability of the theory at every loop, we need to discuss the possible contributions of the perturbation $E$ to the
{\it vacuum polarization}
\begin{equation} \label{vacuum-polarization}
\begin{tikzpicture}[baseline={([yshift=-2mm]current bounding box.center)}]
\coordinate[label=left:$a$] (e1) at (-.9,0);
\coordinate[label=right:$b$] (e2) at (.9,0);
\coordinate (aux) at (0,0);
\coordinate (aux1) at (-.85,.25);
\coordinate (aux2) at (-.4,.25);
\coordinate[label=above:$q$] (label) at ($(aux1)!0.5!(aux2)$);
\node [circle,draw,inner sep=2.5mm,pattern=north east lines] (blob) at (aux) {};
\path [sines/.style={
        line join=round, 
        draw=black, 
        decorate, 
        decoration={complete sines, number of sines=2, amplitude=2mm}},postaction={sines}] (e1) -- (blob);
\path [sines/.style={
        line join=round, 
        draw=black, 
        decorate, 
        decoration={complete sines, number of sines=2, amplitude=2mm}},postaction={sines}] (blob) -- (e2);
\draw [->] (aux1) -- (aux2);        
\end{tikzpicture}
\eqdef i\Pi^{ab} (q) \,,
\end{equation}
which collects the one-particle-irreducible radiative corrections to the photon propagator, to the {\it fermion self-energy}
\begin{equation} \label{fermion-self-energy}
\begin{tikzpicture}[baseline={([yshift=-2mm]current bounding box.center)}]
\coordinate[label=left:$\alpha$] (e1) at (-.9,0);
\coordinate[label=right:$\beta$] (e2) at (.9,0);
\coordinate (aux) at (0,0);
\coordinate[label=above:$p$] (label) at (.6,0);
\node [circle,draw,inner sep=2.5mm,pattern=north east lines] (blob) at (aux) {};
\draw[fermion] (e2) -- (blob);
\draw[fermion] (blob) -- (e1);
\end{tikzpicture}
\eqdef -i  \big[\Sigma(p) \big]^\alpha_{\; \beta} \,,
\end{equation}
which collects the one-particle-irreducible radiative corrections to the fermion propagator, and to the {\it proper vertex}
\begin{equation} \label{proper-vertex}
\begin{tikzpicture}[baseline={(current bounding box.center)}]
\coordinate (aux) at (0,0);
\coordinate [label=left:$p'$,label=right:$\alpha$] (e1) at (-1,.8);
\coordinate [label=right:$p$,label=left:$\beta$] (e2) at (1,.8);
\coordinate [label=below:$n$] (e3) at (0,-.8);
\coordinate (arr1) at (.3,-.8);
\coordinate (arr2) at (.3,-.4);
\coordinate [label=right:$q$] (arrlab) at ($(arr1)!0.5!(arr2)$);
\node [circle,draw,inner sep=2.5mm,pattern=north east lines] (blob) at (aux) {};
\draw [fermion] (e2) -- (blob);
\draw [fermion] (blob) -- (e1);
\path [sines/.style={
        line join=round, 
        draw=black, 
        decorate, 
        decoration={complete sines, number of sines=1.5, amplitude=2mm}},postaction={sines}] (blob) -- (e3);
\draw[->] (arr1) -- (arr2);
\end{tikzpicture}
\eqdef -i Q\pose \big[\Gamma^n (p',p) \big]^\alpha_{\;\;\beta} \,,
\end{equation}
which collects the one-particle-irreducible radiative corrections to the vertex.
To this end, we first notice that every term already appearing in standard QED may now appear multiplied by $E^{ij}{}_{ij}$.
This, in turn, implies that the renormalization constants $Z_\psi$, $Z_m$ and $Z_{\pose}$ may now also contain a part linear in $E^{ij}{}_{ij}$.
The analysis of the other possible contributions is greatly simplified if we replace $E_{arbs}$ in the photon propagator~(\ref{photon-propagator}) with $(E_{arbs}+E_{bras})/2$. This substitution does not change the value of the propagator, by virtue of the symmetries~(\ref{area-metric-symmetries}).
We begin by writing all possible Lorentz-covariant terms, using as building blocks the momenta of the external legs, the gamma matrices and the components  $E^{abcd}$ in the particular combination described above.
We then simplify each of them by use of the symmetries~(\ref{area-metric-symmetries}) and the first restriction ~(\ref{pertconditionsb}) on the perturbation.

In the vacuum polarization~(\ref{vacuum-polarization}), there is a new divergent term proportional to $E^{arbs} q_r q_s$, which can be eliminated through a renormalization of $E^{abcd}$ and yields the value of $Z_E$.
In contrast to the other renormalization constants, $Z_E$ cannot depend on $E^{ij}{}_{ij}$ because such a contribution would be of second order in $E$.
The fermion self-energy~(\ref{fermion-self-energy}) does not contain any other new term.
The proper vertex~(\ref{proper-vertex}) is related with the fermion self-energy through the Ward identity
\begin{equation} \label{Ward-identity-vertex-fermion-self-energy}
 \Gamma^n(p,p)=\gamma^n - \frac{\partial \Sigma (p)}{\partial p_n} \,,
\end{equation}
which holds also in our theory, since the gauge-fixing and BRST source sectors of the extended action~(\ref{extended-action}) are the same as in metric QED. Thus, also the proper vertex does not contain any other new contribution.

At this point, we are ready to prove the renormalizability of the theory at every loop. For it can be shown that since the bare extended action $S_B$ satisfies the master equation \mbox{$(S_B,S_B)=0$}, also the bare generating functional of the one-particle-irreducible diagrams $\Gamma_B$ does, so that \mbox{$(\Gamma_B,\Gamma_B)=0$}.
The proof then proceeds using induction on the number of loops in order to show that the divergences can be removed preserving the master equation, as in the standard case  \cite{Anselmi:renormalization,Weinberg:1996kr}.
The only difference arises when considering all the possible divergent gauge-invariant local terms of dimension less than or equal to 4. Other than the standard terms $F_{ab} F^{ab}$, $\bar{\psi} \psi$ and $\bar{\psi} \slashed{D} \psi$, which can be eliminated by a suitable choice of $Z_\psi$, $Z_e$ and $Z_m$, we also need to include $E^{abcd} F_{ab} F_{cd}$, which can be eliminated by a suitable choice of $Z_E$.
Thus, general linear electrodynamics can be renormalized at every loop in a gauge-invariant way. Note that the restrictions~(\ref{pertconditionsb}) on the perturbations were central to the background of our proof. If they are not heeded in the renormalization process, new interactions in the loop corrections of the fermion sector are generated, see~\cite{Santos:2015koa}.

\subsection{Renormalization of the \new{fermion propagator and the vertex} in the on-shell scheme} \label{sec:one-loop-renormalization-constants}
\new{So far we have dealt with the foundations  and the general aspects of the renormalization of the theory.
For the future application to the the scattering of an electron in an external magnetic field, in section~\ref{sec:electron-external-magnetic-field}, we still need to compute the explicit value of the renormalization constants involved in the vertex.
For this reason,} using dimensional regularization in $D=4-\varepsilon$ complex dimensions, we now work out the relevant one-loop renormalization constants in the on-shell scheme, where the renormalized propagators have poles in the physical masses with a residue of 1, and the renormalized electric charge is the physical one.

\new{The on-shell renormalization of the vertex is achieved by requiring that the non-relativistic potential of a charged particle in a quasi-static and uniform electric field, obtained by means of the Born approximation, is the renormalized electric charge in the on-shell scheme times the electrostatic potential of the external field.
This is obtained by imposing the condition
\begin{equation*}
 \lim_{q \rightarrow 0} \bar{u}(p') \big[ -iQ\pose \Gamma^n (p',p) \big] u(p) = \bar{u}(p') \big[-iQ\pose \gamma^n \big] u(p) \,,
\end{equation*} 
on the proper vertex in renormalized perturbation theory, i.e., with the counterterms included.
By virtue of the Ward identity~(\ref{Ward-identity-vertex-fermion-self-energy}), the above condition is satisfied as long as $Z_e^{\, \text{o.s.}}=\big( Z_A^{\, \text{o.s.}}\big)^{-1/2}$, from which follows that the renormalization constant for the vertex is $Z_\psi^{\, \text{o.s.}} Z_e^{\, \text{o.s.}} \big( Z_A^{\, \text{o.s.}}\big)^{1/2}=Z_\psi^{\, \text{o.s.}}$.
Thus, for the purpose of renormalizing the vertex, it is enough to compute the renormalization constant of the fermion wave function $Z_\psi^{\, \text{o.s.}}$. 
In practice,  one needs to consider the fermion self energy in renormalized perturbation theory and impose the on-shell condition that the ensuing fermion propagator has a pole in the physical mass with a residue of 1.
This translates into the two conditions 
\[
 \Sigma(p)|_{\slashed{p}=m}=0
 \qquad \text{and} \qquad
 \left[ \frac{\partial \Sigma(p)}{\partial \slashed{p}} \right]_{\slashed{p}=m}=0 
\]
on the fermion self-energy $-i\Sigma(p)$ in renormalized perturbation theory.
From these, one obtains the one-loop renormalization constants
\begin{equation} \label{renormalization-constants-fermion-propagator}
 Z_\psi^{\, \text{o.s.}}=1+ \left[ \frac{\partial \Sigma_{\text{1-loop}}(p)}{\partial \slashed{p}} \right]_{\slashed{p}=m}
 \qquad \text{and} \qquad
 Z_m^{\, \text{o.s.}}=1- \frac{ \Sigma_{\text{1-loop}}(p)|_{\slashed{p}=m}}{m} \,.
\end{equation}

After the above considerations, the only} diagram to be taken under consideration is the {\it one-loop fermion self-energy}
\[
\begin{tikzpicture}[baseline={([yshift=-2mm]current bounding box.center)}]
\coordinate[label=left:$\alpha$] (e1) at (-1.3,0);
\coordinate[label=right:$\beta$] (e2) at (1.3,0);
\coordinate (i1) at (-.75,0);
\coordinate (i2) at (.75,0);
\coordinate (aux) at (0,0);
\coordinate[label=above:$p$] (label) at (1.1,0);
\draw[draw=black, postaction={decorate}, decoration={markings,mark=at position 0.5 with {\arrow[scale=.7,black]{triangle 45}}}] (e2) -- (i2);
\draw[draw=black, postaction={decorate}, decoration={markings,mark=at position 0.7 with {\arrow[scale=.7,black]{triangle 45}}}] (i1) -- (e1);
\draw[fermion] (i2) -- (i1);
\draw[draw=none,sines/.style={
        line join=round, 
        draw=black, 
        decorate, 
        decoration={complete sines, number of sines=5.5, amplitude=2mm}},postaction={sines}] (i1) arc (180:0:.75cm);
\end{tikzpicture}
\eqdef -i  \big[\Sigma_{\text{1-loop}}(p) \big]^\alpha_{\; \beta} \,,
\]
which now depends on the perturbation $E$, due to the presence of the photon propagator.
In order to avoid infrared divergences, it is useful to replace the photon propagator~(\ref{photon-propagator}) with
\begin{equation} \label{photon-propagator-IR-regulated}
\begin{tikzpicture}[baseline={([yshift=-2mm]current bounding box.center)}]
\coordinate[label=left:$a$] (e1) at (-.75,0);
\coordinate[label=right:$b$] (e2) at (.75,0);
\coordinate[label=above:$q$] (lab) at (0,0);
\path [sines/.style={
        line join=round, 
        draw=black, 
        decorate, 
        decoration={complete sines, number of sines=5, amplitude=2mm}},postaction={sines}] (e1) -- (e2);
\end{tikzpicture}
 =-\frac{i}{q^2-\delta^2+i\epsilon}\left[ \eta_{ab}+ (\lambda-1) \frac{q_a q_b}{q^2-\lambda\delta^2+i\epsilon}
 -E_{arbs} \frac{q^r q^s}{q^2-\delta^2+i\epsilon} \right]\,,
\end{equation}
where $\delta$ is a small photon mass used as an infrared regulator.
The Feynman prescription $i \epsilon$ will not be shown explicitly in the following equations.
The one-loop fermion self-energy
\begin{equation*}
 -i \Sigma_{\text{1-loop}}(p)\!=\!\big(iQ\pose \mu^{\varepsilon/2}\, \big)^2\! \int\! \frac{d^D k}{(2\pi)^D} \gamma^a \frac{i(\slashed{p}+\slashed{k}+m)}{(p+k)^2-m^2} \gamma^b
 \left\{\!-\frac{i}{k^2\!-\!\delta^2} \left[ \eta_{ab}\!+\!(\lambda\!-\!1) \frac{k_a k_b}{k^2\!-\!\lambda \delta^2}\!-\!E_{arbs} \frac{k^r k^s}{k^2\!-\!\delta^2}  \right] \right\}
\end{equation*}
can be split into a sum of the zeroth order contribution in $E$, denoted by $-i \Sigma_{\text{1-loop}}^{0}(p)$, and the first order contribution, denoted by $-i \Sigma_{\text{1-loop}}^{E}(p)$.
The former is the standard QED contribution
\begin{equation} \label{fermion-self-energy-divergent-QED}
\begin{aligned}
 -i \Sigma_{\text{1-loop}}^{0}(p) =
 & -3 \frac{iQ^2 \pose^2}{(4\pi)^2} \left[ \frac{2}{\varepsilon}-\gamma_E+\log \left( \frac{4\pi\mu^2}{m^2} \right) +\frac{4}{3} \right]m\\
 & +(\slashed{p}-m) \frac{iQ^2 \pose^2}{(4\pi)^2}
 \left\{ \lambda \left[ \frac{2}{\varepsilon}-\gamma_E+\log \left( \frac{4\pi\mu^2}{m^2} \right) +4-2 \log \left( \frac{m^2}{\delta^2} \right) \right] \right. \\
 &\left.-(\lambda-1)\left[ 3+\frac{\lambda \log \lambda}{\lambda-1} -3\log \left( \frac{m^2}{\delta^2} \right) \right]\right\}
 +\bigo \big((\slashed{p}-m)^2 \big)\,.
\end{aligned}
\end{equation}
On the other hand, the contribution linear in $E$ is
\[
\begin{aligned}
 -i \Sigma_{\text{1-loop}}^{H}(p) = &
  \frac{iQ^2 \pose^2}{4(4\pi)^2}\, E^{ij}_{\;\;\; ij} \left[ \frac{2}{\varepsilon}-\gamma_E+\log \left( \frac{4\pi\mu^2}{m^2} \right) +\frac{5}{2} \right]m\\
 &+ \frac{iQ^2 \pose^2}{4(4\pi)^2}\, E^{ij}_{\;\;\; ij} \left[ \log \left( \frac{m^2}{\delta^2} \right)-2 \right](\slashed{p}-m)  +\bigo \big((\slashed{p}-m)^2 \big)\,.
 \end{aligned}
\]
Using the above result, one finds the one-loop values
\begin{multline} \label{Zpsi-on-shell}
 Z_\psi^{\,\text{o.s.}}=1-\frac{Q^2 \pose^2}{(4\pi)^2}
 \left\{ \lambda \left[ \frac{2}{\varepsilon}-\gamma_E+\log \left( \frac{4\pi\mu^2}{m^2} \right) +4-2 \log \left( \frac{m^2}{\delta^2} \right) \right] \right. \\
 \left.-(\lambda-1)\left[ 3+\frac{\lambda \log \lambda}{\lambda-1} -3\log \left( \frac{m^2}{\delta^2} \right) \right]
 -\frac{E^{ij}_{\;\;\; ij}}{4} \left[ 2-\log \left( \frac{m^2}{\delta^2} \right) \right] \right\}
\end{multline}
and
\begin{equation} \label{Zm-on-shell}
 Z_m^{\,\text{o.s.}}=1-\frac{3Q^2 \pose^2}{(4 \pi)^2 }
 \left\{ \left(1-\frac{E^{ij}_{\;\;\; ij}}{12}  \right)\! \left[\frac{2}{\varepsilon}-\gamma_E+\log \left( \frac{4\pi\mu^2}{m^2} \right) \right] +\frac{4}{3} -\frac{5}{24}E^{ij}_{\;\;\; ij} \right\}
\end{equation}
for the other two renormalization constants.

\new{We have thus determined the renormalization constants $Z_\psi^{\,\text{o.s.}}$ and $Z_m^{\,\text{o.s.}}$ at one-loop in the on-shell scheme. 
This was possible analyzing the the one-loop fermion self-energy, which contains a new term linear in the full trace of the perturbation $E^{ij}{}_{ij}$.

The one-loop on-shell renormalization of  electrodynamics on an effectively flat region of a general linear background} can find its application also in the photon sector of the SME~\cite{SME-flat-space,SME-curved-spacetime}. In this context, indeed, the one-loop renormalization was worked out considering the divergent part of the relevant diagrams~\cite{SME-one-loop-renormalization}, without the finite contributions necessary for the on-shell scheme.

\section{Scattering and anomalous magnetic moment} \label{sec:scatterings}
We now calculate potentially measurable effects using the Feynman rules for the photon propagator (\ref{photon-propagator}), fermion propagator (\ref{fermion-propagator}) and vertex (\ref{vertex}) for general linear quantum electrodynamics.
This is pursued by analizing the effects of the perturbation~(\ref{pert}) in three different scattering processes: $e^+ e^-\rightarrow \bar{f}f$, Bhabha scattering and the scattering of an electron in an external magnetic field.
Several scattering processes with similar backgrounds, such as SME, have been considered in the literature: see~\cite{SME-scattering} for the study of $e^+ e^- \rightarrow \gamma\gamma$ with a perturbation close to~(\ref{pert}) and~\cite{SME-Bhabha} for the study of Bhabha scattering with a modified Dirac sector, but a standard photon sector in the action.

\subsection{$e^+ e^-\rightarrow \bar{f}f$ scattering} \label{sec:e+e-fermion-antifermion}
The first process taken into consideration is the scattering $e^+ e^-\rightarrow \bar{f}f$, for a fermion $f$ different from $e^\pm$, but otherwise generic.
This process contains only fermions as external legs for which we can use the usual free Dirac spinors $u(\vect{p},s)$ and $v(\vect{p},s)$, since the Dirac sector in~(\ref{action-gled-fermion}) and the Dirac algebra remain unchanged by the perturbation~(\ref{pert}), as already mentioned in section~\ref{sec_BVquant}. At tree level, the scattering consists in the diagram
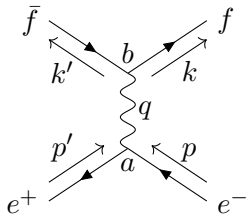
\begin{center}
 \begin{tikzpicture}[baseline={(current bounding box.center)}]
\coordinate[label=right:$q$] (c) at (0,0);
\coordinate[label=below:$a$] (v1) at (0,-.5);
\coordinate[label=above:$b$] (v2) at (0,+.5);
\coordinate[label=left:$e^+$] (e1) at ($(-1.05,-.7)+(v1)$);
\coordinate[label=right:$e^-$] (e2) at ($(+1.05,-.7)+(v1)$);
\coordinate[label=left:$\bar{f}$] (f1) at ($(-1.05,.7)+(v2)$);
\coordinate[label=right:$f$] (f2) at ($(+1.05,.7)+(v2)$);

\coordinate[above=2.5mm of e1] (a11);
\coordinate (a12) at ($(a11)+.7*(v1)-.7*(e1)$);
\coordinate[label=above left:$p'$] (a1) at ($(a11)!0.5!(a12)+(+.1,-.1)$);

\coordinate[above=2.5mm of e2] (a21);
\coordinate (a22) at ($(a21)+.7*(v1)-.7*(e2)$);
\coordinate[label=above right:$p$] (a1) at ($(a21)!0.5!(a22)+(-.1,-.1)$);

\coordinate[below=2.5mm of f1] (b12);
\coordinate (b11) at ($(b12)+.7*(v2)-.7*(f1)$);
\coordinate[label=below left:$k'$] (b1) at ($(b11)!0.5!(b12)+(+.1,+.1)$);

\coordinate[below=2.5mm of f2] (b22);
\coordinate (b21) at ($(b22)+.7*(v2)-.7*(f2)$);
\coordinate[label=below right:$k$] (b2) at ($(b21)!0.5!(b22)+(-.1,+.1)$);

\draw[fermion] (e2) -- (v1);
\draw[fermion] (v1) -- (e1);
\draw[fermion] (f1) -- (v2);
\draw[fermion] (v2) -- (f2);
\path [sines/.style={
        line join=round, 
        draw=black, 
        decorate, 
        decoration={complete sines, number of sines=3, amplitude=2mm}},postaction={sines}] (v1) -- (v2);
        
\draw[->] (a11) -- (a12);
\draw[->] (a21) -- (a22);
\draw[->] (b11) -- (b12);
\draw[->] (b21) -- (b22);
\end{tikzpicture}
.
\end{center}
Here, the electron has charge $-\pose$ and mass $m$, while $f$ has charge $Q\pose$ and mass $M$. The tree-level amplitude in the Feynman gauge ($\lambda=1$) is
\[
 i \amplitude = \bar{u}_f (k) \big[-iQ\pose\gamma^b \big] v_f (k') \bar{v}_e (p') \big[i\pose\gamma^a \big] u_e (p)  \frac{-i}{q^2} \left[ \eta_{ab} -E_{ambn}\frac{q^m q^n}{q^2} \right] \,,
\]
where $\varepsilon=0$, since the tree-level process is not divergent.
In order to simplify the results, we move to the center of mass frame through a Lorentz transformation.
One should only keep in mind that $E^{abcd}$ needs to be transformed accordingly with a Lorentzian change of frame.
We further choose the orientation of the axes in such a way that
\begin{gather*}
 \vect{p} = -\vect{p}'= |\vect{p}| \hat{z} \quad \text{and} \\
 \vect{k} = -\vect{k}'= |\vect{k}| \hat{r} \eqdef |\vect{k}| (\sin \theta \cos \varphi \,\hat{x} + \sin \theta \sin \varphi \,\hat{y} + \cos \theta \,\hat{z} ) \,.
\end{gather*}
Since spin polarizations are often not measured experimentally, we then take the average of the initial spin polarizations, the sum of the final spin polarizations and compute the differential cross section, which, with the help of the first restriction~(\ref{pertconditionsb}) on the perturbation, can be put in the form
\begin{equation} \label{cross-section-muons}
 \begin{aligned}
  \frac{d \sigma}{d \Omega} &= \frac{Q^2 \alpha^2}{4s} \sqrt{\frac{s-4 M^2}{s-4 m^2}} \left\{
  1\!+\! \frac{4(m^2\!+\!M^2)}{s}+\! \left( 1- \frac{4 m^2}{s} \right)\!\! \left( 1- \frac{4 M^2}{s} \right)\! \cos^2 \theta + \right. \\
  &-2 \sin^2\! \theta \! \left( 1-\frac{4 M^2}{s} \right)\! \big( \cos^2\! \varphi\, E^{0101}\!\!  + \sin^2\! \varphi \, E^{0202}\!+2\sin \varphi \cos \varphi \,E^{0102} \big) + \\
  &-2 \left[ \left( 1-\frac{4 m^2}{s} \right) + \frac{4 m^2}{s} \left( 1-\frac{4 M^2}{s} \right) \cos^2 \theta \right]  E^{0303} -\frac{1}{2} E^{ij}_{\;\;\; ij}+\\
  &-2 \left.  \sin \theta \cos \theta \left( 1+\frac{4 m^2}{s} \right) \left( 1-\frac{4 M^2}{s} \right) \big( \cos \varphi \, E^{0103}+\sin \varphi \, E^{0203}  \big)  \right\} \,,
 \end{aligned}
\end{equation}
where $\alpha=\pose^2/(4\pi)$ is the fine structure constant and $s \eqdef (p+p')^2$ is one of Mandelstam's variables.

We notice that the factor presented by the curly brackets in~(\ref{cross-section-muons}) is not symmetric in $m$ and $M$.
However, this does not imply a violation of time-reversal invariance when computing the inverse scattering process $\bar{f} f \rightarrow e^+ e^-$ cross section and comparing the result with the time-reversed one of the direct process.
This is seen as follows.
The cross section of the inverse process can be indeed inferred when switching $m$ with $M$, but is expressed in the frame where $\vect{k}$ is along $\hat{z}$ and $\vect{p}$ along $\hat{r}$.
The cross-section of the time-reversed process instead is expressed in the frame where $\vect{p}$ is along $-\hat{z}$ and $\vect{k}$ along $-\hat{r}$, and does not change by flipping the sing of all the spatial momenta.
The two frames are related by the rotation that interchanges $\vect{k}/|\vect{k}|$ and $\vect{p}/|\vect{p}|$.
If we express the components of $E$ in the one frame in terms of its components in the other, we indeed find the differential cross section of the inverse process starting from the one of the direct process, implying that time-reversal is not violated.

Even if the detector does not resolve different angles $\varphi$, we still have an effect. Indeed, 
integrating out $\varphi$ in the cross section and using again the first restriction ~(\ref{pertconditionsb}) on the perturbation, we find
\begin{align*}
  \frac{d \sigma}{d \cos \theta} &= \frac{\pi Q^2 \alpha^2}{2s}  \sqrt{\frac{s-4 M^2}{s-4 m^2}} \left\{
  1+ \frac{4(m^2+M^2)}{s}+ \left( 1- \frac{4 m^2}{s} \right) \left( 1- \frac{4 M^2}{s} \right) \cos^2 \theta + \right. \\
  &- \left[ \left( 1-\frac{8 m^2}{s}+\frac{4 M^2}{s} \right) + \left( 1+ \frac{8 m^2}{s} \right) \left( 1-\frac{4 M^2}{s} \right) \cos^2 \theta \right] E^{0303} + \\
  &\left. - \frac{1}{4} \left[ \left(1+\frac{4M^2}{s}\right)+\left(1-\frac{4M^2}{s}\right) \cos^2\theta \right] E^{ij}_{\;\;\; ij} \right\} \,.
\end{align*}

\subsection{Bhabha scattering} \label{sec:Bhabha}
The second process we take into consideration is the Bhabha scattering $e^+ e^- \rightarrow e^+ e^-$,
which is of direct interest for the determination of luminosity~\cite{Bhabha-colliders}, both in past~\cite{LEP-Bhabha} and in future $e^+ e^-$ colliders~\cite{TLEP-Bhabha,ILC-Bhabha}.
As for the previous scattering, only fermions appear as external legs and we can use for them the usual free Dirac spinors $u(\vect{p},s)$ and $v(\vect{p},s)$, since the Dirac sector in~(\ref{action-gled-fermion}) and the Dirac algebra remain unchanged by the perturbation~(\ref{pert}). At tree level, the scattering consists in the sum of the two diagrams
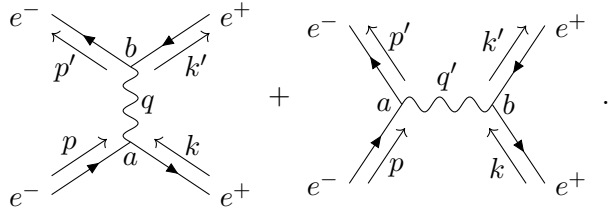
\[
\begin{tikzpicture}[baseline={(current bounding box.center)}]
\coordinate[label=right:$q$] (c) at (0,0);
\coordinate[label=below:$a$] (v1) at (0,-.5);
\coordinate[label=above:$b$] (v2) at (0,+.5);
\coordinate[label=left:$e^-$] (e1) at ($(-1.05,-.7)+(v1)$);
\coordinate[label=right:$e^+$] (e2) at ($(+1.05,-.7)+(v1)$);
\coordinate[label=left:$e^-$] (f1) at ($(-1.05,.7)+(v2)$);
\coordinate[label=right:$e^+$] (f2) at ($(+1.05,.7)+(v2)$);
\coordinate[above=2.5mm of e1] (a11);
\coordinate (a12) at ($(a11)+.7*(v1)-.7*(e1)$);
\coordinate[label=above left:$p$] (a1) at ($(a11)!0.5!(a12)+(+.1,-.1)$);
\coordinate[above=2.5mm of e2] (a21);
\coordinate (a22) at ($(a21)+.7*(v1)-.7*(e2)$);
\coordinate[label=above right:$k$] (a1) at ($(a21)!0.5!(a22)+(-.1,-.1)$);
\coordinate[below=2.5mm of f1] (b12);
\coordinate (b11) at ($(b12)+.7*(v2)-.7*(f1)$);
\coordinate[label=below left:$p'$] (b1) at ($(b11)!0.5!(b12)+(+.1,+.1)$);
\coordinate[below=2.5mm of f2] (b22);
\coordinate (b21) at ($(b22)+.7*(v2)-.7*(f2)$);
\coordinate[label=below right:$k'$] (b2) at ($(b21)!0.5!(b22)+(-.1,+.1)$);
\draw[fermion] (v1) -- (e2);
\draw[fermion] (e1) -- (v1);
\draw[fermion] (v2) -- (f1);
\draw[fermion] (f2) -- (v2);
\path [sines/.style={
        line join=round, 
        draw=black, 
        decorate, 
        decoration={complete sines, number of sines=3, amplitude=2mm}},postaction={sines}] (v1) -- (v2);
\draw[->] (a11) -- (a12);
\draw[->] (a21) -- (a22);
\draw[->] (b11) -- (b12);
\draw[->] (b21) -- (b22);
\end{tikzpicture}
+
\begin{tikzpicture}[baseline={(current bounding box.center)}]
\coordinate (c) at (0,0);
\coordinate[label=above:$q'$] (labelq) at (0,+.05);
\coordinate[label=left:$a$] (v1) at (-.6,0);
\coordinate[label=right:$b$] (v2) at (+.6,0);
\coordinate[label=left:$e^-$] (e1) at ($(-.7,-1.05)+(v1)$);
\coordinate[label=right:$e^+$] (e2) at ($(+.7,-1.05)+(v2)$);
\coordinate[label=left:$e^-$] (f1) at ($(-.7,+1.05)+(v1)$);
\coordinate[label=right:$e^+$] (f2) at ($(+.7,+1.05)+(v2)$);
\coordinate[right=2.5mm of e1] (a11);
\coordinate (a12) at ($(a11)+.7*(v1)-.7*(e1)$);
\coordinate[label=below right:$p$] (a1) at ($(a11)!0.5!(a12)+(-.1,+.1)$);
\coordinate[left=2.5mm of e2] (a21);
\coordinate (a22) at ($(a21)+.7*(v2)-.7*(e2)$);
\coordinate[label=below left:$k$] (a1) at ($(a21)!0.5!(a22)+(+.1,+.1)$);
\coordinate[right=2.5mm of f1] (b12);
\coordinate (b11) at ($(b12)+.7*(v1)-.7*(f1)$);
\coordinate[label=above right:$p'$] (b1) at ($(b11)!0.5!(b12)+(-.1,-.1)$);
\coordinate[left=2.5mm of f2] (b22);
\coordinate (b21) at ($(b22)+.7*(v2)-.7*(f2)$);
\coordinate[label=above left:$k'$] (b2) at ($(b21)!0.5!(b22)+(+.1,-.1)$);
\draw[fermion] (v1) -- (f1);
\draw[fermion] (e1) -- (v1);
\draw[fermion] (f2) -- (v2);
\draw[fermion] (v2) -- (e2);
\path [sines/.style={
        line join=round, 
        draw=black, 
        decorate, 
        decoration={complete sines, number of sines=3, amplitude=2mm}},postaction={sines}] (v1) -- (v2);
\draw[->] (a11) -- (a12);
\draw[->] (a21) -- (a22);
\draw[->] (b11) -- (b12);
\draw[->] (b21) -- (b22);
\end{tikzpicture} \,.
\]
The charge of the electron is $-\pose$ and its mass is $m$. The tree-level amplitude in Feynman gauge ($\lambda=1$) is
\begin{align*}
 i \amplitude &= \bar{v} (k) \big(i\pose\gamma^a \big) u (p) \bar{u} (p') \big(i\pose\gamma^b \big) v (k')   \frac{-i}{q^2} \left[ \eta_{ab} -E_{ambn}\frac{q^m q^n}{q^2} \right] +\\
 &-\bar{u} (p') \big(i\pose\gamma^a \big) u (p) \bar{v} (k) \big(i\pose\gamma^b \big) v (k')   \frac{-i}{q'^2} \left[ \eta_{ab} -E_{ambn}\frac{q'^m q'^n}{q'^2} \right] \,,
\end{align*}
where $\varepsilon=0$, since the tree-level process is not divergent.
As before, we move to the center of mass frame with the axes oriented in such a way that
\begin{gather*}
 \vect{p} = -\vect{k}= |\vect{p}| \hat{z} \quad \text{and} \\
 \vect{p}' = -\vect{k}'= |\vect{p}| (\sin \theta \cos \varphi \,\hat{x} + \sin \theta \sin \varphi \,\hat{y} + \cos \theta \,\hat{z} ) \,.
\end{gather*}
Since spin polarizations are often not measured experimentally, we can take the average of the initial spin polarizations, the sum of the final spin polarizations and compute the differential cross section.
In the ultra-relativistic limit ($s \gg m^2$), using the first restriction ~(\ref{pertconditionsb}) on the perturbation, one obtains
\begin{equation} \label{cross-section-Bhabha}
 \begin{aligned}
  \frac{d \sigma}{d \Omega} &=\frac{\alpha^2}{2 s} \left[ \frac{1}{2} (1+\cos^2 \theta)+\frac{1+\cos^4 (\theta/2)}{\sin^4 (\theta/2)} -
  2\frac{\cos^4 (\theta/2)}{\sin^2 (\theta/2)}  +  \right. \\
  &- \frac{(3+\cos^2\theta) \sin^2 \theta}{4 \sin^4(\theta/2)} \big(\cos^2\varphi \, E^{0101}\! +\sin^2\varphi \, E^{0202}\! +2\sin\varphi \cos\varphi \, E^{0102} \big) +\\ 
  &- \frac{(7+\cos^2 \theta) \sin \theta \cos \theta}{4 \sin^4 (\theta/2)} \big(\cos \varphi \, E^{0103}+\sin \varphi \, E^{0203}\big) + \\
  &- \left.\frac{3+5 \cos^2 \theta}{4 \sin^4 (\theta/2)}  E^{0303} -\frac{5+3 \cos^2\theta}{16 \sin^4(\theta/2)} E^{ij}_{\;\;\; ij} \right] \,.
 \end{aligned}
\end{equation}
Again there is an effect, even if the detector does not resolve different angles $\varphi$.
Integrating out $\varphi$ and using again the first restriction~(\ref{pertconditionsb}) on the perturbation, we also find
\begin{multline*}
 \frac{d \sigma}{d \cos \theta} = \frac{\pi \alpha^2}{s} \left[ \frac{1}{2} (1+\cos^2 \theta)+\frac{1+\cos^4 (\theta/2)}{\sin^4 (\theta/2)} -
 2\frac{\cos^4 (\theta/2)}{\sin^2 (\theta/2)} + \right.  \\
 \left. - \frac{\cos^4 \theta +12 \cos^2 \theta +3}{8 \sin^4 (\theta/2)} E^{0303} - \frac{(1+\cos^2 \theta)(7+\cos^2 \theta)}{32 \sin^4 (\theta/2)}\, E^{ij}_{\;\;\; ij} \right] \,.
\end{multline*}
We have thus shown, for two prototypical tree-level processes, how the Feynman rules derived in section~\ref{sec_BVquant} can produce measurable effects in the tree-level cross-sections, both when the detector is sensitive to the angle $\varphi$ and when it is not. We now continue with the investigation of one-loop processes.

\subsection{Electron in an external magnetic field} \label{sec:electron-external-magnetic-field}
We now study the scattering of an electron in a quasi-static and uniform external magnetic field.
At tree level, this process consists of merely the vertex~(\ref{vertex}) and, thus, it does not contain any correction of the perturbation $E^{abcd}$.
For this reason, we study one-loop radiative corrections to the vertex, which will provide us with the anomalous magnetic moment of the electron on an effectively flat region of a general linear background. 

The said process, up to one-loop, is the sum of two diagrams: the tree-level vertex and the one-loop contribution. We thus write
\begin{equation*}
\begin{tikzpicture}[baseline={(current bounding box.center)}]
\coordinate (aux) at (0,0);
\coordinate [label=left:$p'$] (e1) at (-1,.8);
\coordinate [label=right:$p$] (e2) at (1,.8);
\coordinate [label=below:$n$] (e3) at (0,-1);
\coordinate (arr1) at (.3,-.95);
\coordinate (arr2) at (.3,-.55);
\coordinate [label=right:$q$] (arrlab) at ($(arr1)!0.5!(arr2)$);
\node [circle,draw,inner sep=3.5mm,pattern=north east lines] (blob) at (aux) {};
\draw [fermion] (e2) -- (blob);
\draw [fermion] (blob) -- (e1);
\path [sines/.style={
        line join=round, 
        draw=black, 
        decorate, 
        decoration={complete sines, number of sines=1.5, amplitude=2mm}},postaction={sines}] (blob) -- (e3);
\draw[->] (arr1) -- (arr2);
\end{tikzpicture}
=
\begin{tikzpicture}[baseline={(current bounding box.center)}]
\coordinate (aux) at (0,0);
\coordinate[label=left:$p'$] (e1) at (-1,.8);
\coordinate[label=right:$p$] (e2) at (1,.8);
\coordinate[label=below:$n$] (e3) at (0,-1);
\coordinate[above right=.2 and .25 of e3] (arr1);
\coordinate[below right=.2 and .25 of aux] (arr2);
\coordinate[label=right:$q$] (arrlab) at ($(arr1)!0.5!(arr2)$);
\draw[fermion] (aux) -- (e1);
\draw[fermion] (e2) -- (aux);
\path [sines/.style={
        line join=round, 
        draw=black, 
        decorate, 
        decoration={complete sines, number of sines=2.5, amplitude=2mm}},postaction={sines}] (aux) -- (e3);
\draw[->] (arr1) -- (arr2);
\end{tikzpicture}
+
\begin{tikzpicture}[baseline={(current bounding box.center)}]
\coordinate (aux3) at (0,0);
\coordinate[label=left:$p'$] (e1) at (-1,.8);
\coordinate[label=right:$p$] (e2) at (1,.8);
\coordinate (aux1) at ($0.75*(e1)$);
\coordinate (aux2) at ($0.75*(e2)$);
\coordinate[label=below:$n$] (e3) at (0,-1);
\coordinate[above right=.2 and .25 of e3] (arr1);
\coordinate[below right=.2 and .25 of aux3] (arr2);
\coordinate[label=right:$q$] (arrlab) at ($(arr1)!0.5!(arr2)$);
\draw[-] (aux1) -- (e1);
\draw[-] (e2) -- (aux2);
\draw[fermion] (aux2) -- (aux3);
\draw[fermion] (aux3) -- (aux1);
\path [sines/.style={
        line join=round, 
        draw=black, 
        decorate, 
        decoration={complete sines, number of sines=2.5, amplitude=2mm}},postaction={sines}] (aux3) -- (e3);
\path [sines/.style={
        line join=round, 
        draw=black, 
        decorate, 
        decoration={complete sines, number of sines=4.5, amplitude=2mm}},postaction={sines}] (aux1) to[bend left] (aux2);
\draw[->] (arr1) -- (arr2);
\end{tikzpicture}
\eqdef
i \pose  \Gamma^n (p,p') \,,
\end{equation*}
where $p$ and $p'$ are the momenta of the ingoing and outgoing on-shell electrons, respectively, $q=p'-p$ is $\hbar$ times the frequency of the oscillating external  potential and $-\pose$ is the charge of the electron.
Since the external magnetic field is quasi-static and uniform, $q$ is very small.
Thus it is reasonable to expand the scattering amplitude in powers of $q$ and neglect second and higher orders.

The tree-level contribution above is
\begin{equation} \label{tree-level-vertex}
 \bar{u} (p') \big[  i\pose \Gamma^n_{\text{tree}} (p,p') \big] u(p)=\bar{u} (p') \big[  i\pose \mu^{\varepsilon/2}\, \gamma^n \big] u(p) \,,
\end{equation}
while the one-loop contribution $I^n \eqdef \bar{u}(p')\big[ i\pose \Gamma^n_{\text{1-loop}}(p',p) \big] u(p)$ is 
\begin{multline*}
 I^n =(i\pose\mu^{\varepsilon/2})^3  \int \frac{d^{\, D} k}{(2 \pi)^D} \left[ -\frac{i}{k^2-\delta^2+i\epsilon} \left( \eta_{ab}-E_{arbs}\frac{k^r k^s}{k^2-\delta^2+i\epsilon} \right) \right] \times	\\
 \times \bar{u}(p') \gamma^a \frac{i(\slashed{p}' -\slashed{k} +m)}{(p'-k)^2 -m^2 +i\epsilon} \gamma^n 
 \frac{i(\slashed{p} -\slashed{k} +m)}{(p-k)^2 -m^2 +i\epsilon} \gamma^b u(p) \,,
\end{multline*}
where we have chosen the Feynman gauge ($\lambda=1$) and used the photon propagator~(\ref{photon-propagator-IR-regulated}) with a small photon mass $\delta$ in order to avoid infrared divergences.
The ultraviolet divergences, on the other hand, are treated using dimensional regularization, where integrals are performed in $D=4-\varepsilon$ complex dimensions and an arbitrary mass scale $\mu$ is introduced in order to preserve the correct physical dimension of the electric charge.
The Feynman prescription $i\epsilon$ will not be shown explicitly in the equations below.
The integral $I^n$ can be split into a sum consisting of a zeroth order contribution in $H$, denoted by $I^n_0$, and a first order contribution in $H$, denoted by $I^n_H$.
The former gives the standard QED contribution
\begin{equation} \label{1-loop-vertex-qed}
 I^n_0=i\pose\mu^{\varepsilon/2}\, \frac{\alpha}{2 \pi} \bar{u}(p') \left\{ \gamma^n
 \left[ \frac{1}{\varepsilon}+\frac{1}{2} \log \left( \frac{4\pi \mu^2}{m^2} \right) -\frac{\gamma_E}{2} +2 -\log \left( \frac{m^2}{\delta^2} \right) \right]+
 \frac{i\sigma^{n\ell} q_{\ell}}{2m}\right\} u(p) +\bigo(q^2) \,,
\end{equation}
where $\gamma_E$ is the Euler-Mascheroni constant, $\alpha$ the fine structure constant, $\sigma^{n \ell} \eqdef \frac{i}{2}[\gamma^n,\gamma^\ell]$ and the expression has already been expanded in powers of $q$, neglecting second and higher orders.

We start the computation of $I^n_E$ by using twice the Feynman parametrization to simplify the denominators, and obtain
\begin{equation} \label{1-loop-integral}
 I^n_E=  -6 \pose^3  \mu^{3\varepsilon/2}  \int_0^1 dx \int_0^1 dy\, y(1-y) 
 \int \frac{d^{\, D} k}{(2\pi)^D} \frac{\bar{u}(p') E_{arbs} k^r k^s \gamma^a (\slashed{k}-\slashed{p}'-m) \gamma^n (\slashed{k}-\slashed{p}-m) \gamma^b u(p) }{\Big[ (k-y\, p_x)^2 -\big( y^2 p_x^2+(1-y) \delta^2 \big) \Big]^{4}} \,,
\end{equation}
having defined $p_x \eqdef x p+(1-x) p'$. We can now simplify the numerator by use of $\slashed{p} u(p)=mu(p)$, $\bar{u} (p') \slashed{p}'=\bar{u}(p') m$ and the first restriction ~(\ref{pertconditionsb}) on the perturbation.
Changing the integration variable $k'=k-y\, p_x$ and dropping all terms containing odd powers of $k'$, the numerator $N=N_A+N_B+N_C$ consists of a part
\[
 N_A = \bar{u}(p') \left[ -\frac{E^{ij}_{\;\;\; ij}}{D} k'^2 \slashed{k}' \gamma^n \slashed{k}'+ 2 E_{a}^{\; rns} k'_r k'_s \slashed{k}' \gamma^a \slashed{k}' \right] u(p) \,,
\]
which is quartic in the integrated momentum $k'$, another part
\begin{align*}
 N_B &=\bar{u}(p') \Big\{-y^2 \frac{E^{ij}_{\;\;\; ij}}{D} \big[ k'^2 \slashed{p}_x \gamma^n \slashed{p}_x+p_x^2\, \slashed{k}' \gamma^n \slashed{k}' 
 +4k'\cdot p_x(k'^n \slashed{p}_x +p_x^n \slashed{k}') -4(k'\cdot p_x)^2\gamma^n \big]  + \\
 &+2 y^2 E_a^{\; rns} \big[ k'_r k'_s \slashed{p}_x \gamma^a \slashed{p}_x +p_{xr} p_{xs} \slashed{k}' \gamma^a \slashed{k}'+2(k'_r p_{xs}+p_{xr} k'_s) (p_x^a \slashed{k}' +k'^a \slashed{p}_x -k' \cdot p_x \gamma^a)  \big] + \\
 &+2y E_a^{\; rbs} \big[ k'_r k'_s (p_b \slashed{p}_x \gamma^a \gamma^n+p'_b \gamma ^n \gamma^a \slashed{p}_x) + 
  (k'_r p_{xs}+p_{xr} k'_s) (p_b \slashed{k}' \gamma^a \gamma^n+p'_b \gamma ^n \gamma^a \slashed{k}')  \big]+ \\
 & +4 E^{arbs} k'_r k'_s p_ap'_b \gamma^n \Big\} u(p) \,,
\end{align*}
which is quadratic in the integrated momentum $k'$, and a third part
\begin{align*}
 N_C &=\bar{u}(p') \bigg\{ -y^4 \frac{E^{ij}_{\;\;\; ij}}{D} p_x^2\, \slashed{p}_x \gamma^n \slashed{p}_x+
 2 y^4 E_a^{\; rns} p_{xr} p_{xs} \slashed{p}_x \gamma^a \slashed{p}_x+ \\
 &+2 y^3 E_a^{\; rbs} p_{xr} p_{xs} (p_b \slashed{p}_x \gamma^a \gamma^n+p'_b \gamma^n \gamma^a \slashed{p}_x)
 +4 y^2 E^{arbs}  p_{xr} p_{xs} p_a p'_b \gamma^n \bigg\} u(p) \,,
\end{align*}
which does not depend on the integrated momentum $k'$ at all.

The contribution of $N_A$ to the integral~(\ref{1-loop-integral}) is divergent a priori, but in fact finite, yielding
\[
 (I_E^n)_A = \frac{i \mu^{\varepsilon/2} \pose^3}{8(4 \pi)^2} E^{ij}_{\;\;\; ij} \gamma^n \,.
\]
The contributions $N_B$ and $N_C$, which we denote by $(I^n_E)_B$ and $(I^n_E)_C$ respectively, are ultraviolet finite and can thus be computed at $D=4$, yielding
\begin{align*}
 (I_E^n)_B = \frac{i\mu^{\varepsilon/2} \pose^3}{12(4\pi)^2} &\left\{ \frac{E^{ij}_{\;\;\; ij}}{2} \gamma^n \left[ 4-3\int_0^1 \!dx \frac{m^2}{p_x^2} 
 +6 \int_0^1 \!dx \frac{p\cdot p'}{p_x^2} \left( \log\!\left( \frac{p_x^2}{\delta^2} \right)\!-2 \right) \right] + \right. \\
 &-\frac{E^{ij}_{\;\;\; ij}}{2}m(p+p')^n \int_0^1 \!dx \frac{1}{p_x^2} -16 E_a^{\;\;rns} \int_0^1 \!dx\, \frac{p_{xr} p_{xs}}{p_x^2} + \\
 & \left. -3 (E_{arbs}+E_{abrs}) p^r p'^s (\gamma^n \gamma^a \gamma^b + \gamma^a \gamma^b \gamma^n) \int_0^1 \!dx \frac{1}{p_x^2} \right\}
\end{align*}
and
\begin{align*}
 (I_E^n)_C=-\frac{i\mu^{\varepsilon/2} \pose^3}{(4\pi)^2} &\left\{ \frac{E^{ij}_{\;\;\; ij}}{24} \gamma^n -\frac{E^{ij}_{\;\;\; ij}}{24}m (p+p')^n \int_0^1 \!\!dx \frac{1}{p_x^2}
 -\frac{1}{3} E_a^{\;\;rns} \gamma^a \int_0^1 \!\!dx\, \frac{p_{xr} p_{xs}}{p_x^2}+ \right. \\
 &+ E_a^{\;\;rbs} \int_0^1 \!dx\, \frac{p_{xr} p_{xs}}{p_x^2} (p_b \slashed{p}_x \gamma^a \gamma^n+ p'_b \gamma^n \gamma^a \slashed{p}_x)+ \\
 & \left. +2E^{arbs} p_a p'_b \gamma^n \int_0^1 \!dx\, \frac{p_{xr} p_{xs}}{(p_x^2)^2} \left[ \log\left( \frac{p_x^2}{\delta^2} \right)-3 \right] \right\} \,.
\end{align*}
Expanding in powers of $q$ neglecting second and higher orders, summing all contributions and replacing $(p+p')^n$ with $2m \gamma^n-i\sigma^{n\ell} q_\ell$ by use of the Gordon identity, we then find
\begin{equation} \label{1-loop-vertex-h}
\begin{aligned} 
 I_E^n =& - i \pose\mu^{\varepsilon/2} \frac{\alpha}{4\pi} \bar{u}(p') \left[\,\frac{1}{2} E^{ij}_{\;\;\; ij} \gamma^n -\frac{1}{4} E^{ij}_{\;\;\; ij} \gamma^n \log\left(\frac{m^2}{\delta^2} \right) \right.+\\
  &+\frac{i \sigma^{ab}}{8m^3} E_{a}^{\;\; rns} (p'+p)_r (p'+p)_s q_b + 2 \frac{i \sigma^{na}}{8m^2} E_{arbs} (p'+p)^b (p'+p)^r q^s +  \\
  &\left. + \frac{1}{2} \frac{E_{arbs}+E_{abrs}}{4m^2} (p'+p)^r q^s (\gamma^a \gamma^b \gamma^n + \gamma^n \gamma^a \gamma^b )  \right] u(p) \,.
\end{aligned}
\end{equation}
Taking the sum of the tree-level contribution~(\ref{tree-level-vertex}), the one-loop corrections~(\ref{1-loop-vertex-qed}), (\ref{1-loop-vertex-h}) and the one-loop counter-term in the on-shell scheme,
\[
\bar{u} (p') \left[
\begin{tikzpicture}[baseline={(current bounding box.center)}]
\coordinate (aux) at (0,0);
\coordinate[label=left:$p'$] (e1) at (-.7,.5);
\coordinate[label=right:$p$] (e2) at (.7,.5);
\coordinate[label=below:$n$] (e3) at (0,-.6);
\coordinate[above right=.1 and .25 of e3] (arr1);
\coordinate[below right=.1 and .25 of aux] (arr2);
\coordinate[label=right:$q$] (arrlab) at ($(arr1)!0.5!(arr2)$);
\node [circle,draw,inner sep=.5mm,fill] (dot) at (aux) {};
\draw[fermion] (dot) -- (e1);
\draw[fermion] (e2) -- (dot);
\path [sines/.style={
        line join=round, 
        draw=black, 
        decorate, 
        decoration={complete sines, number of sines=1.5, amplitude=1.5mm}},postaction={sines}] (dot) -- (e3);
\draw[->] (arr1) -- (arr2);
\end{tikzpicture}
\right] u(p)=
i\pose \mu^{\varepsilon/2} (Z_\psi^{\text{o.s.}}-1) \bar{u} (p') \gamma^n u(p) \,,
\]
where the renormalization constant $Z_\psi^{\,\text{o.s.}}$ is given by~(\ref{Zpsi-on-shell}), after setting $\lambda=1$ and $Q=-1$, we finally obtain the relevant renormalized quantity
\begin{equation} \label{1-loop-vertex-complete}
 \begin{aligned}
  i\pose &\bar{u} (p') \Gamma^n_{\text{o.s.}} u(p) = i\pose \bar{u} (p') \left[ F_1\, \gamma^n + F_2\, \frac{i \sigma^{n\ell} q_\ell}{2m}+   \right. \\
  &+ F_3\, \frac{i \sigma^{ab}}{8m^3} E_{a}^{\; rns} (p'+p)_r (p'+p)_s q_b + F_4\, \frac{i \sigma^{na}}{8m^3} E_{arbs} (p'+p)^b (p'+p)^r q^s +  \\
  & \left. +F_5\, \frac{E_{arbs}+E_{abrs}}{4m^2}  (p'+p)^r q^s (\gamma^a \gamma^b \gamma^n + \gamma^n \gamma^a \gamma^b )  \right] u(p) +\bigo(q^2) \,,
 \end{aligned}
\end{equation}
where we calculated the form factors to take the values
\begin{equation} \label{form-factors}
 F_1=1 \,,	\qquad F_2=\frac{\alpha}{2\pi} \,,	\qquad F_3= - \frac{\alpha}{4\pi} \,,	\qquad F_4= - \frac{\alpha}{2\pi} \,,	\qquad F_5= - \frac{\alpha}{8\pi} \,.
\end{equation}

This concludes the computation of the one-loop proper vertex.
In the next subsection, we analyze the physical implications.

\subsection{Emergence of a qualitatively new spin-magnetic coupling} \label{sec:anomalous-magnetic-moment}
We now use the renormalized one-loop vertex~(\ref{1-loop-vertex-complete}) in order to determine the non-relativistic potential of an electron in a quasi-static and uniform external magnetic field.
This will provide us with a correction of the anomalous magnetic moment of the electron---which result we will however significantly refine in the next subsection---and a new small interaction between the spin and the magnetic field.
Both are corrections to the standard QED case.

Multiplying the one-loop vertex~(\ref{1-loop-vertex-complete}) by the Fourier transform of the electromagnetic external potential $\tilde{A}_n(q)$, one obtains the scattering amplitude
\begin{equation*}
 i \mathcal{M} = \bar{u} (p') \big[ i\pose \Gamma_{\text{o.s.}}^n (p,p') \big] u(p) \tilde{A}_n (q) \,.
\end{equation*}
Then taking the non-relativistic limit $|\vect{p}| \ll m$, by neglecting all contributions quadratic in $\vect{p}$ and all those linear in both $\vect{p}$ and $H$, and using the Born approximation
\[
 V(\vect{x})=- \frac{1}{2m}\int \frac{d^{\,3} q}{(2\pi)^3} e^{-i q_\alpha x^\alpha} \mathcal{M}(q) \,,
\]
one derives the non-relativistic interaction potential
\begin{equation*} 
 \begin{aligned}
  V(\vect{x}) = \left[ 2 (F_1+F_2)+ \frac{E^{ij}_{\;\;\; ij}}{4}(F_3-F_4+6F_5)\right] \left\langle \frac{\pose \vect{s}}{2m} \right\rangle \cdot \vect{B}
  +(F_3-F_4+6F_5) E^{0 \alpha 0 \beta} \left\langle \frac{\pose s_\alpha}{2m} \right\rangle  B_\beta \,,
 \end{aligned}
\end{equation*}
where $\langle \vect{s} \rangle \eqdef \xi' \vect{s}\, \xi$, with $\xi$ denoting the polarization of the ingoing electron and  $\xi'$ that of the outgoing electron.
Expanding $V(\vect{x})$ in terms of the form factors~(\ref{form-factors}) and the tracefree tensor
\begin{equation}\label{widehatE}
 \widehat{E}^{\alpha \beta}\eqdef E^{0 \alpha 0 \beta}+\frac{E^{ij}_{\;\;\; ij}}{12} \delta^{\alpha \beta} 
\end{equation}
(which will make another appearance in section \ref{subsec:full}, where it will be seen to control a splitting of the triplet of the hyperfine structure of hydrogen),
the non-relativistic potential for the electron in an external quasi-static and uniform magnetic field becomes
\begin{equation} \label{potential}
  V(\vect{x})= 2 \left[ 1+\frac{\alpha}{2\pi} \left(1 -\frac{E^{ij}_{\;\;\; ij}}{12}\right) \right] \left\langle \frac{\pose \vect{s}}{2m} \right\rangle \cdot \vect{B}-
  \frac{\alpha}{2\pi} \widehat{E}^{\alpha \beta} \left\langle \frac{\pose s_\alpha}{2m} \right\rangle  B_\beta \,.
\end{equation}

The first summand in $V(\vect{x})$ is the  minimal coupling of the spin with the magnetic field, featuring a however slightly modified anomalous magnetic moment of the electron
\begin{equation} \label{anomalous-magnetic-moment}
 a \eqdef \frac{g-2}{2}= (F_1+F_2-1)+ \frac{E^{ij}_{\;\;\; ij}}{12}(F_3-F_4+6F_5) = \frac{\alpha}{2\pi}\left( 1-\frac{E^{ij}_{\;\;\; ij}}{12} \right) \,,
\end{equation}
which is compatible with what is obtained by Carone \textit{et al.}~\cite{Carone:2006tx} for the special case of an isotropic background.
On the other hand, the second summand in~(\ref{potential}) presents a qualitatively new interaction
\begin{equation} \label{potential-correction}
 \delta V(\vect{x}) = - \frac{\alpha}{2\pi}\widehat{E}^{\alpha \beta} \left\langle \frac{\pose s_\alpha}{2m} \right\rangle  B_\beta
\end{equation}
 between the spin and the magnetic field.

This concludes the calculation of the scattering of an electron in an external electromagnetic field on an effectively flat region of a general linear background.
As for the tree-level scatterings, we found only very subtle differences with respect to QED, both in the modification of the Schwinger correction to the anomalous magnetic moment~(\ref{anomalous-magnetic-moment}) and in the appearance of a quantitatively small, but qualitatively new term in the potential~(\ref{potential-correction}).

\subsection{Anomalous magnetic moment to any order in general linear QED} \label{sec:AMM-beyond-one-loop}
In section~\ref{sec:electron-external-magnetic-field}, we have computed the one-loop radiative corrections to the vertex and used them in section~\ref{sec:anomalous-magnetic-moment} to derive the one-loop anomalous magnetic moment~(\ref{anomalous-magnetic-moment}) of the electron. Now we show how to deduce the anomalous magnetic moment of the electron at every loop order, once it is known in metric QED.

Since the renormalization constant for the vertex and the anomalous magnetic moment are scalar, they depend on the perturbation $E^{abcd}$ only through its full trace $E^{ij}_{\;\;\; ij}$.
Thus when computing the anomalous magnetic moment, we can replace the perturbation $E^{abcd}$ with any other perturbation $\tilde{E}^{abcd}$, on the condition that both the perturbation have the same full trace, $E^{ij}_{\;\;\; ij}=\tilde{E}^{ij}_{\;\;\; ij}$.
In particular, we can take
\begin{equation} \label{particular-perturbation}
 \tilde{E}^{abcd}=\frac{E^{ij}_{\;\;\; ij}}{D(D-1)} (\eta^{ac}\eta^{bd}-\eta^{ad}\eta^{bc}) \,.
\end{equation}
Moreover, since the physical quantities do not depend on the gauge parameter $\lambda$, we choose the \emph{Landau gauge} to set it to zero.
Then the photon propagator~(\ref{photon-propagator}) for the perturbation~(\ref{particular-perturbation}) becomes
\begin{equation} \label{photon-propagator-particular}
\begin{tikzpicture}[baseline={([yshift=-2mm]current bounding box.center)}]
\coordinate[label=left:$a$] (e1) at (-.75,0);
\coordinate[label=right:$b$] (e2) at (.75,0);
\coordinate[label=above:$q$] (lab) at (0,0);
\path [sines/.style={
        line join=round, 
        draw=black, 
        decorate, 
        decoration={complete sines, number of sines=5, amplitude=2mm}},postaction={sines}] (e1) -- (e2);
\end{tikzpicture}
=- \frac{i}{q^2+i\epsilon}\left( \eta_{ab} - \frac{q_a q_b}{q^2+i\epsilon}\right) \left[ 1-\frac{E^{ij}_{\;\;\; ij}}{D(D-1)} \right] \,,
\end{equation}
which, up to the scalar factor in square brackets, is the usual photon propagator for standard QED in Landau gauge.

The topology of the diagrams implies that the $\ell$-loop vertex radiative correction $i\pose\, \Gamma^n_{\ell}(p,p',\alpha)$ contains $\ell$ photon propagators and $(2\ell+1)$ vertices. Here we chose to denote the momentum of the ingoing fermion by $p$ and that of the outgoing by $p'$, to leave explicit the dependence on the fine structure constant $\alpha$ and to factor out $\pose$.
The $(2\ell+1)$ vertices carry a factor $\pose\alpha^{\,\ell}$, while the $\ell$ photon propagators carry the factor in square brackets in~(\ref{photon-propagator-particular}) to the power of $\ell$.
Thus, we can write
\[
 i\pose\, \Gamma^n_{\ell}(p,p',\alpha)=i\pose\, \alpha^{\,\ell} \left[ 1-\frac{E^{ij}_{\;\;\; ij}}{D(D-1)} \right]^{\ell} I^n_{\ell}(p,p') \,,
\]
where $I^n_{\ell}(p,p')$ is defined through the relation $\Gamma^n_{\ell,\text{\tiny{QED}}}(p,p',\alpha)=\alpha^{\,\ell} I^n_{\ell}(p,p')$.
We then find, for the particular perturbation~(\ref{particular-perturbation}) in Landau gauge, the relation
\begin{equation} \label{vertex-loops-relation}
 i\pose\, \Gamma^n_{\ell}(p,p',\alpha)=i\pose\, \Gamma^n_{\ell,\text{\tiny{QED}}}(p,p',\tilde{\alpha}) \,,
 \quad\text{where}\quad
 \tilde{\alpha}\eqdef \alpha \left[ 1-\frac{E^{ij}_{\;\;\; ij}}{D(D-1)} \right] \,,
\end{equation}
from which it follows that the $\ell$-loop anomalous magnetic moment for quantum electrodynamics on an effectively flat region of a general linear background is given by
\begin{equation} \label{AMM-loops-relation}
 a_\ell(\alpha)=a_{\ell,\text{\tiny{QED}}}(\tilde{\alpha}) \,,
 \quad\text{where}\quad
 \tilde{\alpha}\eqdef \alpha \left( 1-\frac{E^{ij}_{\;\;\; ij}}{12} \right) \,,
\end{equation}
having set $D=4$, since the quantity is finite.
In principle, ultraviolet and infrared regulators, as the photon mass employed in section~\ref{sec:electron-external-magnetic-field}, could break relation~(\ref{vertex-loops-relation}), but this is merely an effect of the chosen regularization technique, which disappears after renormalization and does not affect relation~(\ref{AMM-loops-relation}) between physical quantities.

The anomalous magnetic moment of the electron is measured with high precision~\cite{Hanneke:2008tm} and can be used, in metric QED, to determine the value of the fine structure constant~\cite{Aoyama:2012wj}.
With the same procedure, we can determine the value
\begin{equation} \label{alpha-tilde-value}
  \alpha \left( 1-\frac{E^{ij}_{\;\;\; ij}}{12} \right)=1/137.035\, 999\, 2(2) \,,
\end{equation}
where the uncertainty in brackets comes from having neglected, in the theoretical calculation, the hadronic contribution, whose order of magnitude is $a^{\text{hadron}}\simeq 2\times 10^{-12}$, and the electroweak contribution, whose order of magnitude is even smaller.

It is thus shown to be possible to determine, with high precision, the value of a particular combination of the fine structure constant $\alpha$ and the double trace of the perturbation $E^{ij}_{\;\;\; ij}$. In order to determine their individual values, it is necessary to find at least another independent measurable quantity depending on them.


\section{Bound states and spectrum of hydrogen} \label{sec:hydrogen}
We obtain the bound states of the hydrogen atom on an effectively flat region of a general linear background. This is done by employing the Born-Oppenheimer approximation and a classical treatment of the  
electromagnetic field generated by the proton's charge and magnetic moment, followed by a second order quantum mechanical perturbation around the known exact solution of the relativistic hydrogen atom. In particular, we find the effect of a weak general linear background on the hyperfine structure of hydrogen, which plays a significant role in astrophysics.  

\subsection{The proton's electromagnetic field}\label{sec:solutions}
The classical electromagnetic potential produced by the proton has two constituents: one from the electric charge of the proton and one due to the proton's magnetic moment. We calculate now both contributions from the electromagnetic field equations. 

First, we obtain the refinement of the general time-independent solution of the field equations. On an effectively flat region of a general linear background, and after choosing there the Lorenz gauge $\eta^{ab}\partial_a A_b = 0$, the field equations \labelcref{eq:motion} for the gauge potential $A$ become \begin{equation} \label{eq:field}
\left(\eta^{ad}\Box - E^{abcd}\partial_b\partial_c\right)A_d = j^a
\end{equation}
with $\Box := \eta^{ab}\partial_a\partial_b$.
Solutions to these field equations are given by convolution of the Green's function $\Delta_{de}$ with the source term $j^a$. The Green's function is the solution of
\begin{equation*}
\left(\eta^{ad}\Box - E^{abcd}\partial_b\partial_c\right)\Delta_{de}(x) = \delta^a_e\delta^{(4)}(x)
\end{equation*}
and for its Fourier transform one finds
\begin{equation*}
\left(\mathcal{F}\Delta_{de}\right)(q) = -\frac{1}{q^2}\left(\eta_{de} + E_{d\hphantom{bc}e}^{\hphantom{d}bc} \frac{q_bq_c}{q^2}\right)\,,
\end{equation*}
with $q^2 := \eta^{ab}q_a q_b$. For source terms $j(\vect{x},t) = j(\vect{x})$ without time dependence, as are provided by a resting proton's charge and magnetic moment, we can perform an inverse Fourier transform to find the refined Biot-Savart law
\begin{equation}\label{eq:Biot-Savart}
A_a = \int\dd^3 x'\frac{j^b(\vect{x}')}{4\pi|\vect{x}'-\vect{x}|}\left(\eta_{ab} + \frac{1}{2}E_{a\mu\nu b}\left(\eta^{\mu\nu}+\frac{(x'-x)^\mu(x'-x)^\nu}{|\vect{x}'-\vect{x}|^2}\right)\right)\,,
\end{equation}
where $|\vect{x}'-\vect{x}| := \sqrt{-\eta_{\mu\nu}(x'-x)^\mu(x'-x)^\nu}$. With this expression at hand, it is now a simple exercise to calculate the two contributions to the proton's potential.

The proton's charge $\pose$ sources the electromagnetic field equations by
\begin{equation*}
j^0(\vect{x},t) = \pose\delta^{(3)}(\vect{x}) \qquad \text{and} \qquad j^\beta = 0\,,
\end{equation*}
which trivializes the integration in the Biot-Savart law and thus yields as the modification of the standard Coulomb potential on an effectively flat region of a general linear background
\begin{equation}\label{eq:pointcharge}
A^\textrm{\tiny charge}_a(\vect{x}) = \frac{Q\pose}{4\pi r}\left(\eta_{a0} +\frac{1}{2} E_{a\mu\nu 0}\left(\eta^{\mu\nu}+\frac{x^\mu x^\nu}{r^2}\right)\right)\,,
\end{equation}
where $r^2 := -\eta_{\mu\nu}x^\mu x^\nu$. The second contribution to the electromagnetic field of the proton is due to its magnetic dipole moment. Classically, a magnetic dipole is a circular current in the limit of vanishing radius. This amounts to the electromagnetic potential of a magnetic dipole $\vect{M}$ being, by definition, the solution of the field equations \labelcref{eq:field} with the source term
\begin{equation*}
j^0 = 0 \qquad \text{and} \qquad \vect{j} = \frac{|\vect{M}|}{\pi a^2} \delta(r - a)\delta(\vect{M}\cdot\vect{x})\vect{M}\times\vect{x}
\end{equation*}
in the limit $a \rightarrow 0$. For tangibility we used the intuitive notations $\vect{M}\cdot\vect{x}$ and $\vect{M}\times\vect{x}$ which are defined as
\begin{equation}\label{eq:dotncross}
\vect{M}\cdot\vect{x} := -\eta_{\mu\nu}M^\mu x^\nu \qquad \text{ and }\qquad \left(\vect{M}\times\vect{x}\right)^\mu := -\eta^{\mu\nu}\epsilon_{\nu\rho\sigma}M^\rho x^\sigma\,.
\end{equation}
Inserting this into the Biot-Savart law \labelcref{eq:Biot-Savart}, expanding in powers of $\frac{a}{r}$, calculating the occurring integrals with the help of Mathematica, and then letting $a\rightarrow 0$, one finds the electromagnetic potential of the magnetic dipole to be
\begin{equation} \label{eq:dipole}
A^\textrm{\tiny magnetic}_a = -\frac{M^\mu x^\nu}{4\pi r^3}\left(\epsilon_{a\mu\nu} + \frac{1}{2}\epsilon_{\beta\mu\nu} E_{a\rho\sigma}{}^\beta \left(\eta^{\rho\sigma} + 3\frac{x^\rho x^\sigma}{r^2}\right) + \epsilon_{\beta\mu\rho}\eta_{\nu\sigma} E_a{}^{(\rho\sigma)\beta}\right)\,.
\end{equation}
Note that for a vanishing perturbation $E$ this reproduces the standard result $\vect{A} = \frac{1}{4\pi r^3}\vect{M}\times\vect{x}$.

\subsection{Refined Hamiltonian of hydrogen}\label{subsec:setup}
We consider the hydrogen atom quantum mechanically as a two-body system composed of a spin-$\frac{1}{2}$ proton of mass  $m_p$ and charge $\pose$ and a spin-$\frac{1}{2}$ electron of mass $m_e$ and charge $-\pose$ that moves in the electromagnetic potential generated by the charge and the magnetic moment of the proton. The much larger mass of the proton allows for a  Born-Oppenheimer approximation, so that the state of the entire atom can be assumed to be a simple tensor product
$$\Psi^\textrm{\tiny atom} = \psi^\textrm{\tiny electron}\otimes\psi^\textrm{\tiny proton}\,,$$
rather than a linear combination of such. Additionally considering the frame where the proton is at rest, we may further describe the positive energy states of the proton simply by $\psi^\textrm{\tiny proton} \in \mathbb{C}^2$. The electron's wave function, however, is a full Dirac spinor field $\psi^\textrm{\tiny electron}$ with reduced Dirac mass $m_\text{\tiny red} := \frac{m_e m_p}{m_e + m_p}$, whose negative energy states will be discarded later on by hand in this first quantized treatment. 

With the above assumptions, the fully time-dependent quantum mechanical equation of motion for the hydrogen atom reads
\begin{equation*}
(\ii \gamma^a\partial_a + \pose\gamma^a A_a - m_\text{\tiny red})\Psi^\textrm{\tiny atom} = 0\,,
\end{equation*}
if all operators, with the notable exception of $A_a$, are agreed to act non-trivially only on the first factor of the tensor product $\psi^\textrm{\tiny electron}\otimes\psi^\textrm{\tiny proton}$, e.g., $\gamma^a (\psi^\textrm{\tiny electron}\otimes\psi^\textrm{\tiny proton}) := (\gamma^a\psi^\textrm{\tiny electron})\otimes\psi^\textrm{\tiny proton}$. The electromagnetic potential, in contrast, being dependent on the proton's spin through its magnetic moment operator
\begin{equation*}
{M}^\mu (\psi^\textrm{\tiny electron}\otimes\psi^\textrm{\tiny proton}) := \frac{g_p \pose}{2m_p} \psi^\textrm{\tiny electron} \otimes (\tfrac{1}{2} \sigma^\mu \psi^\textrm{\tiny proton})\qquad\textrm{(with gyromagnetic ratio } g_p \approx 5.59\textrm{)} \,,
\end{equation*}
acts non-trivially on both factors. The complete electromagnetic potential generated by the proton is given by the sum of the two contributions $A^{\text{\tiny charge}}$ and $A^{\text{\tiny magnetic}}$ of \cref{eq:pointcharge,eq:dipole} respectively, such that
\begin{equation*}
A_a = \frac{\pose}{4\pi r}\left(\eta_{a0} +\frac{1}{2} E_{a\mu\nu 0}\frac{x^\mu x^\nu}{r^2}\right) - \frac{{M}^\mu x^\nu}{4\pi r^3}\left(\epsilon_{a\mu\nu} + \frac{1}{2}\epsilon_{\beta\mu\nu}E_{a\rho\sigma}{}^\beta \left(\eta^{\rho\sigma} + 3\frac{x^\rho x^\sigma}{r^2}\right) + \epsilon_{\beta\mu\rho}\eta_{\nu\sigma} E_a{}^{(\rho\sigma)\beta}\right)\,.
\end{equation*}
Rewriting the above equation of motion in Schr\"odinger form $i\partial_t\Psi^\textrm{\tiny atom}  = H \Psi^\textrm{\tiny atom} $, which one obtains by acting on both sides with $\gamma^0$ and rearranging terms, one identifies the Hamiltonian 
$${H} := -\ii\gamma^0\gamma^\mu\partial_\mu - \pose\gamma^0\gamma^a A_a + \gamma^0 m_\text{\tiny red}$$
of the hydrogen atom in an effectively flat region of a general linear background.

For stationary states $\Psi^\textrm{\tiny atom} (\vect{x},t) = \Psi^\textrm{\tiny atom} (\vect{x})\ee^{-\ii Et}$, solving the Dirac equation thus reduces to the eigenvalue problem $E\Psi^\textrm{\tiny atom} (\vect{x}) = {H}\Psi^\textrm{\tiny atom} (\vect{x})$ .
With regards to the perturbative determination of the spectrum of $H$, we choose to display the Hamiltonian as a sum 
\begin{equation}\label{hamiltonian}
{H} = {H}_0 + {H}_{\textrm{\tiny hfs}} + {H}_{\text{\tiny abs}} +  {H}_{\text{\tiny rel}}
\end{equation}
of individual contributions (controlled by the fine structure constant  $\alpha := \pose^2/(4\pi)$ and the hyperfine parameter $\mu := g_p/(2m_p r_\text{\tiny B})$, where $r_\text{\tiny B} := 1/(m_\text{\tiny red}\alpha)$ denotes the Bohr radius) 
\begin{align*}
{H}_{0}  :={}& -\ii\gamma^0\gamma^\mu\partial_\mu - \frac{\alpha}{r} + \gamma^0 m_\textrm{\tiny red}\,,\\
{H}_{\textrm{\tiny hfs}} :={}& \mu \frac{\alpha}{r}\frac{r_B}{r}\gamma^0\gamma^\alpha\hat{S}^\mu \frac{x^\nu}{r}\epsilon_{\alpha\mu\nu}\,,\\
{H}_{\textrm{\tiny abs}}  :={}& - \frac{\alpha}{2r}\gamma^0\gamma^a E_{a\mu\nu 0}\frac{x^\mu x^\nu}{r^2}\,,\\
{H}_{\textrm{\tiny rel}}  :={}& \mu \frac{\alpha}{r}\frac{r_B}{r}\gamma^0\gamma^a\hat{S}^\mu \frac{x^\nu}{r}\left(\frac{1}{2}\epsilon_{\beta\mu\nu} E_{a\rho\sigma}{}^\beta \left(\eta^{\rho\sigma} + 3\frac{x^\rho x^\sigma}{r^2}\right) + \epsilon_{\beta\mu\rho}\eta_{\nu\sigma} E_a{}^{(\rho\sigma)\beta}\right)\,.
\end{align*}
The part $H_0$ is known to be exactly solvable and thus provides the basis for a perturbative treatment of the other contributions. Of these, only  $H_\textrm{\tiny hfs}$ appears in the calculation of the standard hyperfine structure, while the two remaining parts are entirely due to general linear deviations from a metric geometry.


\subsection{The need for a second order perturbation analysis}\label{subsec:secondorder}
Compared to the Coulomb potential $\frac{\alpha}{r}$ in ${H}_0$, the three correction operators in the full Hamiltonian (\ref{hamiltonian}) are suppressed by the hyperfine parameter $\mu$, general linear perturbation components $E^{abcd}$, or both. More precisely, 
\begin{equation}\label{eq:mags}
{H}_{\text{\tiny hfs}} \sim \mu H_0\,,\qquad {H}_{\text{\tiny abs}} \sim E^{abcd} H_0\,,\qquad  {H}_{\text{\tiny rel}} \sim \mu E^{abcd} H_0\,.
\end{equation}
With the hyperfine parameter   
$\mu = g_p/(2m_p r_\text{\tiny B}) \approx 1.11\times 10^{-5}$
being small compared to unity, and the eleven-parameter family of small deviations $E^{abcd}$ to be considered only to linear order in our approach, we will be able to derive the hydrogen spectrum as a twelve-parameter perturbation of the eigenvalue problem for $H_0$, whose exact solutions are known and for the conceptual discussion in this section will be denoted by $|E_n,\alpha\rangle$ with the eigenenergy $E_n$ and a degeneracy label $\alpha$.

Following usual quantum mechanical perturbation theory, the corrections to the energy levels $E_n$ are given by the eigenvalues of the transition matrix $T_n$, whose elements $\langle E_n,\alpha | T_n | E_n,\beta \rangle$ are given by the obvious first order contribution 
$\langle E_n,\alpha | {H}_{\textrm{\tiny hfs}} + {H}_{\text{\tiny abs}} +  {H}_{\text{\tiny rel}} | E_n,\beta \rangle$
plus the second order contribution  
$$\sum_{m\neq n,\,\gamma}\frac{\langle E_n,\alpha | {H}_{\textrm{\tiny hfs}} + {H}_{\text{\tiny abs}} +  {H}_{\text{\tiny rel}} | E_m,\gamma \rangle\langle E_m,\gamma | {H}_{\textrm{\tiny hfs}} + {H}_{\text{\tiny abs}} +  {H}_{\text{\tiny rel}} | E_n,\beta \rangle}{E_m - E_n}$$
plus higher order contributions, which we will not need to consider.
Bearing in mind the relative magnitudes \labelcref{eq:mags} of the single operators, it is apparent that for the first order correction due to $H_{\text{\tiny rel}}$, there is a second order term of comparable magnitude from the product of $H_\text{\tiny hfs}$ and $H_\text{\tiny abs}$. It is thus inevitable to also calculate these second order terms. However, we will not need to calculate all the second order contributions, but only those comparable in magnitude to the first order terms. Taking into account the phenomenological fact that the spectral corrections from extra geometric degrees of freedom are significantly weaker than those from the magnetic moment of the proton, these relevant terms are the second order correction to the usual hyperfine structure, $H_\text{\tiny hfs}^2$, and the above mentioned products $H_\text{\tiny hfs}H_\text{\tiny abs}$ and $H_\text{\tiny abs}H_\text{\tiny hfs}$. 

In the following sections, all these contributions to the transition matrix $T_n$ are calculated. After having fixed notation in a most concise review of the exact solution of the eigenproblem of $H_0$ in \cref{subsec:exact}, we calculate the usual hyperfine structure up to second order in \cref{subsec:nonbiref}. From then on we turn to the investigation of the effects of the general linear deviations from a metric. While \cref{subsec:lambda} is concerned with the purely non-metric corrections, in \cref{subsec:lambdamu} we investigate the combined effects of the general linear deviations from a metric background and proton spin, including the relevant second order terms identified above. The full transition matrix is then of course obtained by summing all these contributions. Because of its immediate phenomenological relevance, we investigate in \cref{subsec:full} the hyperfine structure of hydrogen on an effectively flat region of a general linear background.

\subsection{Exact solution without proton spin and without deviation from a metric background}\label{subsec:exact}
The eigenproblem of $H_0$, the unperturbed relativistic hydrogen problem, is exactly solvable \cite{dirac1928quantum,pidduck1929laguerre} and can be found in many textbooks on advanced quantum mechanics \cite{sakurai1967advanced,merzbacher1998quantum}. We will thus only shortly revisit its most essential features, and thereby introduce the notation used in the sections to follow.

The fully relativistic energy spectrum of the hydrogen atom is given by
\begin{equation*}
E(n,\kappa,m,s) := E(n,|\kappa|) = \frac{m_\text{\tiny red}}{\sqrt{1+\left(\frac{\alpha}{n + \sqrt{\kappa^2 - \alpha^2}}\right)^2}}\,, 
\end{equation*}
where $n\in\mathbb{N}$, $\kappa\in\, - \mathbb{N}$ (for $n=0$) or $\kappa\in\mathbb{Z}\backslash\lbrace 0\rbrace$ (for $n>0$), $m = -|\kappa|+\tfrac{1}{2}, \dots, |\kappa|-\tfrac{1}{2}$ and $s=-\tfrac{1}{2}, \tfrac{1}{2}$,
and the corresponding eigenstates $|n\kappa ms\rangle$ explicitly read
\begin{equation*}
\langle \vect{x} | n\kappa ms\rangle = \mathcal{N}\left(2m_\text{\tiny red}\alpha\right)^{\frac{3}{2}}\ee^{-\frac{y}{2w}} y^{K-1} \begin{pmatrix} \left[L_n^{2K}\left(\frac{y}{w}\right) - \frac{w+\kappa}{n}L_{n-1}^{2K}\left(\frac{y}{w}\right)\right]\mathcal{Y}_\kappa^m (\vect{n}) \\[2ex] \frac{w-W}{\ii \alpha}\left[L_n^{2K}\left(\frac{y}{w}\right) + \frac{w+\kappa}{n}L_{n-1}^{2K}\left(\frac{y}{w}\right)\right]\mathcal{Y}_{-\kappa}^m (\vect{n}) \end{pmatrix} \otimes \chi^{(s)}\,,
\end{equation*}
where we split the coordinates $\vect{x}$ into radial and angular components according to $x^\mu = \frac{1}{2}r_\text{\tiny B} y n^\mu$ such that $|\vect{n}| = 1$ and the radial variable $y$ is dimensionless. Furthermore we abbreviated frequent combinations of quantum numbers (and the finestructure constant $\alpha$) as
\begin{equation*}
K := \sqrt{\kappa^2 - \alpha^2}\,,\qquad W := n + K\,,\qquad w := \sqrt{W^2 + \alpha^2}\,.
\end{equation*}
The definitions of the Laguerre functions $L_n^\lambda(z)$ and the two component spherical spinors $\mathcal{Y}^m_\kappa (\vect{n})$, and the precise form of the normalization constant $\mathcal{N}$ are given in the appendix. 
The $\chi^{(s)}$ denote the unit spinors for the proton, with $\chi^{(+\tfrac{1}{2})} = (1,\,0)$, $\chi^{(-\tfrac{1}{2})} = (0,\,1)$.

Since we wish to inspect how general linear deviations from a merely metric background impact the 21~cm line, we are especially interested in the level splitting of the ground state(s). These we denote by $|ms\rangle:= |0, -1, m, s\rangle$ with both the electron spin $m$ and and the proton spin $s$ taking values either $-\tfrac{1}{2}$ or $+\tfrac{1}{2}$.  With all these prerequisites at hand, we can now begin to calculate all the relevant contributions to the 4-by-4 ground state transition matrix $\langle m's' | T_g | ms \rangle$. By the outline at the end of \cref{subsec:secondorder}, the latter is given by
\begin{align*}
\langle m's' | T_g | ms \rangle ={}& \langle m's' | T_\text{\tiny\labelcref{subsec:nonbiref}} |ms\rangle + \langle m's' | T_\text{\tiny\labelcref{subsec:lambda}} |ms\rangle + \langle m's' | T_\text{\tiny\labelcref{subsec:lambdamu}} | ms \rangle\,,
\end{align*}
where the single contributions explicitly read
\begin{align*}
\langle m's' | T_\text{\tiny\labelcref{subsec:nonbiref}} | ms \rangle ={}& \langle m's' | H_\text{\tiny hfs} | ms\rangle + \sum_{n'',\kappa'',m'',s''} \frac{\langle m's' | H_\text{\tiny hfs} | n''\kappa''m''s''\rangle\langle n''\kappa''m''s'' | H_\text{\tiny hfs} | ms \rangle}{E(n'',|\kappa''|)-E_g}\,, \\
\langle m's' | T_\text{\tiny\labelcref{subsec:lambda}} | ms \rangle ={}& \langle m's' | H_\text{\tiny abs} | ms\rangle\,, \phantom{\sum_{n''}}\\
\langle m's' | T_\text{\tiny\labelcref{subsec:lambdamu}} | ms \rangle ={}& \langle m's' | H_\text{\tiny rel} | ms\rangle +2 \sum_{n'',\kappa'',m'',s''} \frac{\langle m's' | H_\text{\tiny hfs} | n''\kappa''m''s''\rangle\langle n''\kappa''m''s'' | H_\text{\tiny abs} | ms \rangle}{E(n'',|\kappa''|)-E_g}\,.
\end{align*}
Remember that the sums in the second order terms run over only those combinations of quantum numbers for which $E(n'',|\kappa''|)$ is different from the ground state energy $E_g$, such that the denominator is non-zero. The second order term in the last line is a cross-term from the square, thus the factor of two. More precisely, this would have to be two terms, $H_\text{\tiny hfs}H_\text{\tiny abs}$ and $H_\text{\tiny abs}H_\text{\tiny hfs}$, which a priori are not equal. However, our explicit calculation below shows that this in fact is the case, which brings about the two.

\subsection{Corrections from proton spin, but without deviation from a metric background}\label{subsec:nonbiref}

This section is dedicated to the corrections that have nothing to do with the non-metric perturbation, namely the usual hyperfine structure, to first and second order perturbation theory in $\mu$. To this end, we consider the first order correction matrix elements $\langle n'\kappa'm's'| H_\text{\tiny hfs} |n\kappa ms\rangle$, which take the explicit form
\begin{align}
& \mu m_\text{\tiny red}\alpha\mathcal{N}'\mathcal{N}\notag\\
&\times \left[\left(w-W+w'-W'\right)\left(\tIoonk - \tIiink\right) + \left(w-W-w'+W'\right)\left(\tIoink - \tIionk\right)\right]\notag\\
&\times \left[\frac{\delta_{-\kappa', \kappa - 1}}{|2\kappa - 1|}\left(\tau_{-\kappa}^\mu\right)_{m'm} + \frac{4\kappa \delta_{\kappa',\kappa}}{4\kappa^2 - 1}\left(\sigma_\kappa^\mu\right)_{m'm} - \frac{\delta_{-\kappa',\kappa+1}}{|2\kappa+1|}\left(\tau_\kappa^\mu\right)_{m'm}\right]\left(\sigma_\mu\right)_{s's}\,. \label{eq:hfs}
\end{align}
The precise form of the integrals $\Ink{a}{ij}$ and the matrices $\sigma_\kappa^\mu$ and $\tau_\kappa^\mu$ are given in the appendix. Of immediate relevance, however, is that the $\sigma_{\pm 1}^\mu$ coincide with the three standard Pauli sigma matrices $\sigma^\mu$.
Using the above expression, we can now straightforwardly calculate the individual terms in $T_\text{\tiny\labelcref{subsec:nonbiref}}$ for the ground states. The first one is simply $\langle m's' | H_\text{\tiny hfs} | ms \rangle$, for which one obtains
\begin{equation*}
-\frac{2}{3}\frac{\mu m_\text{\tiny red}\alpha^3}{2-2\alpha^2-\sqrt{1-\alpha^2}}\left(\sigma^\mu\otimes\sigma_\mu\right)_{m's',ms} =: \frac{1}{4}\Delta E_\text{\tiny hfs}\begin{pmatrix} 1&&&\\&-1&2&\\&2&-1&\\&&&1 \end{pmatrix}_{m's',ms}\,.
\end{equation*}
This is the usual first order hyperfine transition matrix. Indeed, it has the three-fold degenerate eigenvalue $+\frac{1}{4}\Delta E_\text{\tiny hfs}$ corresponding to the hyperfine structure triplet state and the non-degenerate eigenvalue $-\frac{3}{4}\Delta E_\text{\tiny hfs}$ corresponding to the singlet state, given in terms of the hyperfine transition energy
\begin{equation*}
\Delta E_\text{\tiny hfs} \approx 9.41\times 10^{-25} \text{ J} = \frac{hc}{21.1\text{ cm}}\,.
\end{equation*}

Let us go on to the second term in $T_\text{\tiny\labelcref{subsec:nonbiref}}$. For generic states $|n\kappa ms\rangle$, $|n'\kappa'm's'\rangle$ of same energy, that is $n' = n$ and $|\kappa'| = |\kappa|$, we have the second order correction
\begin{equation*}
\sum_{n'',\kappa'',m'',s''}\frac{\langle n\kappa'm's'| H_\text{hfs} |n''\kappa'' m''s''\rangle \langle n''\kappa''m''s''| H_\text{hfs} |n\kappa ms\rangle}{E(n'',|\kappa''|) - E(n,|\kappa|)}\,,
\end{equation*}
for which one explicitly obtains
\begin{align*}
 4m_\text{\tiny red}&\alpha^2\mu^2\delta_{\kappa',\kappa} \Bigg\lbrace \delta_{m'm}\delta_{s's}\left[ \frac{\kappa-1}{2\kappa - 1} A_{n\kappa}(-\kappa+1) + \frac{4\kappa^2}{4\kappa^2 - 1} A_{n\kappa}(\kappa) + \frac{\kappa+1}{2\kappa+1} A_{n\kappa}(-\kappa-1)\right] \notag\\
 & + \left(\sigma_\kappa^\mu\otimes\sigma_\mu\right)_{m's',ms}\bigg[ \frac{\kappa-1}{(2\kappa-1)^2}A_{n\kappa}(-\kappa+1) + \frac{8\kappa^2}{(4\kappa^2-1)^2}A_{n\kappa}(\kappa) 
 - \frac{\kappa + 1}{(2\kappa + 1)^2}A_{n\kappa}(-\kappa-1)\bigg]\Bigg\rbrace\,,
\end{align*}
with the function $A_{n\kappa}(\kappa'')$ defined in the appendix. For the ground states $|ms\rangle$, this becomes
\begin{equation*}
a_1 \Delta E_\text{\tiny hfs}\, \delta_{m'm}\delta_{s's} + \frac{a_2}{4}\Delta E_\text{\tiny hfs} \begin{pmatrix} 1&&& \\ &-1&2& \\ &2&-1& \\ &&&1 \end{pmatrix}_{m's',ms}
\end{equation*}
with
\begin{align*}
a_1 :&= \frac{\mu}{\alpha}(2-2\alpha^2-\sqrt{1-\alpha^2})(A_{0,-1}(2)+2A_{0,-1}(-1)) \approx 2.0\times 10^{-8}\,,\\
a_2 :&= \frac{4\mu}{3\alpha}(2-2\alpha^2-\sqrt{1-\alpha^2})(A_{0,-1}(2)-4A_{0,-1}(-1)) \approx -5.4\times 10^{-8}\,.
\end{align*}
The defintions are exact, but the approximate numeric values are given only as a rough estimate. They have been obtained numerically by truncating the infinite sum over $n''$ occuring in $A_{n\kappa}$ and will need to be replaced by sufficiently precise estimates before they are used to compare predictions to data.

The results of this subsection yield the contribution $T_\text{\tiny\labelcref{subsec:nonbiref}}$ to the full transition matrix. In addition to the usual first order hyperfine structure, we obtained the second order terms, which qualitatively, however, yield nothing new: the $a_1$ term is proportional to unity, $\delta_{m'm}\delta_{s's}$, and thus just shifts all the states together, while $a_2$ constitutes a small correction to the hyperfine transition energy $\Delta E_\text{\tiny hfs}$. We had to calculate them, however, in order to be consistent with the introduction of all corrections linear in $E$ in the next two subsections.

\subsection{Corrections due to deviations from a metric background, but without proton spin}\label{subsec:lambda}
In this section we calculate the purely non-metric contribution to the transition matrix, that is $\langle m's' | H_\text{\tiny abs} | ms\rangle$. Since we will need to use them again for $T_\text{\tiny\labelcref{subsec:lambdamu}}$, let us start by calculating all the matrix elements of $H_\text{\tiny abs}$. For $ \langle n'\kappa'm's' | H_\text{\tiny abs} | n\kappa m s\rangle$ one gets
\begin{eqnarray}
 & & \frac{m_\text{\tiny red}\alpha}{4}\mathcal{N}'\mathcal{N}\delta_{s's} \notag\\
& &\times\Bigg\lbrace \bigg[\left(\alpha + \frac{(w'-W')(w-W)}{\alpha}\right)\left(\iIoonk + \iIiink\right) 
- \left(\alpha - \frac{(w'-W')(w-W)}{\alpha}\right)\left(\iIoink + \iIionk\right)\bigg] \notag\\
& &\qquad\times \bigg[\frac{\delta_{\kappa',\kappa-2}}{4(\kappa-1)^2 - 1}\left(\tau_{-\kappa+1}^\mu\tau_{-\kappa}^\nu\right)_{m'm} + \frac{\delta_{\kappa',\kappa+2}}{4(\kappa+1)^2 - 1}\left(\tau_{\kappa+1}^\mu\tau_{\kappa}^\nu\right)_{m'm} + \frac{4\delta_{-\kappa',\kappa-1}}{|2\kappa-3|(4\kappa^2-1)}\left(\tau_{-\kappa}^\mu\sigma_{-\kappa}^\nu\right)_{m'm} \notag\\
& &\qquad\qquad + \frac{4\delta_{-\kappa',\kappa+1}}{|2\kappa+3|(4\kappa^2-1)}\left(\tau_{\kappa}^\mu\sigma_{\kappa}^\nu\right)_{m'm} 
- \frac{2\delta_{\kappa'\kappa}}{4\kappa^2-1}\left(\sigma_\kappa^\mu\sigma_\kappa^\nu - (4\kappa^2-1)\eta^{\mu\nu}\mathbb{1}\right)_{m'm}\bigg]E_{\mu 0\nu 0}\notag\\
& &\quad +\frac{1}{2} \bigg[\left(w-W+w'-W'\right)\left(\iIoonk - \iIiink\right) + \left(w-W-w'+W'\right)\left(\iIoink - \iIionk\right)\bigg] \notag\\
& &\qquad\quad\times \bigg[ \frac{\delta_{-\kappa',\kappa-2}}{4(\kappa-1)^2 - 1}\left(\tau_{-\kappa+1}^\mu\tau_{-\kappa}^\nu\right)_{m'm} - \frac{\delta_{-\kappa',\kappa+2}}{4(\kappa+1)^2 - 1}\left(\tau_{\kappa+1}^\mu\tau_{\kappa}^\nu\right)_{m'm} \notag\\
& &\qquad\qquad + \frac{2\delta_{\kappa',\kappa-1}}{|2\kappa+1|(2\kappa-3)}\left(\tau_{-\kappa}^\mu\sigma_{-\kappa}^\nu\right)_{m'm} + \frac{2\delta_{\kappa',\kappa+1}}{|2\kappa-1|(2\kappa+3)}\left(\tau_{\kappa}^\mu\sigma_{\kappa}^\nu\right)_{m'm} \bigg] \epsilon_{\mu\rho\sigma}E^{\rho\sigma}{}_{\nu 0} \Bigg\rbrace\,, \label{eq:abs}
\end{eqnarray}
which for the ground states reduces to
\begin{equation*}
\frac{m_\textrm{\tiny red}\alpha^2 E^{ab}{}_{ab}}{12\sqrt{1-\alpha^2}}\delta_{m'm}\delta_{s's}\,.
\end{equation*}
Note that, just like the $a_1$ term in $T_\text{\tiny\labelcref{subsec:nonbiref}}$, this contribution is proportional to the identity $\delta_{m'm}\delta_{s's}$; it does not contribute to any splitting of the states, but merely yields an absolute shift of all the states together. Thus it is irrelevant to the hyperfine structure, but contributes to a shift in the ionization energy (\ref{Eionization}) below.

\subsection{Combined corrections from proton spin and deviations from a metric background}\label{subsec:lambdamu}
We now come to the calculation of $\langle m's' | T_\text{\tiny\labelcref{subsec:lambdamu}} | ms \rangle$, which presents the non-metric correction to the hyperfine structure. As before, for generality we will start with arbitrary states of same energy ($n' = n$, $|\kappa'| = |\kappa|$) and only then specialize to the ground states $|ms\rangle$.

The matrix elements of $H_\text{\tiny rel}$ between states of same energy, $\langle n\kappa'm's' | H_\text{\tiny rel} | n\kappa ms\rangle$, are
\begin{align*}
 -\frac{1}{2}&\frac{m_\text{\tiny red}\alpha^2}{4\kappa^2 - 1}\mathcal{N}^2\sqrt{\frac{w-\kappa'}{w-\kappa}} (\sigma_\mu)_{s's} \notag\\
&\times\Bigg\lbrace \delta_{\kappa',-\kappa} \bigg[\left(\alpha + \frac{(w-W)^2}{\alpha}\right)\left(\tIoonk + \tIiink\right) %
- \left(\alpha - \frac{(w-W)^2}{\alpha}\right)\left(\tIoink + \tIionk\right)\bigg] \notag\\
&\qquad\quad\,\,\times \left[\eta^{\rho\sigma}\sigma_\kappa^\mu - \frac{9}{2(4\kappa^2-9)}\left(\sigma_\kappa^{(\mu}\sigma_\kappa^\rho\sigma_\kappa^{\sigma)} + (4\kappa^2-5)\eta^{(\rho\sigma}\sigma_\kappa^{\mu)}\right)\right]_{m'm} \epsilon_{\rho\alpha\beta}E^{\alpha\beta}{}_{\sigma 0} \notag\\
&\quad\quad + \delta_{\kappa',\kappa}\frac{4\kappa(w-W)}{4\kappa^2-9}\left(\tIoonk - \tIiink\right) \notag\\
&\qquad\quad\,\,\times \big[4(4\kappa^2-9)\eta^{\mu\rho}\sigma_\kappa^{\sigma} + (4\kappa^2+3)\eta^{\rho\sigma}\sigma_\kappa^\mu + 3\sigma_\kappa^\mu\sigma_\kappa^\rho\sigma_\kappa^\sigma - 6\ii\epsilon^{\mu\rho}{}_\nu\sigma_\kappa^\sigma\sigma_\kappa^\nu\big]_{m'm} E_{\rho 0\sigma 0} \notag\\
&\quad\quad - \delta_{\kappa',\kappa}(w-W)\left(\tIoink-\tIionk\right)\ii\epsilon^{\mu\rho}{}_\nu\lbrace\sigma_\kappa^\sigma, \sigma_\kappa^\nu\rbrace_{m'm}E_{\rho 0\sigma 0} \Bigg\rbrace\,,
\end{align*}
which for the ground states reduces to
\begin{equation*}
\frac{1}{4}\Delta E_\text{\tiny hfs}\left(\frac{7}{10}\widehat{E}^\mu{}_\nu + \frac{1}{12}E^{ab}{}_{ab}\delta^\mu_\nu\right)(\sigma_\mu\otimes\sigma^\nu)_{m's',ms}\,,
\end{equation*}
and thus directly yields the first order term in $T_\text{\tiny\labelcref{subsec:lambdamu}}$. Note that the tracefree tensor (\ref{widehatE}) makes another appearance here. As before, where $\widehat{E}^{\mu\nu}$ controlled a qualitatively new spin-magnetic coupling, also here the term $\widehat{E}^\mu{}_\nu(\sigma_\mu\otimes\sigma^\nu)$ produces a qualitatively new effect: it lifts the degeneracy of the triplet states in the hyperfine structure of the hydrogen atom, as we will see in detail in the next section.

For the remaining second order term in $T_\text{\tiny\labelcref{subsec:lambdamu}}$ we take recourse to the matrix elements of $H_\text{\tiny hfs}$ and $H_\text{\tiny abs}$, calculated in (\ref{eq:hfs}) and (\ref{eq:abs}), respectively. Between states of equal energy, we find the second order contribution
\begin{equation*}
2\sum_{n'',\kappa'',m'',s''}\frac{\langle n\kappa'm's'| H_\text{\tiny hfs} |n''\kappa'' m''s''\rangle \langle n''\kappa''m''s''| H_\text{\tiny abs} |n\kappa ms\rangle}{E(n'',|\kappa''|) - E(n,|\kappa|)}
\end{equation*}
to be given explicitly by
\begin{align*}
 & - \frac{2\mu m_\text{\tiny red}\alpha^2\left(\sigma_\mu\right)_{s's}}{4\kappa^2-1} \notag\\
 &\times\Bigg\lbrace \delta_{\kappa'\kappa}\bigg[\frac{B_{n\kappa}(-\kappa+1)}{4(\kappa-1)^2-1}\Big(\sigma_\kappa^\mu\sigma_\kappa^{\alpha}\sigma_\kappa^{\beta} + \left(4(\kappa - 1)^2 -1\right)\eta^{\mu\alpha}\sigma_\kappa^{\beta} + 2(2\kappa-1)\eta^{\alpha\beta}\sigma_\kappa^\mu - (2\kappa-1)\ii\epsilon^{\mu\alpha}{}_\tau\sigma_\kappa^\beta\sigma_\kappa^\tau\Big)_{m'm} \notag\\
&\qquad - \frac{B_{n\kappa}(-\kappa-1)}{4(\kappa+1)^2 -1}\Big(\sigma_\kappa^\mu\sigma_\kappa^\alpha\sigma_\kappa^\beta + \left(4(\kappa+1)^2 - 1\right)\eta^{\mu\alpha}\sigma_\kappa^\beta
- 2(2\kappa+1)\eta^{\alpha\beta}\sigma_\kappa^\mu + (2\kappa+1)\ii\epsilon^{\mu\alpha}{}_\tau\sigma_\kappa^\beta\sigma_\kappa^\tau\Big)_{m'm}\notag\\
&\qquad + 2\kappa B_{n\kappa}(\kappa)\Big(\sigma_\kappa^\mu\sigma_\kappa^\alpha\sigma_\kappa^\beta - (4\kappa^2-1)\eta^{\alpha\beta}\sigma_\kappa^\mu\Big)_{m'm} \bigg]E_{\alpha 0\beta 0} \notag\\
&\quad -\frac{1}{4}\delta_{\kappa',-\kappa}\bigg[\frac{C_{n\kappa}(\kappa-1)}{2\kappa-3}\Big(\sigma_\kappa^\mu\sigma_\kappa^{\alpha}\sigma_\kappa^{\beta} + \left(4(\kappa - 1)^2 -1\right)\eta^{\mu\alpha}\sigma_\kappa^{\beta} 
+ 2(2\kappa-1)\eta^{\alpha\beta}\sigma_\kappa^\mu - (2\kappa-1)\ii\epsilon^{\mu\alpha}{}_\tau\sigma_\kappa^\beta\sigma_\kappa^\tau\Big)_{m'm} \notag\\
&\qquad -\frac{C_{n\kappa}(\kappa+1)}{2\kappa+3}\Big(\sigma_\kappa^\mu\sigma_\kappa^\alpha\sigma_\kappa^\beta + \left(4(\kappa+1)^2 - 1\right)\eta^{\mu\alpha}\sigma_\kappa^\beta 
- 2(2\kappa+1)\eta^{\alpha\beta}\sigma_\kappa^\mu + (2\kappa+1)\ii\epsilon^{\mu\alpha}{}_\tau\sigma_\kappa^\beta\sigma_\kappa^\tau\Big)_{m'm}  \bigg]\epsilon_{\alpha\rho\sigma}E^{\rho\sigma}{}_{\beta 0}\!\Bigg\rbrace\,,
\end{align*}
with $B_{n\kappa}(\kappa'')$ and  $C_{n\kappa}(\kappa'')$ defined in the appendix. 
For the ground states, this reduces to
\begin{equation*}
-\frac{\lambda}{4}\Delta E_\text{\tiny hfs} (b_2\widehat{E}^\mu{}_\nu + \frac{b_1}{12} E^{ab}{}_{ab}\delta^\mu_\nu)(\sigma_\mu\otimes\sigma^\nu)_{m's',ms}\,,
\end{equation*}
with
\begin{align*}
b_1 :={}& \frac{24}{\alpha}(2-2\alpha^2-\sqrt{1-\alpha^2})B_{0,-1}(-1) \approx 1.74\,,\\
b_2 :={}& \frac{4}{5\alpha}(2-2\alpha^2-\sqrt{1-\alpha^2})B_{0,-1}(2) \approx -2.9\times 10^{-4}\,.
\end{align*}
Again, the second order does not bring qualitatively new features. Both terms appearing in the formula above are already present in the first order correction due to $H_\text{\tiny rel}$. The precise impact of all these terms on the energy levels will be the topic of the next section.

\subsection{Hyperfine structure and ionization energy of hydrogen}\label{subsec:full}
We find that three out of the eleven general linear deviations from a metric background affect the hyperfine structure. In particular, the triplet states are no longer degenerate in general, as illustrated in \cref{fig:boat1}. Their splitting is controlled by the three cubic roots of a complex number that encodes two of the three effective degrees of freedom. The remaining relevant parameter is the full contraction $E^{ab}{}_{ab}$, which shifts both, the energy of ionization and of the hyperfine transition. The latter plays a central role for the detection of hydrogen and its state of motion in astrophysical measurements.

\begin{figure}[h]
\begin{tikzpicture}

\newcommand{\levellength}{2}
\newcommand{\levelseph}{1}
\newcommand{\levelline}{ultra thick}
\newcommand{\conline}{dashed}
\newcommand{\tripletv}{3}
\newcommand{\tripletsplit}{.0}
\newcommand{\cosii}{.966}
\newcommand{\cosiii}{-.707}
\newcommand{\cosi}{-.259}
\newcommand{\cosfactor}{1}
\newcommand{\fsplit}{.5}
\newcommand{\xUnpert}{0}
\newcommand{\yUnpert}{0}
\newcommand{\xf}{\xUnpert+\levellength+\levelseph}
\newcommand{\xM}{\xf+\levellength+\levelseph}
\newcommand{\confU}{\xUnpert+\levellength}
\newcommand{\conff}{\xf}
\newcommand{\conMM}{\xM}
\newcommand{\conMU}{\xf+\levellength}
\newcommand{\skipAxis}{.5}
\newcommand{\axisBottom}{\yUnpert-3*\fsplit-\skipAxis}
\newcommand{\ylabels}{\axisBottom-\skipAxis}

\draw[thick,->] (\xUnpert-.5*\levelseph,\axisBottom) -- (\xUnpert-.5*\levelseph,\yUnpert+\tripletv+\cosfactor*\cosii+2*\skipAxis) node[left,label={[label distance = 0cm, rotate = 90]left:Energy}]{};

\draw[\levelline] (\xUnpert,\yUnpert) node[above,label={[distance=0cm]right:singlet}]{} 							-- (\xUnpert+\levellength,\yUnpert);
\draw[\levelline] (\xUnpert,\yUnpert+\tripletv) 				-- (\xUnpert+\levellength,\yUnpert+\tripletv);
\draw[\levelline] (\xUnpert,\yUnpert+\tripletv+\tripletsplit) node[above,label={[distance=0cm]right:triplet}]{} 	-- (\xUnpert+\levellength,\yUnpert+\tripletv+\tripletsplit);
\draw[\levelline] (\xUnpert,\yUnpert+\tripletv-\tripletsplit) 	-- (\xUnpert+\levellength,\yUnpert+\tripletv-\tripletsplit);

\draw[\conline] (\confU,\yUnpert) 							-- (\conff,\yUnpert-3*\fsplit);
\draw[\conline] (\confU,\yUnpert+\tripletv) 				-- (\conff,\yUnpert+\tripletv+\fsplit);
\draw[\conline] (\confU,\yUnpert+\tripletv+\tripletsplit) 	-- (\conff,\yUnpert+\tripletv+\tripletsplit+\fsplit);
\draw[\conline] (\confU,\yUnpert+\tripletv-\tripletsplit) 	-- (\conff,\yUnpert+\tripletv-\tripletsplit+\fsplit);

\draw[\levelline] (\xf,\yUnpert-3*\fsplit)								-- (\xf+\levellength,\yUnpert-3*\fsplit);
\draw[\levelline] (\xf,\yUnpert+\tripletv+\fsplit) 						-- (\xf+\levellength,\yUnpert+\tripletv+\fsplit);
\draw[\levelline] (\xf,\yUnpert+\tripletv+\tripletsplit+\fsplit) 		-- (\xf+\levellength,\yUnpert+\tripletv+\tripletsplit+\fsplit);
\draw[\levelline] (\xf,\yUnpert+\tripletv-\tripletsplit+\fsplit) 		-- (\xf+\levellength,\yUnpert+\tripletv-\tripletsplit+\fsplit);

\draw[\conline] (\conMU,\yUnpert-3*\fsplit) 						-- (\conMM,\yUnpert-3*\fsplit);
\draw[\conline] (\conMU,\yUnpert+\tripletv+\fsplit) 				-- (\conMM,\yUnpert+\tripletv+\cosfactor*\cosi+\fsplit);
\draw[\conline] (\conMU,\yUnpert+\tripletv+\tripletsplit+\fsplit) 	-- (\conMM,\yUnpert+\tripletv+\cosfactor*\cosii+\fsplit);
\draw[\conline] (\conMU,\yUnpert+\tripletv-\tripletsplit+\fsplit) 	-- (\conMM,\yUnpert+\tripletv+\cosfactor*\cosiii+\fsplit);

\draw[\levelline] (\xM,\yUnpert-3*\fsplit) 								-- (\xM+\levellength,\yUnpert-3*\fsplit);
\draw[\levelline] (\xM,\yUnpert+\tripletv+\cosfactor*\cosi+\fsplit) 	-- (\xM+\levellength,\yUnpert+\tripletv+\cosfactor*\cosi+\fsplit);
\draw[\levelline] (\xM,\yUnpert+\tripletv+\cosfactor*\cosii+\fsplit) 	-- (\xM+\levellength,\yUnpert+\tripletv+\cosfactor*\cosii+\fsplit);
\draw[\levelline] (\xM,\yUnpert+\tripletv+\cosfactor*\cosiii+\fsplit) 	-- (\xM+\levellength,\yUnpert+\tripletv+\cosfactor*\cosiii+\fsplit);

\node at (\xUnpert+.5*\levellength,\ylabels) {\parbox{\levellength cm}{$E^{ab}{}_{ab} = 0$, $\widehat{E}_{\mu\nu} = 0$}};
\node at (\xf+.5*\levellength,\ylabels) {\parbox{\levellength cm}{$E^{ab}{}_{ab} \neq 0$, $\widehat{E}_{\mu\nu} = 0$}};
\node at (\xM+.5*\levellength,\ylabels) {\parbox{\levellength cm}{$E^{ab}{}_{ab} \neq 0$, $\widehat{E}_{\mu\nu} \neq 0$}};

\end{tikzpicture}
  \caption{Hyperfine structure of hydrogen in the presence of general linear deviations from a mere metric background. Left: the degenerate triplet for $E^{ab}{}_{ab} = 0$, $\widehat{E}_{\mu\nu} = 0$. Note that this is the case if the entire perturbation $E^{abcd} $ vanishes, but that the converse does not hold: there may still be components $E^{\alpha\beta\gamma 0} \neq 0$, but those do not affect the ground states. Middle: the triplet remains degenerate, but the hyperfine transition energy is modified if $E^{ab}{}_{ab} \neq 0$, $\widehat{E}_{\mu\nu}=0$. Right: the triplet is resolved, but with an unchanged center of weight, if and only if $\widehat{E}_{\mu\nu}\neq 0$.}
  \label{fig:boat1}
\end{figure}

Collecting all relevant terms from sections \ref{subsec:nonbiref}, \ref{subsec:lambda} and \ref{subsec:lambdamu}, we obtain for the ground state transition matrix
\begin{equation*}
\langle m's' | T_g | ms \rangle = \langle m's' | T_\text{\tiny\labelcref{subsec:nonbiref}} |ms\rangle + \langle m's' | T_\text{\tiny\labelcref{subsec:lambda}} |ms\rangle + \langle m's' | T_\text{\tiny\labelcref{subsec:lambdamu}} | ms \rangle
\end{equation*}
the sum
\begin{align}
\left(a_1\Delta E_\text{\tiny hfs} \!+\! \frac{m_\text{\tiny red}\alpha^2 E^{ab}{}_{ab}}{12\sqrt{1-\alpha^2}}\right)&\delta_{m'm}\delta_{s's}\nonumber\\
- \frac{1}{4}\Delta E_\text{\tiny hfs}&\left[\left(1\!+\!a_2\!-\!\frac{1\!-\!b_1}{12}E^{ab}{}_{ab}\right)\delta^\mu_\nu\!-\!\left(\frac{7}{10}\!-\!b_2\right)\widehat{E}^\mu{}_\nu\right]\!(\sigma_{\mu} \otimes\sigma^\nu)_{m'\!s'\,,ms}\,.\label{eq:transitionMatrix}
\end{align}
In order to calculate the eigenvalues $\Delta E_k$, $k\in\lbrace 0,1,2,3 \rbrace$, we note that the transition matrix has two constituents: the first line is proportional to $\delta_{m'm}\delta_{s's}$, which is the identity on the space of ground states, diagonal in every basis, whereas the second line has non-trivial eigenvalues. Thus we may make the ansatz
\begin{equation}\label{eq:DE}
\Delta E_k = \left(a_1\Delta E_\text{\tiny hfs} + \frac{m_\text{\tiny red}\alpha^2 E^{ab}{}_{ab}}{12\sqrt{1-\alpha^2}}\right) + \frac{1}{4}\Delta E_\text{\tiny hfs}\, \xi_k\,.
\end{equation}
Up to a factor $\frac{1}{4}\Delta E_\text{\tiny hfs}$, the $\xi_k$ controlling these eigenvalues are given by the roots of the characteristic polynomial of the second line in \cref{eq:transitionMatrix}. This is a quartic polynomial, and thus algebraically solvable. One quickly finds the first root, and thus the first hyperfine energy level through (\ref{eq:DE}), 
\begin{equation}
\xi_0 = -3\left(1+a_2+ \frac{b_1-1}{12}E^{ab}{}_{ab}\right)\,,\label{eq:xisa}
\end{equation}
which reduces the problem of finding the remaining three hyperfine structure energy levels to the solution of a cubic polynomial. Then using the known solution formula one also finds the remaining three ($\kappa=1,2,3$) roots 
\begin{equation}
  \xi_{\kappa} = 1+a_2 + \frac{b_1 - 1}{12}E^{ab}{}_{ab} + 2^{\frac{5}{3}}\left(\frac{7}{10} - b_2\right) \,\text{Re}\sqrt[3,\kappa]{\text{det }\widehat{E} + \ii\sqrt{\frac{1}{54}(\text{tr }\widehat{E}^2)^3 - (\text{det }\widehat{E})^2}}\,, \label{eq:xisb}
\end{equation}
featuring the four numerical constants $a_1$, $a_2$, $b_1$, $b_2$ determined in the two subsections \ref{subsec:nonbiref} and \ref{subsec:lambdamu} above.
Note in particular, that the last term involves a cubic root, whose three possible choices of complex phase distinguish the three $\xi_{\kappa}$. 
Suppressing all deviations from a metric background and second order terms ($a_{1/2} = b_{1/2} = 0$), we are left with
\begin{align*}
\Delta E_0 ={}& -\frac{3}{4}\Delta E_\text{\tiny hfs}& \textrm{and}&& \Delta E_{\kappa} ={}& \frac{1}{4}\Delta E_\text{\tiny hfs}\,,
\end{align*}
which properly recovers the standard first order hyperfine structure, see the left of fig. \ref{fig:boat1}. From (\ref{eq:xisa}) and (\ref{eq:xisb}), we see that this standard splitting is now stretched to 
\begin{equation}\label{stretching}
  \Delta E_\text{\tiny hfs}^\textrm{\tiny AM} = \left(1+a_2+ \frac{b_1-1}{12}E^{ab}{}_{ab}\right)  \Delta E_\text{\tiny hfs}
\end{equation}
for weak area metric deviations of the background geometry from a Lorentzian spacetime. Note that the stretching factor can be both larger and smaller than unity. This reveals a modification of the hyperfine splitting $\Delta E_\text{\tiny hfs}$, but yields not a qualitatively new effect, see the middle of fig. \ref{fig:boat1}. 

Qualitatively new, however, is the splitting of the triplet states, induced by the cubic root in \cref{eq:xisb}. The real, symmetric and tracefree perturbation components $\widehat{E}^\mu{}_\nu$ feature only two invariant perturbation degrees of freedom, while the remaining three encode the spatial orientation of the eigenbasis, but do not affect the energy levels of the hydrogen atom. This can be deduced from the fact that $\text{det }\widehat{E}$ as well as $\text{tr }\widehat{E}^2$ can be written completely in terms of the eigenvalues of $\widehat{E}$. Doing so, one also quickly sees that
$(\text{tr }\widehat{E}^2)^3 \geq 54(\text{det }\widehat{E})^2$, 
where equality holds when two eigenvalues coincide. The expression under the cubic root thus takes values in the entire complex plane, up to the fact that the sign of the imaginary part is determined by the choice of sign for the square root. Precisely this sign, however, is irrelevant, since only the real part of the three cubic roots contributes to (\ref{eq:xisb}). Picturing the cubic root as an equilateral triangle in the complex plane, whose corners are projected onto the real axis, it is evident that the unweighted average of the three levels remains unaltered, since the sum of the three possible roots adds up to zero. Orientation and size of the mentioned equilateral triangle correspond to the two effective degrees of freedom of $\widehat{E}^\mu{}_\nu$, see the prototypical situation on the right of fig. \ref{fig:boat1}.

With the energy level of the singlet state now known, we finally collect all terms that contribute to the ionization energy 
\begin{equation}\label{Eionization}
E^\textrm{\tiny AM}_\textrm{\tiny ionization} = m_\textrm{\tiny red}\left(1-\sqrt{1-\alpha^2}\right) + \left(\frac{3+3a_2}{4} - a_1\right) \Delta E_\textrm{\tiny hfs} - \frac{1}{12}\left(\frac{m_\textrm{\tiny red}\alpha^2}{\sqrt{1-\alpha^2}}+\frac{3-3b_1}{4} \Delta E_\textrm{\tiny hfs}\right) E^{ab}{}_{ab}
\end{equation}
of a hydrogen atom on an effectively flat region of an area metric background. As in the scaling (\ref{stretching}) of the $21.1\, \textrm{cm}$ line of hydrogen, it is again the double trace $E^{ab}{}_{ab}$ of the  perturbation that single-handedly controls the shift in the ionization energy.


\section{\new{Scattering and hydrogen atom around a point mass}}
\label{sec:around-a-point-mass}
In the previous sections, we have seen some effects produced on quantum electrodynamics by a weak, but otherwise arbitrary, area metric perturbation~(\ref{perta}) of a Minkowskian background.
In fact, the area metric is of course not arbitrary, but obeys precise dynamics sourced by the matter in the universe.
The comparison between the physical quantities computed in the previous sections and experiments, which we wish now to undertake, requires the knowledge of the area metric perturbation in the system under consideration.
For an experiment on the surface of Earth, for instance, we need to determine the value of the area metric perturbation around Earth and only then compare theoretical quantities with experimental data.

In this section, we now illustrate the results of the previous sections~\ref{sec:scatterings} and~\ref{sec:hydrogen} for an underlying weak gravitational field as given by the particular  solution~(\ref{am-around-point-mass})-(\ref{am-around-point-mass-end}) of the gravitational field equations for an area metric sourced by a point mass.
This yields concrete predictions for these scattering amplitudes and bound state energies for any situation where the gravitational source is, at least in first approximation, a point mass.
This includes, notably, observations performed around Earth, stars, and galaxies.

\subsection{Area metric perturbation produced by a point mass}

In this section, we make use of the area-metric solution around a point mass~(\ref{am-around-point-mass}-\ref{am-around-point-mass-end}).
Following section~\ref{sec:flat-regions}, we assume that the area metric is flat in the region where the experiment is taking place. Moreover, we move to a new frame by means of~(\ref{frame-transf}).
In this new frame, which is an observer frame, the area metric becomes
\begin{equation} \label{area-metric-pm-sf}
  E^{0\alpha0\beta} =	\widetilde V (R)\, \gamma^{\alpha \beta} \,, \qquad
  E^{0\alpha \beta \gamma} = 0	\,, \qquad
  E^{\alpha \beta \gamma \delta} = -\widetilde V (R) \,\big( \gamma^{\alpha \gamma} \gamma^{\beta \delta} - \gamma^{\alpha \delta} \gamma^{\beta \gamma} \big) \,, \qquad
  e=-\frac{3}{2}\, \widetilde V (R) \,,
\end{equation}
for which the conditions~(\ref{pertconditionsa}) are readily satisfied.
Note that, as was to be expected, only area-metric corrections contribute once the quasi-flat frames hav been chosen.
We recall that $\widetilde V (R) \eqdef \frac{M\gamma}{4\pi R} e^{-\sqrt{\mu}R}$, where $\gamma$ and $\mu$ are constants of the theory.

Finally, some of the results of section~\ref{sec:scatterings} and section~\ref{sec:hydrogen} were expressed using the double trace of the perturbation $E^{ab}{}_{ab}$ and the tracefree tensor $\widehat{E}^{\alpha \beta}\eqdef E^{\alpha 0 \beta 0}+\gamma^{\alpha \beta} \, E^{ij}{}_{ij}/12$.
These two quantities can be easily computed from~(\ref{area-metric-pm-sf}), finding
\begin{equation} \label{doubletrace-tracefree-pm}
 E^{ab}{}_{ab}=-12\,\widetilde V (R) \qquad \text{and} \qquad \widehat{E}^{\alpha \beta}=0 \,.
\end{equation}

In the next subsections, we specialize the results concerning scatterings and the hydrogen atom using~(\ref{area-metric-pm-sf}) and~(\ref{doubletrace-tracefree-pm}).

\subsection{$e^+ e^-\rightarrow \bar{f}f$ scattering}
Let us begin with the scattering of an electron and a positron into a fermion and its antifermion, described in section~(\ref{sec:e+e-fermion-antifermion}).
As a consequence of $\widehat{E}^{\alpha \beta}=0$, the differential cross section~(\ref{cross-section-muons}) does not depend on the angle $\varphi$ and we have
\begin{equation} \label{muon-antimuon-pm}
 \frac{d \sigma}{d \Omega}=\Big( 1+2 \widetilde V (R) \Big) \frac{Q^2 \alpha^2}{4s} \sqrt{\frac{s-4 M^2}{s-4 m^2}} \left[ 1+\frac{4(m^2+M^2)}{s}+\left( 1-\frac{4m^2}{s} \right) \left( 1-\frac{4M^2}{s} \right) \cos^2 \theta \right] \,.
\end{equation}
This is precisely the cross section in standard QED times a factor $1+2 \widetilde V (R)$.
Note that, since we are in linear perturbation theory, $1+2 \widetilde V (R)$ is equivalent to the square of $1+ \widetilde V (R)$, whose qualitative behavior is depicted in figure~\ref{fig:one+u}.

\subsection{Bhabha scattering}
Let us continue with the elastic scattering of an electron and a positron, described in section~\ref{sec:Bhabha}.
Also in this case, as a consequence of $\widehat{E}^{\alpha \beta}=0$, the differential cross section~(\ref{cross-section-Bhabha}) does not depend on the angle $\varphi$. We find
\begin{equation} \label{Bhabha-pm}
 \frac{d \sigma}{d \Omega}=\Big( 1+2 \widetilde V (R) \Big)\, \frac{\alpha^2}{2s} \left[ \frac{1}{2} (1+\cos^2 \theta)+\frac{1+\cos^4 (\theta/2)}{\sin^4 (\theta/2)} - 2\frac{\cos^4 (\theta/2)}{\sin^2 (\theta/2)} \right] \,,
\end{equation}
which is again the cross section in standard QED times a factor $1+2 \widetilde V (R)$.

\subsection{Anomalous magnetic moment}
In section~\ref{sec:AMM-beyond-one-loop}, we showed that the anomalous magnetic moment in the presence of an area metric perturbation can be computed starting from that in standard QED $a_{\text{\tiny QED}}(\alpha)$ expressed as a function of the fine-structure constant $\alpha$.
Specializing equation~(\ref{AMM-loops-relation}) for the area metric around a point mass~(\ref{area-metric-pm-sf}), the anomalous magnetic moment becomes
\begin{equation} \label{AMM-pm}
 a_{\text{point mass}}=a_{\text{\tiny QED}} \Big(\alpha \big(1+\widetilde V (R) \big)\Big) 	\,.
\end{equation}
We notice again the dependence on the function $1+\widetilde V (R)$, whose qualitative behavior is depicted is figure~\ref{fig:one+u}.
For instance, at one loop, we find
\[
 a_{\text{point mass}}^{(1-\text{loop})}=\frac{\alpha}{2\pi} \big(1+\widetilde V (R) \big) \,,
\]
which is the usual Schwinger correction times the factor $1+\widetilde V (R)$.

\begin{figure}[t]
  \begin{tikzpicture}
\begin{axis}[scale=3,
          xmin=0,xmax=2,
          ymin=0,ymax=1.1,
          x=1cm,y=1cm,
          hide axis,
          enlargelimits=false
          ]
\addplot[thick,domain=0:2,samples=100]  {1-0.25*pow(e,-x)/(1+x)};
\end{axis}
\node[below] at (0.3,0) {\small$R_0$};
\node[below] at (6,0) {\small$R$};
\node at (5,2.6) {\small $1+\widetilde V (R)$};
\node[left] at (-0.15,0) {\small $0$};
\node[left] at (-0.15,3) {\small $1$};
\draw[->,>=latex] (0,-0.4) -- (0,3.3);
\draw[->,>=latex] (-0.2,0) -- (6,0);
\draw[dashed] (-0.2,3) -- (6,3);
\end{tikzpicture}
\caption{The behavior of the function $1+\widetilde V (R)$ is depicted above. The function increases as the distance from the point mass increases and asymptotically approaches the value of $1$ at large distances.
This way, far away from the point mass, we recover standard QED and any local effect of the area metric disappears.
Note that the minimum distance from the point mass $R_0$ could represent either a real boundary of the system, e.g., the radius of Earth $R_\oplus$ for experiments around Earth, or the distance below which the linear approximation ($E\ll 1$) cease to be valid.}
  \label{fig:one+u}
\end{figure}

\subsection{Hyperfine structure and ionization energy of hydrogen}
Let us move on to the discussion about the hydrogen atom.
The presence of an area metric perturbation generated by a point mass affects in two ways the hyperfine structure of hydrogen.

While triplet does not split in the weak gravitational field around a point mass, since the tracefree tensor $\widehat{E}^{\alpha \beta}$ is zero, the hyperfine splitting of the singlet and the triplet is indeed stretched.
Therefore, in this case, we are in the situation depicted in the middle of figure~\ref{fig:boat1}.
Inserting~(\ref{doubletrace-tracefree-pm}) into the general formula~(\ref{stretching}), we find
\begin{equation}\label{stretching-pm}
  \Delta E_\text{\tiny hfs}^\textrm{\tiny \text{point mass}} = \Big[1+a_2- (b_1-1) \widetilde V (R) \Big]  \Delta E_\text{\tiny hfs} \,,
\end{equation}
where $a_1$ and $b_1$ are the numerical factors determined in sections~\ref{subsec:nonbiref} and~\ref{subsec:lambdamu}.
The behavior of $\Delta E_\text{\tiny hfs}^\textrm{\text{point mass}}/\Delta E_\text{\tiny hfs}$ is depicted in figure~\ref{fig:hfs-energy}.

A further effect caused by the weak area-metric field around a point mass is the shift in the ionization energy.
Inserting~(\ref{doubletrace-tracefree-pm}) into the general formula~(\ref{Eionization}), we find
\begin{equation}\label{Eionization-pm}
E^\textrm{\tiny \text{point mass}}_\textrm{\tiny ionization} = m_\textrm{\tiny red}\left(1-\sqrt{1-\alpha^2}\right) + \left(\frac{3+3a_2}{4} - a_1\right) \Delta E_\textrm{\tiny hfs} +\left(\frac{m_\textrm{\tiny red}\alpha^2}{\sqrt{1-\alpha^2}}+\frac{3-3b_1}{4} \Delta E_\textrm{\tiny hfs}\right) \widetilde V (R) \,.
\end{equation}

\begin{figure}[h]
  \begin{tikzpicture}
\begin{axis}[scale=3, 
          xmin=0,xmax=2,
          ymin=0,ymax=1.3,
          x=1cm,y=1cm,
          hide axis,
          enlargelimits=false
          ]
\addplot[thick,domain=0:2,samples=100]  {1+0.74*0.25*pow(e,-x)/(1+x)};
\end{axis}
\node[below] at (0.3,0) {\small$R_0$};
\node[below] at (6,0) {\small$R$};
\node at (4.8,3.3) {\small $\Delta E_\text{\tiny hfs}^\textrm{\text{point mass}}/\Delta E_\text{\tiny hfs}$};
\node[left] at (-0.15,0) {\small $0$};
\node[left] at (-0.15,3) {\small $1$};
\draw[->,>=latex] (0,-0.4) -- (0,3.9);
\draw[->,>=latex] (-0.2,0) -- (6,0);
\draw[dashed] (-0.2,3) -- (6,3);
\end{tikzpicture}
\caption{The behavior of the function $\Delta E_\text{\tiny hfs}^\textrm{\text{point mass}}/\Delta E_\text{\tiny hfs}$ is depicted above. The function decreases as the distance from the point mass increases and asymptotically approaches the value of $1$ at large distances.
This way, far away from the point mass, we recover standard QED and any local effect of the area metric disappears.
Note that the minimum distance from the point mass $R_0$ could represent either a real boundary of the system, e.g., the radius of Earth $R_\oplus$ for experiments around Earth, or the distance below which the linear approximation ($E\ll 1$) cease to be valid.}
  \label{fig:hfs-energy}
\end{figure}

\newpage
\section{Conclusions} \label{sec:conclusions}
\new{The observation of small effects often requires, not in principle but in practice, prior theoretical predictions about where and when such effects occur. 
The effects resulting from a refinement of Maxwell theory to the most general electrodynamics with still linear field equations, as they are studied in this article, present an extreme and thus instructive example in this regard, since they are purely due to a refinement of the underlying spacetime geometry. The where and when of the ensuing effects thus becomes a question of precisely predicting this refined background geometry by virtue of a correspondingly refined gravitational action, while the actual effect requires a quantum field theoretic treatment of the electrodynamics under these circumstances. Indeed, only the interplay of suitable dynamics for the refined spacetime on the one hand, with the actual quantum electrodynamics formulated thereon on the other hand, provides a sufficient theoretical setting to underpin a promising search for  quantum electrodynamical observable effects. 

The method employed here for general linear electrodynamics, however, applies quite generally: Any diffeomorphism-invariant local matter field action satisfying three elementary causality and energy conditions can be supplemented, by way of the constructive gravitational closure mechanism, with gravitational field equations for the very background geometry they employ. If one is then also able to quantize the matter theory and show its renormalizability, it can be used to derive observable effects on any fixed background that solves the derived gravitational field equations. 

Our quantum field theoretical analysis here focused on the phenomenologically most likely situation where the area metric geometry differs only slightly from a metric background and also slowly enough to allow calculate sufficiently localizable quantum effects for the technical fiction of a locally flat area metric for. Our proof of the renormalizability of the theory in gauge-invariant fashion to any loop order then showed that, under these circumstances, general linear electrodynamics can be employed as a fundamental theory whose physical parameters are determined by finitely many measurements. 

With respect to scattering amplitudes, we found that area metric perturbations make the differential cross sections of both Bhabha and electron-positron into fermion-antifermion scattering generically dependent on the azimuthal angle $\varphi$ with respect to the center-of-momentum axis. For the particular area metric spacetime generated by a gravitating point mass source, however, the amplitudes reduced to the standard QED amplitudes times a radial factor while the said angular dependence completely disappears. This illustrates the importance of having a specific prediction for what effects one may expect in a concrete experimental setting.
Furthermore does the scattering of an electron in an external magnetic field at one loop reveal a qualitatively new interaction~(\ref{potential-correction}) between the electron spin and the external magnetic field.
Because of the special structure of the area metric deviations from a flat metric, we were able to compute the anomalous magnetic moment of the electron up to every loop order at which it is known in standard QED, yielding the particular combination (\ref{alpha-tilde-value}) of the fine structure constant $\alpha$ and the double trace of the eleven-parameter perturbation $E^{abcd}$.
A single further result of comparable precision, relating the same two quantities or containing either one of them as the only experimental unknown, will thus allow to determine both the fine structure constant and the double trace of the non-metric perturbation.
In the very specific situation in which the area metric is sourced by a point mass, the new interaction~(\ref{potential-correction}) between the electron spin and the external magnetic field disappears, whereas the anomalous magnetic moment~(\ref{AMM-pm}) picks up a dependence on the distance from the point mass.

For the bound states of atomic hydrogen, we found that only three out of the eleven perturbative degrees of freedom affect the hyperfine structure, and thus the astrophysically crucially relevant $21.1\,\textrm{cm}$ line.
While two of these source a splitting of the triplet states, the remaining one, namely again the double trace $E^{ab}{}_{ab}$, simultaneously controls two effects: it shifts the ionization energy (\ref{Eionization}) as well as the hyperfine transition energy (\ref{stretching}).
While the ionization energy yields the same functional connection of $\alpha$ and $E^{ab}{}_{ab}$ as the anomalous magnetic moment---and hence does not contribute new information---any sufficiently precise measurement of the modified hyperfine structure will, in principle, provide sufficient new information to determine the fine structure constant and the double trace of a non-metric perturbation.
This would therefore amount to a direct quantum test of a non-metric spacetime structure.
When the area-metric perturbations are sourced by a point mass, the spitting of the triplet does not happen, while the shift in the ionization energy and the stretch of the hyperfine transition energy become dependent on the distance from a point mass.

The pivotal relevance of the hyperfine transition for astrophysical measurements shout out loud for a solution of the gravitational closure equations under the assumption of a spatially homogeneous and isotropic area metric background. This solution of the closure equations by way of symmetry-reduction, as opposed to perturbation theory is well under way. From the resulting refined Friedmann equations for two independent scale factors one then immediately obtains, with the results of the present paper, a correspondingly refined relation between the hyperfine transition energies of hydrogen at different stages of the cosmological evolution, their respective redshifts and indeed the pertinent observational raw data. If indeed the spacetime geometry is area metrically refined, this would reveal a systematic error in the interpretation of astrophysical observations based on hyperfine transition emissions. }

\acknowledgments{The authors thank Marcus Werner and Nadine Stritzelberger for insightful comments and suggestions. FPS thanks Marcus Werner and the Yukawa Institute for Theoretical Physics for their generous invitation for a two-month research visit in September, during which part of this work was completed.}



\appendix
\section*{Appendix: Conventions and notations for special functions and integrals}\label{sec_specialfunctions}
We frequently use an array of special functions and integrals in section \ref{sec:hydrogen}, in connection with the eigenstates $|n\kappa ms\rangle$ of the exactly diagonalizable part of the Hamiltonian of the relativistic hydrogen atom. These are collected in this appendix for the reader's convenience. 

In subsection \ref{subsec:exact}, the normalization constant used in the general expression for the eigenstates $|n\kappa m s\rangle$ is given by
\begin{equation}
\mathcal{N} := \frac{1}{2w^{K+1}}\sqrt{\frac{(w-\kappa)(w+W)n!}{w\Gamma(2K + 1 + n)}}\,,
\end{equation}
using the shorthands $W$, $w$ and $K$ introduced in that subsection for expressions in terms of the quantum numbers $n$, $\kappa$, $m$, $s$.
Moreover, the generalized Laguerre polynomials $L_n^\lambda(z)$ used there are the solutions of Laguerre's equation
\begin{equation}
\left(z\frac{\dd^2}{\dd z^2} + (\lambda + 1 - z)\frac{\dd}{\dd z} + n\right)L^\lambda_n(z) = 0\,,
\end{equation}
which read explicitly \cite{WFLaguerre}
\begin{equation}
L^\lambda_n(z) = \sum_{k=0}^n \frac{\Gamma(\lambda + n + 1)}{\Gamma(\lambda + k + 1)}\frac{(-z)^k}{(n-k)!k!}\,.
\end{equation}
The spinor spherical harmonics $\mathcal{Y}_\kappa^m(\vect{n})$ are defined by
\begin{equation}
\mathcal{Y}_\kappa^m (\vect{n}) := \frac{1}{\sqrt{2d}}\begin{pmatrix} -v\sqrt{d-vm}\,Y_{d-\frac{1}{2}}^{m-\frac{1}{2}}(\vect{n}) \\[2ex] \sqrt{d+vm}\,Y_{d-\frac{1}{2}}^{m+\frac{1}{2}}(\vect{n}) \end{pmatrix}\,,\qquad v := \text{sign } \kappa\,,\qquad d := \left|\kappa + \frac{1}{2}\right|
\end{equation}
with the spherical harmonics  \cite{WFSpherical}
\begin{equation}
Y^m_l(\theta,\varphi) := \sqrt{\frac{(2l+1)(l-m)!}{4\pi(l+m)!}}P^m_l(\cos \theta)\ee^{\ii m\varphi}\,.
\end{equation}

In subsection \ref{subsec:nonbiref}, we use that the spinor spherical harmonics satisfy the identity \cite{Ycalculus} 
\begin{equation}
\sigma^\mu\mathcal{Y}_\kappa^m = \frac{-1}{2\kappa + 1}\sum_{m' = -|\kappa|+\frac{1}{2}}^{|\kappa|+\frac{1}{2}}\mathcal{Y}_\kappa^{m'}\left(\sigma_\kappa^\mu\right)_{m'm} + \frac{1}{|2\kappa + 1|} \sum_{m' = -|\kappa'|+\frac{1}{2}}^{|\kappa'|+\frac{1}{2}}\mathcal{Y}_{\kappa'}^{m'}\left(\tau_\kappa^\mu\right)_{m'm}\,,
\end{equation}
where the matrices $\sigma^\mu_\kappa$ and $\tau^\mu_\kappa$ are given by
\begin{align}
\left(\sigma_\kappa^1\right)_{m'm} :={}& \sqrt{\kappa^2-\bar{m}^2}\left(\delta_{m',m-1} + \delta_{m',m+1}\right)\\
\left(\sigma_\kappa^2\right)_{m'm} :={}& \sqrt{\kappa^2-\bar{m}^2}\left(\ii\delta_{m',m-1} - \ii\delta_{m',m+1}\right)\\
\left(\sigma_\kappa^3\right)_{m'm} :={}& 2\bar{m}\delta_{m'm}\\[3ex]
\left(\tau_\kappa^1\right)_{m'm} :={}& -v\sqrt{(\kappa' + \bar{m})(\bar{m} - \kappa)}\delta_{m',m-1} + v\sqrt{(\kappa + \bar{m})(\bar{m} - \kappa')}\delta_{m',m+1}\\
\left(\tau_\kappa^2\right)_{m'm} :={}& -v\ii\sqrt{(\kappa' + \bar{m})(\bar{m} - \kappa)}\delta_{m',m-1} - v\ii\sqrt{(\kappa + \bar{m})(\bar{m} - \kappa')}\delta_{m',m+1}\\
\left(\tau_\kappa^3\right)_{m'm} :={}& -2\sqrt{-\left(\kappa + \frac{1}{2}\right)\left(\kappa' + \frac{1}{2}\right) - \bar{m}^2}\delta_{m'm}
\end{align}
and $\kappa' := -(\kappa + 1)$, $\bar m := (m'+m)/2$. Note that from the explicit expressions above, one can directly read off that $\sigma_{\pm 1}^\mu = \sigma^\mu$, where the $\sigma^\mu$ are the standard Pauli matrices. Furthermore, the $\sigma_\kappa$ and $\tau_\kappa$ matrices satisfy a series of identities, of which the most important for the purposes of this paper is
\begin{equation}
\left[\sigma_\kappa^\mu, \sigma_\kappa^\nu\right] = 2\ii \varepsilon^{\mu\nu}{}_\rho\sigma_\kappa^\rho\,.
\end{equation}

Throughout subsections \ref{subsec:nonbiref}, \ref{subsec:lambda} and \ref{subsec:lambdamu}, we use the following shorthand symbols for repeatedly appearing radial integrals, in order to lighten the notation.
\begin{equation}
\Ink{a}{ij} := \left(\frac{w'+\kappa'}{n'}\right)^i \left(\frac{w+\kappa}{n}\right)^j \int_0^\infty \dd y\, \ee^{-\frac{w'+w}{2w'w}y} y^{K'+K-a} L_{n'-i}^{2K'}\left(\frac{y}{w'}\right)L_{n-j}^{2K}\left(\frac{y}{w}\right)
\end{equation}
Note that this is well-defined only for $n'-i \geq 0$ and $n-j \geq 0$. In all other cases $\Ink{a}{ij} := 0$. Note also that, since $L_0^\lambda = 1$,
\begin{equation}
\Ink{a}{n'n} = \left(\frac{w'+\kappa'}{n'}\right)^{n'} \left(\frac{w+\kappa}{n}\right)^n \left(\frac{w'+w}{2w'w}\right)^{-(K'+K-a+1)} \Gamma\left(K'+K-a+1\right)\,.
\end{equation}
Furthermore, in the second order corrections, there appear the functions $A_{n\kappa}$, $B_{n\kappa}$, $C_{n\kappa}$, which are defined as
\begin{align*}
A_{n\kappa}(\kappa'') := \sum_{n''}&\frac{w''w}{wW''-w''W}\mathcal{N}^2\mathcal{N''}^2\notag\\
&\times \Big[\left(w-W + w'' - W''\right)\left(\tIoo{n''\kappa''}{n\kappa} - \tIii{n''\kappa''}{n\kappa}\right) \notag\\
&\hspace{2cm} + \left(w-W-w''+W''\right)\left(\tIoi{n''\kappa''}{n\kappa}-\tIio{n''\kappa''}{n\kappa}\right)\Big]^2\,,\\
B_{n\kappa}(\kappa'') := \sum_{n''}&\frac{w''w}{wW''-w''W}\mathcal{N}^2\mathcal{N''}^2 \notag\\
&\times \Big[\left(w-W + w'' - W''\right)\left(\tIoo{n''\kappa''}{n\kappa} - \tIii{n''\kappa''}{n\kappa}\right) \notag\\
&\hspace{2cm}+ \left(w-W-w''+W''\right)\left(\tIoi{n''\kappa''}{n\kappa}-\tIio{n''\kappa''}{n\kappa}\right)\Big]\notag\\
&\times \Big[\left(\alpha+\frac{(w''-W'')(w-W)}{\alpha}\right)\left(\iIoo{n''\kappa''}{n\kappa}+\iIii{n''\kappa''}{n\kappa}\right) \notag\\
&\hspace{2cm} - \left(\alpha - \frac{(w''-W'')(w-W)}{\alpha}\right)\left(\iIoi{n''\kappa''}{n\kappa}+\iIio{n''\kappa''}{n\kappa}\right)\Big]\,,\\
C_{n\kappa}(\kappa'') := \sum_{n''}&\frac{w''w}{wW''-w''W}\mathcal{N}^2\mathcal{N''}^2\sqrt{\frac{w+\kappa}{w-\kappa}} \notag\\
&\times \Big[\left(w-W + w'' - W''\right)\left(\tIoo{n''\kappa''}{n,\,-\kappa} - \tIii{n''\kappa''}{n,\,-\kappa}\right) \notag\\
&\hspace{2cm}+ \left(w-W-w''+W''\right)\left(\tIoi{n''\kappa''}{n,\,-\kappa}-\tIio{n''\kappa''}{n,\,-\kappa}\right)\Big]\notag\\
&\times \Big[\left(w-W+w''-W''\right)\left(\iIoo{n''\kappa''}{n\kappa}-\iIii{n''\kappa''}{n\kappa}\right) \notag\\
&\hspace{2cm}+ \left(w-W-w''+W''\right)\left(\iIoi{n''\kappa''}{n\kappa}-\iIio{n''\kappa''}{n\kappa}\right)\Big]\,,
\end{align*}
where the summation over $n''$ runs over $\mathbb{N}_0$ if $|\kappa''|\neq|\kappa|$ and over $\mathbb{N}_0\backslash\lbrace n\rbrace$ if $|\kappa''| = |\kappa|$.


\let\oldaddcontentsline\addcontentsline
\renewcommand{\addcontentsline}[3]{}
\bibliography{biblio,ext}{}
\let\addcontentsline\oldaddcontentsline

\end{document}